\documentclass[prb,aps,twocolumn,superscriptaddress]{revtex4-2}
\usepackage{amsmath}
\usepackage{subfigure}
\usepackage{color}
\usepackage{bbm}
\usepackage{amssymb}
\usepackage{epsfig}
\usepackage{multirow}
\usepackage{amsbsy}
\usepackage{array}
\usepackage{diagbox}
\usepackage{bm}
\usepackage{extarrows}
\usepackage{graphicx}
\usepackage{appendix}
\usepackage{txfonts}
\usepackage{lipsum} 
\usepackage{tablefootnote}
\usepackage{cancel}
\usepackage{soul}
\graphicspath{{figures/}}
\allowdisplaybreaks[4]
\newcommand{\tabincell}[2]{\begin{tabular}{@{}#1@{}}#2\end{tabular}}

\usepackage[colorlinks=true,linkcolor=blue,citecolor=blue,urlcolor=blue,bookmarks=false]{hyperref}

\begin{document}
	
\title{Interplay of Zeeman field, Rashba spin-orbit interaction, and superconductivity: spin susceptibility}

\author{Chen Pang}
\affiliation {Institute of Physics, Chinese Academy of Sciences, Beijing 100190, China}
\affiliation{School of Physical Sciences, University of Chinese Academy of Sciences, Beijing 100190, China}	

\author{Yi Zhou}
\email{yizhou@iphy.ac.cn}
\affiliation {Institute of Physics, Chinese Academy of Sciences, Beijing 100190, China}

\date{\today}
	
\begin{abstract}
We present a self-consistent theory to calculate the static and uniform spin susceptibility in superconductors under simultaneous Zeeman magnetic fields and Rashba-type spin-orbit coupling (SOC). Employing a single-band Bogoliubov-de Gennes Hamiltonian, we solve the gap equation for both conventional $s$-wave spin-singlet and six representative $p$-wave spin-triplet pairing states, categorized into opposite-spin-pairing (OSP) and equal-spin-pairing (ESP) classes. The Kubo formula, decomposed into intra- and interband particle-hole and particle-particle channels, provides two key constraints: at zero temperature, only particle-particle terms contribute, while at the critical temperature $T_c$, only particle-hole terms remain, ensuring $\chi(T_c^{-}) = \chi_N$ for continuous phase transitions. For $s$-wave pairing, a Zeeman field reduces $T_c$, whereas Rashba SOC preserves $T_c$ but yields a residual zero temperature spin susceptibility $\chi(0)$ which approaches $2\chi_N/3$ in the strong SOC limit; combined fields create a Bogoliubov Fermi surface, resulting in a kink in $\chi(0)$. In contrast, $p$-wave states exhibit strong anisotropy: OSP states mimic spin-singlet pairing behavior for parallel Zeeman fields and ESP for transverse ones, while ESP states show the opposite, with Rashba SOC potentially changing the quasiparticle nodal structure, lowering $T_c$, or causing $\chi_{zz}(0)$ divergences. This framework offers quantitative benchmarks for Knight-shift experiments in non-centrosymmetric superconductors like A$_2$Cr$_3$As$_3$ (A = Na, K, Rb, and Cs), enabling diagnostics to disentangle pairing symmetry, SOC strength, and Zeeman effects.
\end{abstract}
	
\maketitle
\section{Introduction}

Superconductivity, a quintessential demonstration of macroscopic quantum coherence, has been a cornerstone of modern physics since its discovery by H. Kamerlingh Onnes in 1911. A key aspect of this phenomenon is how superconductors respond to external electric and magnetic fields, characterized by zero electrical resistance and perfect diamagnetism, known as the Meissner-Ochsenfeld effect~\cite{tinkham}. These responses not only define superconductivity but also serve as critical tools for investigating the pairing symmetry of Cooper pairs, which is essential for advancing our understanding of the field.

To probe pairing symmetry, nuclear magnetic resonance (NMR) techniques, such as the Knight shift and the spin-lattice relaxation rate $1/T_1$, are particularly powerful, as they reveal spin susceptibility and dynamics in superconducting states. For example, in an $s$-wave spin-singlet pairing, the Knight shift becomes isotropic and decreases below the critical temperature $T_c$, vanishing at zero temperature~\cite{BCS1957}, while a spin-triplet pairing shows anisotropic behavior depending on the magnetic field direction~\cite{Leggett1975}. Similarly, the spin-lattice relaxation rate $1/T_1$ exhibits a Hebel-Slichter coherence peak at $T = T_c$ for sign-unchanged $s$-wave pairings, but this peak is diminished or absent in sign-changed pairings like $p$-wave states~\cite{Leggett1975,AndersonandMorel1961,AndersonandBrinkman1973,Balian1963}. These NMR observations highlight the sensitivity of spin dynamics to pairing symmetry.

However, theoretical analyses often rely on perturbation theory, treating external fields as small disturbances to derive linear response properties. In NMR experiments, however, the required magnetic fields are typically strong, with energy scales comparable to or larger than the superconducting gap. This necessitates recalculating the ground state under the field's influence rather than approximating it as a perturbation, ensuring a more accurate assessment of the system's responses.

Complicating this picture, the spin-orbit interaction, often present due to broken inversion symmetry in crystals, further modifies the superconducting ground state. This interaction, exemplified by the Dresselhaus or Rashba effects~\cite{Dresselhaus,Rashiba1960}, arises from structural asymmetries and can be modeled within symmetry-based band theory~\cite{dresselhaus2007group}. Using $\mathbf{k}\cdot\mathbf{p}$ perturbation theory, the spin-orbit coupling (SOC) is typically expressed as $H_{\text{SO}} = \mathbf{g}_{\mathbf{k}}\cdot\hat{\sigma}$, with $\mathbf{g}_{\mathbf{k}}$ containing odd-powered momentum terms. In this work, we adopt a simplified form, $H_{\text{SO}} = g\mathbf{k}\cdot\hat{\sigma}$,  which is a 3D extention to  Rashba SOC and has been shown to capture essential effects~\cite{bauer2012non,Yip14,Samohkin15,Smidman17}.

Building on early experimental and theoretical insights, the interplay between magnetic fields and spin-orbit coupling has been a focal point since the late 1950s. NMR studies on BCS superconductors revealed minimal changes in the Knight shift for heavy elements like Hg~\cite{Reif1957} and Sn~\cite{AndroesKnight1959}, prompting key theoretical works by Ferrell~\cite{Ferrell1959}, Martin and Kadanoff~\cite{Kadanoff1959}, Schrieffer~\cite{Schrieffer1959}, and Anderson~\cite{AndersonKnight1959}. Anderson attributed these observations to strong spin-orbit scattering disrupting spin conservation~\cite{AndersonKnight1959,Andersontheoryofdirty1959}. Subsequent research, including Appel's analysis of SOC in non-transition metals~\cite{Appel1965}, Shiba's work on surface effects~\cite{shiba1976effect}, and studies on magnetic field influences~\cite{zhogolev1972magnetic}, has expanded this understanding. More recently, attention has shifted to superconductors with broken inversion symmetry~\cite{Frigeri2004,Frigeri_2004,Samokhin2007,Edelstein2008,ZHOU2017208,Triplet2021}.

Motivated by recent experimental observations in a new family of superconductors, A$_2$Cr$_3$As$_3$ (A = Na, K, Rb, and Cs)~\cite{Bao15,Tang15_1,Tang15_2,Mu18,Triplet2021}, the current authors have theoretically investigated the effects of an external Zeeman field and Rashba spin-orbit interactions on superconductivity, particularly their influence on the superconducting transition temperature and quasiparticle excitations~\cite{pang2025}.  In this context, attention is devoted to investigating the temperature-dependent spin susceptibility of $s$- and multiple $p$-wave pairing states in the presence of varying Zeeman fields and Rashba SOC.

The remainder of the paper is structured as follows. Section II presents the model Hamiltonian and the Kubo formula for spin susceptibility, along with a brief discussion of results and interpretations. Section III examines $s$-wave spin-singlet pairing superconductors, with a focus on spin susceptibility. Section IV explores various $p$-wave spin-triplet pairing superconductors, highlighting their distinct responses to the Zeeman field and/or Rashba SOC. Section V provides a summary and conclusions.

\section{Theoretical formulation}\label{sec:model}

As an extension of our previous work~\cite{pang2025}, we employ a standard mean-field Hamiltonian to describe a superconducting system. This Hamiltonian comprises an unpaired component $H_0$ and a paired component $H_{\text{SC}}$, and is expressed as:
\begin{equation}\label{eq:H}
H = H_0 + H_{\text{SC}}.
\end{equation}
The unpaired part $H_0$ is given by:
\begin{equation}
H_0 = \sum_{\mathbf{k},\alpha,\beta} c_{\mathbf{k},\alpha}^{\dagger} [H_0(\mathbf{k})]_{\alpha\beta} c_{\mathbf{k},\beta},
\end{equation}
where $\alpha, \beta = \uparrow, \downarrow$ represent spin indices, $c_{\mathbf{k},\alpha}^{\dagger}$ ($c_{\mathbf{k},\alpha}$) denotes the creation (annihilation) operator for an electron with wave vector $\mathbf{k}$ and spin $\alpha$, and $H_0(\mathbf{k})$ is a $2\times2$ Hermitian matrix.

The paired part $H_{\text{SC}}$, which accounts for the superconducting pairing, takes the form:
\begin{equation}\label{eq:HSC}
H_{\text{SC}} = \frac{1}{2} \sum_{\mathbf{k},\alpha,\beta} \left[ c_{\mathbf{k},\alpha}^{\dagger} \Delta_{\alpha\beta}(\mathbf{k}) c_{-\mathbf{k},\beta}^{\dagger} + \text{h.c.} \right].
\end{equation}
Here, the pairing function $\Delta(\mathbf{k})$ is a $2\times2$ matrix characterized by an isotropic component $\Delta$ and a d-vector $\mathbf{d}(\mathbf{k})$, and is defined as:
\begin{equation} \label{dels}
\Delta(\mathbf{k}) = i \left[ \Delta \sigma_0 + \mathbf{d}(\mathbf{k}) \cdot \hat{\sigma} \right] \sigma_y,
\end{equation}
where $\sigma_0$ is the identity matrix, and $\hat{\sigma}$ is a vector comprising the three Pauli matrices~\cite{Balian1963}.

In the normal state, the unpaired Hamiltonian $H_0$ governs the system, incorporating Rashba spin-orbit coupling with strength $g$ and an external Zeeman field $\mathbf{H}$, and reads
\begin{equation} 
H_0(\mathbf{k}) = \xi_{\mathbf{k}} \sigma_0 + \mu_B \mathbf{H} \cdot \hat{\sigma} + g \mathbf{k} \cdot \hat{\sigma},
\label{hamilnormal}
\end{equation} 
where $\xi_{\mathbf{k}}$ is the electron energy relative to the chemical potential $\mu$, and $\mu_B$ is the Bohr magneton. 
For the numerical calculations, we adopt a parabolic dispersion \( \xi_k = \frac{\hbar^2 k^2}{2m} - \mu \) with a spherical Fermi surface. 
The eigenvalues of $H_0(\mathbf{k})$ are given by:
\begin{subequations}\label{eq:eigenH}
\begin{equation}\label{eq:xi0}
\xi_{\mathbf{k}\pm} = \xi_{\mathbf{k}} \pm \left| \mu_B \mathbf{H} + g \mathbf{k} \right|,
\end{equation}
with the corresponding eigenstates expressed as:
\begin{equation}\label{eigennewp}
\left| \mathbf{k}, + \right\rangle = \left| \mathbf{k} \right\rangle \otimes \begin{pmatrix}
\cos \frac{\Theta_{\mathbf{k}}}{2} \\
\sin \frac{\Theta_{\mathbf{k}}}{2} \mathrm{e}^{i \Phi_{\mathbf{k}}}
\end{pmatrix}, \quad
\left| \mathbf{k}, - \right\rangle = \left| \mathbf{k} \right\rangle \otimes \begin{pmatrix}
\sin \frac{\Theta_{\mathbf{k}}}{2} \\
- \cos \frac{\Theta_{\mathbf{k}}}{2} \mathrm{e}^{i \Phi_{\mathbf{k}}}
\end{pmatrix}.
\end{equation}
\end{subequations}
The angles $\Theta_{\mathbf{k}}$ and $\Phi_{\mathbf{k}}$, which define the direction of the vector $\mu_B \mathbf{H} + g \mathbf{k}$, are specified as:
\begin{equation}
\frac{\mu_B \mathbf{H} + g \mathbf{k}}{\left| \mu_B \mathbf{H} + g \mathbf{k} \right|} = (\sin \Theta_{\mathbf{k}} \cos \Phi_{\mathbf{k}}, \sin \Theta_{\mathbf{k}} \sin \Phi_{\mathbf{k}}, \cos \Theta_{\mathbf{k}}).
\end{equation}
Explicitly, these angles are given by:
\begin{equation}\label{eq:Thetak}
\begin{split}
\cos\Theta_{\mathbf{k}} &= \frac{\mu_B H_z + g |\mathbf{k}| \cos \theta_{\mathbf{k}}}{\left| \mu_B \mathbf{H} + g \mathbf{k} \right|}, \\
\tan \Phi_{\mathbf{k}} &= \frac{\mu_B H_y + g |\mathbf{k}| \sin \theta_{\mathbf{k}} \sin \varphi_{\mathbf{k}}}{\mu_B H_x + g |\mathbf{k}| \sin \theta_{\mathbf{k}} \cos \varphi_{\mathbf{k}}},
\end{split}
\end{equation}
where $\theta_{\mathbf{k}}$ and $\varphi_{\mathbf{k}}$ are the polar and azimuthal angles of the wave vector $\mathbf{k}$, respectively~\cite{pang2025}.

\subsection{Pairing interaction and self-consistent equation}

The formation of Cooper pairs in a superconductor is driven by a four-fermion interaction, described by the Hamiltonian:
\begin{equation}\label{eq:Hint}
H_{\text{int}} = -\frac{1}{2} \sum_{\mathbf{k},\mathbf{k}^{\prime},\alpha,\beta,\alpha^{\prime},\beta^{\prime}} V_{\beta\alpha\alpha^{\prime}\beta^{\prime}}(\mathbf{k},\mathbf{k}^{\prime}) c_{\mathbf{k},\alpha}^\dagger c_{-\mathbf{k},\beta}^\dagger c_{\mathbf{k}^{\prime},\alpha^{\prime}} c_{-\mathbf{k}^{\prime},\beta^{\prime}},
\end{equation}
where $V_{\beta\alpha\alpha^{\prime}\beta^{\prime}}(\mathbf{k},\mathbf{k}^{\prime})$ is the pairing potential matrix element, and $\alpha, \beta, \alpha^{\prime}, \beta^{\prime} = \uparrow, \downarrow$ are spin indices.

Assuming a spherical Fermi surface and rotational symmetry, the matrix element $V(\mathbf{k}, \mathbf{k}')$ can be expanded as~\cite{Leggett1975,Sigrist1991}:
\begin{equation}\label{eq:Vl}
V(\mathbf{k},\mathbf{k}') = \sum_{l=0}^{\infty} (2l+1) V_l P_l \left( \cos \theta_{\mathbf{k}, \mathbf{k}'} \right),
\end{equation}
where $\theta_{\mathbf{k},\mathbf{k}'}$ is the angle between the wave vectors $\mathbf{k}$ and $\mathbf{k}'$, and $P_l(x)$ are Legendre polynomials. A positive $V_l$ indicates an attractive interaction; specifically, the $l=0$ channel with $V_0 > 0$ corresponds to $s$-wave pairing, while the $l=1$ channel with $V_1 > 0$ leads to $p$-wave pairing.

Our model is defined by the pairing potential $V(\mathbf{k},\mathbf{k}')$ or its channel-specific parameters $V_l$ from Eq.~\eqref{eq:Vl}. Applying a mean-field decomposition to the interaction $H_{\text{int}}$ in the pairing channel yields the superconducting Hamiltonian $H_{\text{SC}}$ as given in Eq.~\eqref{eq:HSC}. The pairing function $\Delta(\mathbf{k})$ is then determined self-consistently using the following equation~\cite{Sigrist1991}:
\begin{equation}\label{gapeqwhole}    
\Delta_{\alpha\beta}(\mathbf{k}) = -\sum_{\mathbf{k}^{\prime},\alpha^{\prime},\beta^{\prime}} V_{\beta\alpha\alpha^{\prime}\beta^{\prime}}(\mathbf{k},\mathbf{k}^{\prime}) \langle c_{\mathbf{k}^{\prime},\alpha^{\prime}} c_{-\mathbf{k}^{\prime},\beta^{\prime}} \rangle.
\end{equation}

To solve this, we employ the Bogoliubov transformation, detailed in Appendix~\ref{app:BTM}, which is expressed as:
\begin{equation}\label{eq:BT}
c_{\mathbf{k},\alpha} = \sum_{s=\pm} \left( u^{\alpha s}_{\mathbf{k}} \psi_{\mathbf{k},s} + v_{\mathbf{k}}^{\alpha s} \psi_{-\mathbf{k},s}^{\dagger} \right).
\end{equation}
This transformation diagonalizes the mean-field Hamiltonian $H$ from Eq.~\eqref{eq:H}, allowing us to iteratively solve Eq.~\eqref{gapeqwhole}. Here, $\psi_{\mathbf{k},s}$ and $\psi_{\mathbf{k},s}^{\dagger}$ are Bogoliubov quasiparticle operators, with $s = \pm$ denoting the quasiparticle band index. This index differs from that in the normal state (as defined in Eq.~\eqref{eigennewp}). Additionally, the quasiparticle energy in the superconducting state is denoted as $E_{\mathbf{k}\pm}$, to distinguish it from the normal state energy in Eq.~\eqref{eq:xi0}.

\subsection{Spin susceptibility in superconducting states}

After self-consistently solving for the pairing function $\Delta(\mathbf{k})$ from Eq.~\eqref{gapeqwhole}, the temperature-dependent spin susceptibility $\chi_{\mu\nu}(T)$ is calculated using the Kubo formula, with details provided in Appendix~\ref{app:Kubo}. For clarity, $\chi_{\mu\nu}$ is decomposed into different channels:
\begin{subequations}\label{eq:chikd}
\begin{equation}
\chi_{\mu\nu} = \chi_{\mu\nu}^{ph} + \chi_{\mu\nu}^{pp},
\end{equation}
where $\chi_{\mu\nu}^{ph}$ and $\chi_{\mu\nu}^{pp}$ represent contributions from particle-hole ($ph$) and particle-particle/hole-hole ($pp$) processes, respectively. These can be further broken down as
\begin{equation}
\chi_{\mu\nu}^{ph} = \chi_{\mu\nu}^{ph+} + \chi_{\mu\nu}^{ph-} \quad \text{and} \quad \chi_{\mu\nu}^{pp} = \chi_{\mu\nu}^{pp+} + \chi_{\mu\nu}^{pp-},
\end{equation}
with the superscripts $+$ and $-$ indicating intraband and interband contributions.

Then we focus on the diagonal components, which are explicitly given by:
\begin{widetext}
\begin{eqnarray}
\chi_{\mu\mu}^{ph+}(T) & = & -\mu_B^2 \sum_{\mathbf{k}} \sum_{s=\pm} \left[ (u_{\mathbf{k}}^\dagger \sigma_\mu u_{\mathbf{k}})^{ss} - (v_{-\mathbf{k}}^\dagger \sigma_\mu v_{-\mathbf{k}})^{ss} \right]^2 \frac{df(E_{\mathbf{k}s})}{dE_{\mathbf{k}s}}, \label{eq:chi_k1} \\
\chi_{\mu\mu}^{ph-}(T) & = & -2 \mu_B^2 \sum_{\mathbf{k}} \left| (u_{\mathbf{k}}^\dagger \sigma_\mu u_{\mathbf{k}})^{+-} - (v_{-\mathbf{k}}^\dagger \sigma_\mu v_{-\mathbf{k}})^{-+} \right|^2 \frac{f(E_{\mathbf{k}+}) - f(E_{\mathbf{k}-})}{E_{\mathbf{k}+} - E_{\mathbf{k}-}}, \label{eq:chi_kph} \\
\chi_{\mu\mu}^{pp+}(T) & = & -\mu_B^2 \sum_{\mathbf{k}} \sum_{s=\pm} \left| (u_{\mathbf{k}}^\dagger \sigma_\mu v_{\mathbf{k}})^{ss} - (u_{-\mathbf{k}}^\dagger \sigma_\mu v_{-\mathbf{k}})^{ss} \right|^2 \frac{f(E_{\mathbf{k}s}) + f(E_{-\mathbf{k}s}) - 1}{E_{\mathbf{k}s} + E_{-\mathbf{k}s}}, \label{eq:chi_ppp} \\
\chi_{\mu\mu}^{pp-}(T) & = & -\mu_B^2 \sum_{\mathbf{k}} \sum_{s=\pm} \left| (u_{\mathbf{k}}^\dagger \sigma_\mu v_{\mathbf{k}})^{s\bar{s}} - (u_{-\mathbf{k}}^\dagger \sigma_\mu v_{-\mathbf{k}})^{\bar{s}s} \right|^2 \frac{f(E_{\mathbf{k}s}) + f(E_{-\mathbf{k}\bar{s}}) - 1}{E_{\mathbf{k}s} + E_{-\mathbf{k}\bar{s}}}, 
\label{eq:chi_kpp}
\end{eqnarray}
\end{widetext}
\end{subequations}
where $f(E)$ is the Fermi-Dirac distribution function, and $u_{\mathbf{k}}$ and $v_{\mathbf{k}}$ are Bogoliubov transformation matrices.
All the four components, $\chi_{\mu\mu}^{ph(pp)+(-)}$, are semi-positive definite, as $df(E)/dE < 0$ and $f(0) = 1/2$. Additionally, the coefficients satisfy a sum rule:
\begin{equation}\label{eq:sum}
\begin{split}
\sum_{s_1,s_2} & \left[ \left| (u_{\mathbf{k}}^\dagger \sigma_\mu u_{\mathbf{k}})^{s_1s_2} - (v_{-\mathbf{k}}^\dagger \sigma_\mu v_{-\mathbf{k}})^{s_2s_1} \right|^2 \right. \\
& \left. + \left| (u_{\mathbf{k}}^\dagger \sigma_\mu v_{\mathbf{k}})^{s_1s_2} - (u_{-\mathbf{k}}^\dagger \sigma_\mu v_{-\mathbf{k}})^{s_2s_1} \right|^2 \right] = 2,
\end{split}
\end{equation}
with the derivation provided in Appendix~\ref{app:BTM}.

\begin{figure}[tb]
\centering
\includegraphics[width=1.0\linewidth]{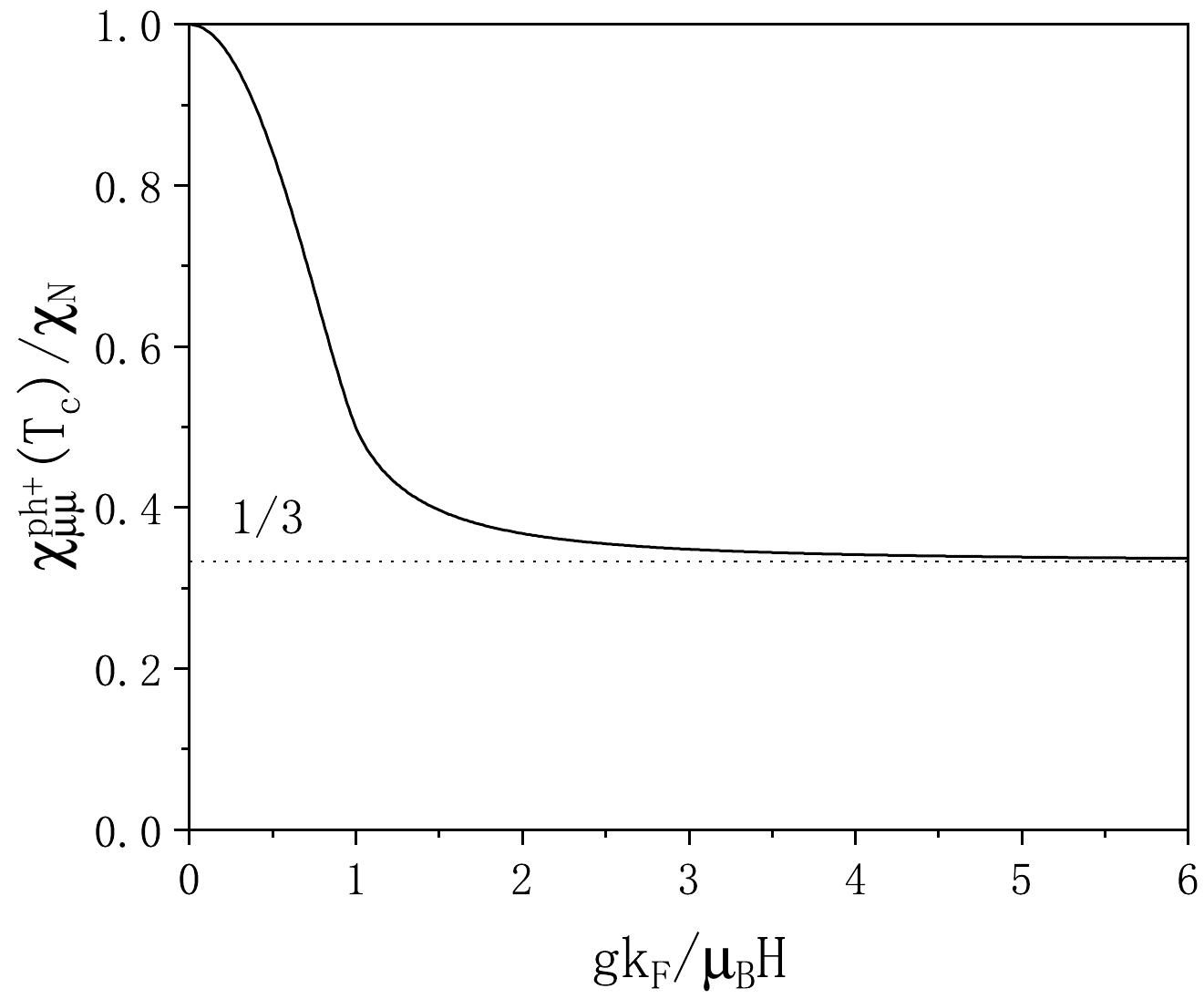} 
\caption{$\chi_{\mu\mu}^{ph+}(T_c) / \chi_N = 1 - \chi_{\mu\mu}^{ph-}(T_c) / \chi_N$ plotted against the ratio $g k_F / \mu_B H$, assuming a continuous superconducting phase transition.}
\label{chi1_chiN_magsoc}
\end{figure}

Before examining specific pairing states, we highlight some universal features of $\chi_{\mu\mu}$ and its decomposition.

\begin{enumerate}
\item{} At zero temperature ($T = 0$), only the $\chi_{\mu\mu}^{pp}$ component contributes, giving:
\begin{equation}\label{eq:xT0}
\chi_{\mu\mu}(T=0) = \mu_B^2 \sum_{\mathbf{k}} \sum_{s_1,s_2} \frac{\left| (u_{\mathbf{k}}^\dagger \sigma_\mu v_{\mathbf{k}})^{s_1s_2} - (u_{-\mathbf{k}}^\dagger \sigma_\mu v_{-\mathbf{k}})^{s_2s_1} \right|^2}{E_{\mathbf{k}s_1} + E_{-\mathbf{k}s_2}}.
\end{equation}
This follows from the vanishing or negative terms in Eq.~\eqref{eq:chikd} as $T \to 0$:
\begin{align*}
\lim_{T \to 0} \frac{df(E)}{dE} &= 0, \\
\lim_{T \to 0} \frac{f(E_{\mathbf{k}+}) - f(E_{\mathbf{k}-})}{E_{\mathbf{k}+} - E_{\mathbf{k}-}} &= 0, \\
\lim_{T \to 0} \frac{f(E_{\mathbf{k}s_1}) + f(E_{-\mathbf{k}s_2}) - 1}{E_{\mathbf{k}s_1} + E_{-\mathbf{k}s_2}} &= -\frac{1}{E_{\mathbf{k}s_1} + E_{-\mathbf{k}s_2}} < 0,
\end{align*}
leading to $\chi_{\mu\mu}^{ph}(T=0) = 0$ and $\chi_{\mu\mu}^{pp}(T=0) \geq 0$.

\item{} Near the critical temperature ($T \to T_c^{-}$), assuming a continuous phase transition, $\Delta(T = T_c^{-}) = 0$, and $\chi_{\mu\mu}(T)$ remains continuous at $T_c$, such that $\chi_{\mu\mu}(T_c^{+}) = \chi_{\mu\mu}(T_c^{-}) = \chi_{\mu\mu}(T_c)$. In this limit:
\begin{equation}
\chi_{\mu\mu}^{pp}(T = T_c^{-}) = 0 \quad \text{and} \quad \chi_{\mu\mu}(T = T_c) = \chi_{\mu\mu}^{ph}(T = T_c).
\end{equation}

Using Eqs.~\eqref{eq:Thetak}, \eqref{eq:xi0}, \eqref{eigennewp}, and \eqref{eq:chi_k1}, we derive the normalized components at $T_c$:
\begin{subequations}\label{eq:chi1_chin}
\begin{eqnarray}
\frac{\chi_{\mu\mu}^{ph+}(T_c)}{\chi_N} & = & \frac{1}{2} \int_0^\pi d\theta_{\mathbf{k}} \sin \theta_{\mathbf{k}} \cos^2 \Theta_{\mathbf{k}}, \\
\frac{\chi_{\mu\mu}^{ph-}(T_c)}{\chi_N} & = & \frac{1}{2} \int_0^\pi d\theta_{\mathbf{k}} \sin \theta_{\mathbf{k}} \sin^2 \Theta_{\mathbf{k}},
\end{eqnarray}
\end{subequations}
where the normal state spin susceptibility is $\chi_N = \chi_{\mu\mu}(T = T_c^{+})$.

The angle $\Theta_{\mathbf{k}}$, defined in Eq.~\eqref{eq:Thetak}, depends solely on the ratio $g k_F / \mu_B H$. Thus, both $\chi_{\mu\mu}^{ph+}(T_c)$ and $\chi_{\mu\mu}^{ph-}(T_c)$ are functions of this ratio:
\begin{itemize}
\item{} In the limit $g k_F / \mu_B H \to 0$, $\Theta_{\mathbf{k}} \to 0$, yielding $$\chi_{\mu\mu}^{ph+}(T_c) \to \chi_N \,\,\mbox{and}\,\,\chi_{\mu\mu}^{ph-}(T_c) \to 0.$$

\item{} In the opposite limit $g k_F / \mu_B H \to \infty$, $\Theta_{\mathbf{k}} \to \theta_{\mathbf{k}}$, resulting in $$\frac{\chi_{\mu\mu}^{ph+}(T_c)}{\chi_N} \to \frac{1}{3}\,\, \mbox{and}\,\, \frac{\chi_{\mu\mu}^{ph-}(T_c)}{\chi_N} \to \frac{2}{3}.$$
 This factor of $1/3$ originates from the spherical average of $\cos^2 \theta_k$ over the 3D Fermi surface, reflecting the geometric factor for a vector component when the spin is locked to the momentum direction.

\item{} For arbitrary ratios, $$\frac{\chi_{\mu\mu}^{ph+}(T_c)}{\chi_N} = 1 - \frac{\chi_{\mu\mu}^{ph-}(T_c)}{\chi_N}$$ is illustrated as a function of $g k_F / \mu_B H$ in Fig.~\ref{chi1_chiN_magsoc}.
\end{itemize}

\end{enumerate}

\section{$s$-wave pairing state}\label{sec:s-wave}

The simplest $s$-wave pairing model is defined by setting $$V_0 > 0\,\,\mbox{and}\,\,V_l = 0\,\,\mbox{for}\,\,l \geq 1,$$ as specified in Eq.~\eqref{eq:Vl}. This leads to an $s$-wave, spin-singlet pairing function, where $\Delta$ is finite and $\mathbf{d}(\mathbf{k}) = 0$ in Eq.~\eqref{dels}. The gap equation from Eq.~\eqref{gapeqwhole} simplifies to:
\begin{equation} \label{eq:sgap}
\Delta = -V_0 \sum_{\mathbf{k}} \langle c_{-\mathbf{k}\downarrow} c_{\mathbf{k}\uparrow} \rangle,
\end{equation}
which can be solved self-consistently to reveal a complex phase diagram influenced by Rashba SOC and a Zeeman field~\cite{pang2025}.

In the absence of Rashba SOC and Zeeman field, the spin susceptibility $\chi(T)$ decreases below the critical temperature $T_c$, and is given by:
\begin{subequations}\label{eq:chis}
\begin{equation}
\chi(T) = \chi_N Y(T),
\end{equation}
where $\chi_N$ is the normal-state spin susceptibility, and $Y(T)$ is the Yosida function:
\begin{equation} 
Y(T) = \int_0^\Lambda d\xi_{\mathbf{k}} \frac{\beta}{2} \left( \cosh \frac{\beta E_{\mathbf{k}}}{2} \right)^{-2},
\label{Yosidas}
\end{equation} 
\end{subequations}
with $\Lambda$ as the energy cutoff and $E_{\mathbf{k}} = \sqrt{\xi_{\mathbf{k}}^2 + |\Delta|^2}$ as the quasiparticle energy.

We now examine how Rashba SOC and the Zeeman field modify the spin susceptibility in an $s$-wave superconductor.

\subsection{Zeeman field effect}\label{sec:s-wave-H}

\begin{figure}[tb]
\centering
\includegraphics[width=\linewidth]{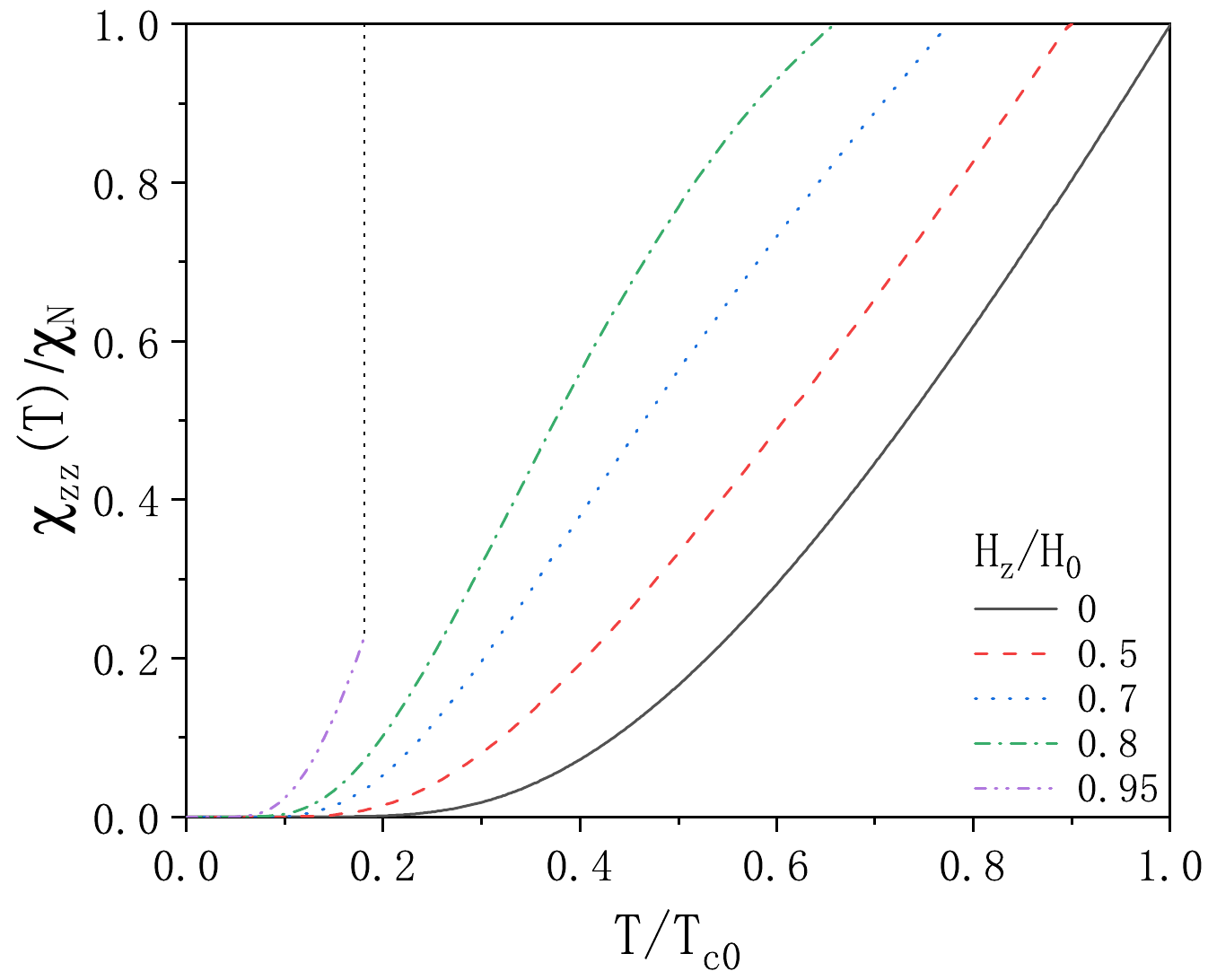} 
\caption{$s$-wave pairing state: Spin susceptibility $\chi(T)$ under a sole Zeeman field $\mathbf{H}=H_z\hat{z}$, where $H_0=H_P=\Delta_0/\sqrt{2}\mu_B$. A second-order phase transition occurs for $T_c > T_0 = 0.556 T_{c0}$, while a first-order transition is seen for $T_c < T_0$, as noted by Maki and Tsuneto~\cite{Maki1964}. 
}
\label{fig:sus_mag}
\end{figure}

We first consider the effect of a Zeeman field alone, with the Hamiltonian in Eq.~\eqref{eq:H} diagonalized via the Bogoliubov transformation. For simplicity, we assume that the Zeeman field is aligned along the $z$-axis, $\mathbf{H} = H_z \hat{z}$, resulting in the quasiparticle energy dispersion:
\begin{equation}\label{eq:Ek-SH}
E_{\mathbf{k}\pm} = E_{\mathbf{k}} \pm \mu_B H_z = \sqrt{\xi_{\mathbf{k}}^2 + |\Delta|^2} \pm \mu_B H_z.
\end{equation}
The Zeeman field induces electron spin polarization, suppressing the formation of spin-singlet Cooper pairs and reducing $T_c$. The maximum $T_c$ occurs at $H_z = 0$, which also defines $T_{c0}=T_{c}(g=0,\mathbf{H}=0)$. If $H_z$ exceeds the Pauli limit~\cite{Clogston1962,Chandrasekhar62}, $H_P = \Delta(T=0)/\sqrt{2} \mu_B$, superconductivity is destroyed, and $T_c$ drops to zero. According to Maki and Tsuneto~\cite{Maki1964}, a critical temperature $T_0 = 0.556 T_{c0}$ determines the phase transition order: second-order for $T_c > T_0$ and first-order for $T_c < T_0$.

By solving the gap function $\Delta(T, H_z)$ self-consistently from Eq.~\eqref{eq:sgap}, the uniform static spin susceptibility $\chi = \chi_{zz}(\mathbf{q} = 0, \omega = 0)$ is computed using Eq.~\eqref{eq:chikd}. As shown in Eq.~\eqref{eq:chizz_swave_mag}, $\chi^{ph-}_{zz}(T) = \chi^{pp}_{zz}(T) = 0$, and
\begin{equation}\label{chimag}  
\chi(T) = \chi_{zz}^{ph+}(T) = -\mu_B^2 \sum_{\mathbf{k}} \left[ \frac{df(E_{\mathbf{k}+})}{dE_{\mathbf{k}+}} + \frac{df(E_{\mathbf{k}-})}{dE_{\mathbf{k}-}} \right],
\end{equation}
with $E_{\mathbf{k}\pm}$ from Eq.~\eqref{eq:Ek-SH}. The ratio $\chi(T) / \chi_N$ is plotted against $T / T_{c0}$ in Fig.~\ref{fig:sus_mag}, showing a decrease to zero as $T \to 0$. For strong $H_z$, a finite drop at $T = T_c$ indicates a first-order transition when $T_c < T_0 = 0.556 T_{c0}$~\cite{Maki1964}.

In Fig.~2 and throughout this work, the parameter $H_0$ is used as a
	characteristic field scale for normalization. Its definition depends on the
	pairing symmetry, in order to account for differences in the condensation
	energy $E_c$ (in the absence of Zeeman field and Rashba SOC) among the
	various superconducting states.	
	The basic definition is
	\begin{equation*}
		E_{c0} = N(0)\mu_B^2 H_0^2,
	\end{equation*}
	where $E_{c0}$ is the condensation energy at $g=0$ and $\mathbf{H}=0$,
	$N(0)$ is the density of states at the Fermi energy, and $\mu_B$ is the
	Bohr magneton.
	
    For small magnetic fields, the Zeeman energy gained by polarizing the
	normal state is
	\begin{equation*}
		E_{\text{Zeeman}} = N(0) \mu_B^2 H^2.
	\end{equation*}
	Meanwhile, the zero-field condensation energy takes the form of
	\begin{align*}
		E_{c0} &= \frac{1}{2} N(0)\Delta_0^2
		\quad &&\text{for s-wave pairing}, \\
		E_{c0} &= \frac{1}{6} N(0)\Delta_0^2
		\quad &&\text{for } k_z\hat{z} \text{ pairing}, \\
		E_{c0} &= \frac{1}{3} N(0)\Delta_0^2
		\quad &&\text{for other } p\text{-wave pairing states}.
	\end{align*}
	For each pairing symmetry, comparing $E_{c0}$ with $E_{\text{Zeeman}}$
	yields a specific form of $H_0$ in terms of the zero-temperature gap
	amplitude $\Delta_0$~\cite{Clogston1962,Chandrasekhar62}:
	\begin{align*}
		\text{s-wave pairing:} \quad
		&H_0 = \frac{\Delta_0}{\sqrt{2}\,\mu_B} \equiv H_P
		&&\text{(Pauli limit)}, \\
		\text{most } p\text{-wave states:} \quad
		&H_0 = \frac{\Delta_0}{\sqrt{3}\,\mu_B}, \\
		k_z\hat{z} \text{ pairing state:} \quad
		&H_0 = \frac{\Delta_0}{\sqrt{6}\,\mu_B}.
	\end{align*}

\subsection{Effect of Rashba SOC}\label{sec:s-wave-g}

\begin{figure}[tb]
\centering
\includegraphics[width=1.0\linewidth]{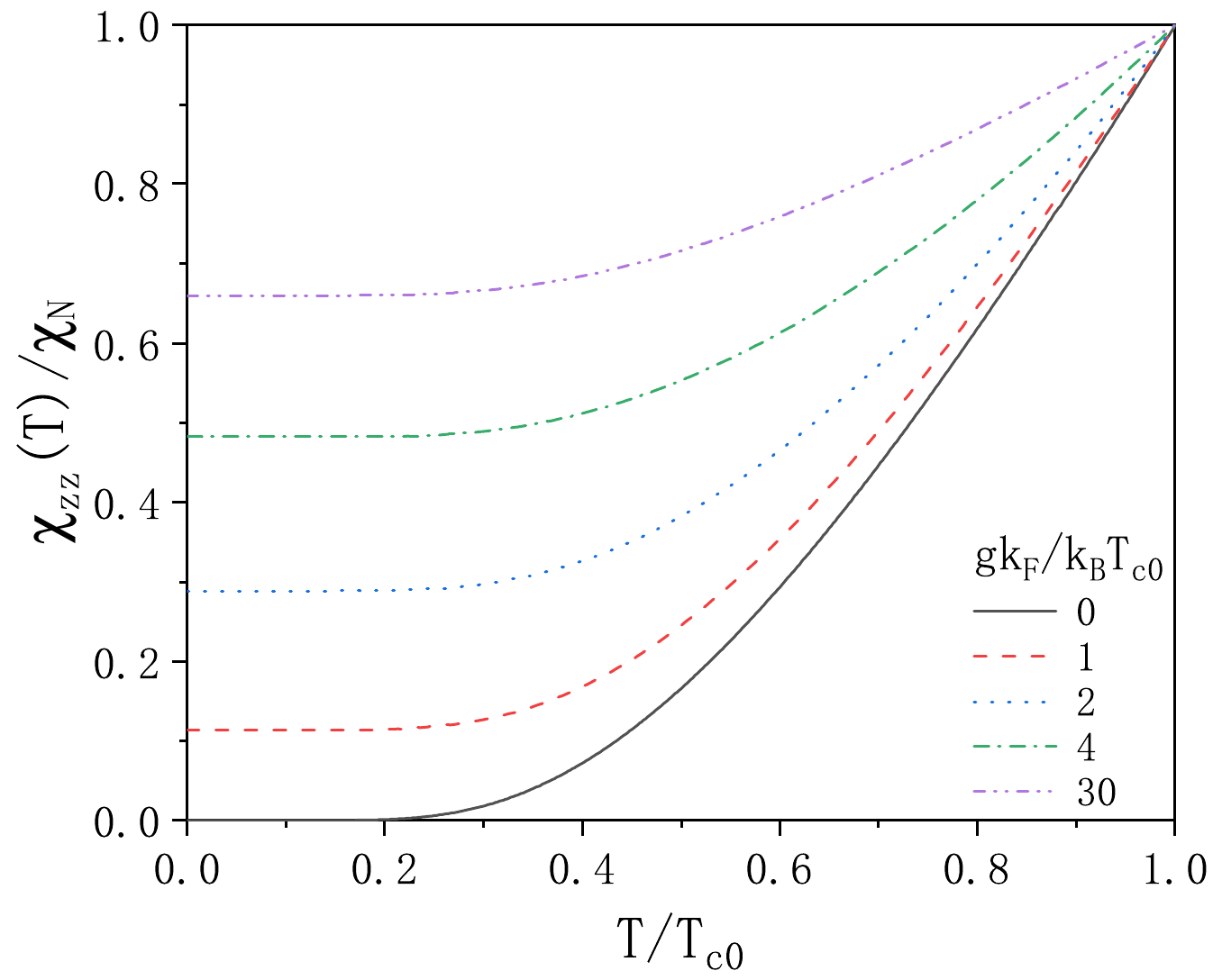} 
\caption{$s$-wave superconductor: Spin susceptibility $\chi(T)$ with Rashba SOC, in the absence of a Zeeman field. }
\label{sus_SOC}
\end{figure}

Next, we activate Rashba SOC while setting the Zeeman field to zero, assuming $g k_F \ll \Lambda$, where $\Lambda$ is the energy cutoff. In this regime, Rashba SOC does not alter the superconducting gap $\Delta(T)$ or $T_c$~\cite{pang2025}, due to preserved time-reversal symmetry and the pairing mechanism~\cite{Andersontheoryofdirty1959}. The quasiparticle energy spectrum is:
\begin{equation}\label{eq:Ek-SOC}
E_{\mathbf{k}\pm} = \sqrt{\xi_{\mathbf{k}\pm}^2 + |\Delta|^2} = \sqrt{(\xi_{\mathbf{k}} \pm g |\mathbf{k}|)^2 + |\Delta|^2}.
\end{equation}
The spin susceptibility $\chi(T) = \chi_{\mu\mu}(T)$ is evaluated using the decomposition in Eq.~\eqref{eq:chikd}:
\begin{widetext}
\begin{subequations}\label{eq:chizz_s_soc}
\begin{eqnarray}
\chi^{ph+}_{\mu\mu}(T) & = & -\mu_B^2 \sum_{\mathbf{k}} \cos^2 \theta_{\mathbf{k}} \left[ \frac{df(E_{\mathbf{k}+})}{dE_{\mathbf{k}+}} + \frac{df(E_{\mathbf{k}-})}{dE_{\mathbf{k}-}} \right], \\
\chi^{ph-}_{\mu\mu}(T) & = & -\mu_B^2 \sum_{\mathbf{k}} \sin^2 \theta_{\mathbf{k}} \left( 1 + \frac{\xi_{\mathbf{k}+} \xi_{\mathbf{k}-} + |\Delta|^2}{E_{\mathbf{k}+} E_{\mathbf{k}-}} \right) \frac{f(E_{\mathbf{k}+}) - f(E_{\mathbf{k}-})}{E_{\mathbf{k}+} - E_{\mathbf{k}-}}, \\
\chi^{pp}_{\mu\mu}(T) & = & -\mu_B^2 \sum_{\mathbf{k}} \sin^2 \theta_{\mathbf{k}} \left( 1 - \frac{\xi_{\mathbf{k}+} \xi_{\mathbf{k}-} + |\Delta|^2}{E_{\mathbf{k}+} E_{\mathbf{k}-}} \right) \frac{f(E_{\mathbf{k}+}) + f(E_{\mathbf{k}-}) - 1}{E_{\mathbf{k}+} + E_{\mathbf{k}-}},
\end{eqnarray} 
\end{subequations}
\end{widetext}
with $E_{\mathbf{k}\pm}$ from Eq.~\eqref{eq:Ek-SOC}. Details of the calculation are in Appendix~\ref{app:BT}.

The effects of sole Rashba SOC are the following: 
\begin{itemize}
\item{} For $g = 0$, $\chi_{\mu\mu}^{pp}(T) = 0$, recovering $\chi(T) = \chi_N Y(T)$ from Eq.~\eqref{eq:chis}. 

\item{} In the strong Rashba SOC limit ($g k_F \gg \Delta$), 
\begin{equation*}
1 - \frac{\xi_{\mathbf{k}+} \xi_{\mathbf{k}-} + |\Delta|^2}{E_{\mathbf{k}+} E_{\mathbf{k}-}} \to 2,
\end{equation*}
leading to $\chi(T) / \chi_N \to 2/3 + (1/3) Y(T)$, consistent with Ref.~\cite{Samokhin2007}.

\item{} For an arbitrary $g$, solving Eq.~\eqref{eq:sgap} self-consistently, $\chi(T) / \chi_N$ is shown in Fig.~\ref{sus_SOC}, where $T_c$ remains $T_{c0}$ despite varying $g$. The reduction reflects intraband and interband contributions, akin to Pauli and Van Vleck susceptibilities~\cite{van1928dielectric}.

\item{} At $T = 0$, only $\chi_{\mu\mu}^{pp}$ contributes, giving:
\begin{equation}\label{eq:chisT0}
\frac{\chi(T=0)}{\chi_N} = \frac{\mu_B^2}{\chi_N} \sum_{\mathbf{k}} \frac{\sin^2 \theta_{\mathbf{k}}}{E_{\mathbf{k}+} + E_{\mathbf{k}-}} \left( 1 - \frac{\xi_{\mathbf{k}+} \xi_{\mathbf{k}-} + |\Delta|^2}{E_{\mathbf{k}+} E_{\mathbf{k}-}} \right) \leq \frac{2}{3},
\end{equation}
with equality at $g k_F / \Delta \to \infty$. 
\end{itemize}

\subsection{Combination of Zeeman field and Rashba SOC}

\begin{figure*}[tb]
\centering
\subfigure[]{
\includegraphics[width=0.48\linewidth]{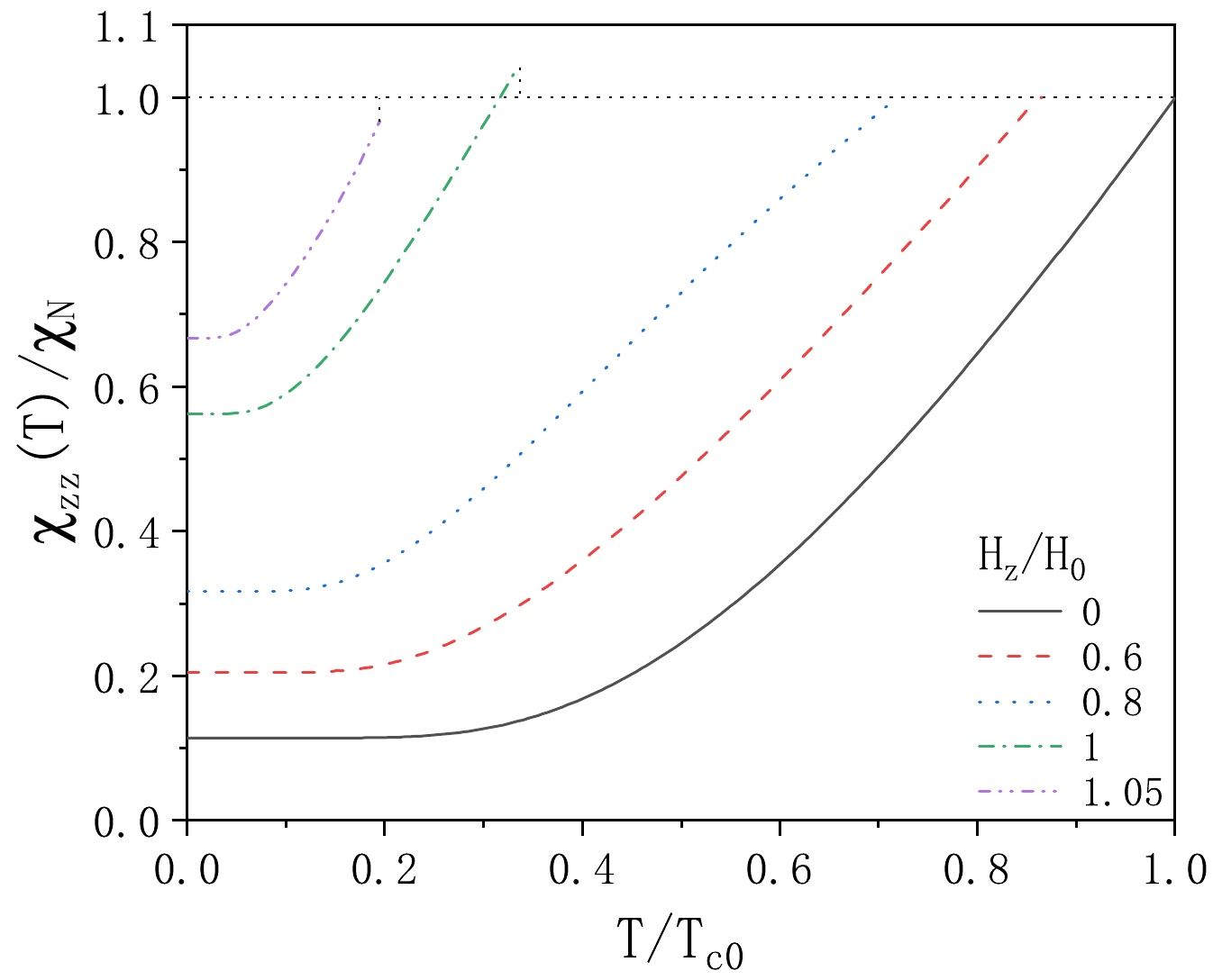} }
\subfigure[]{
\includegraphics[width=0.48\linewidth]{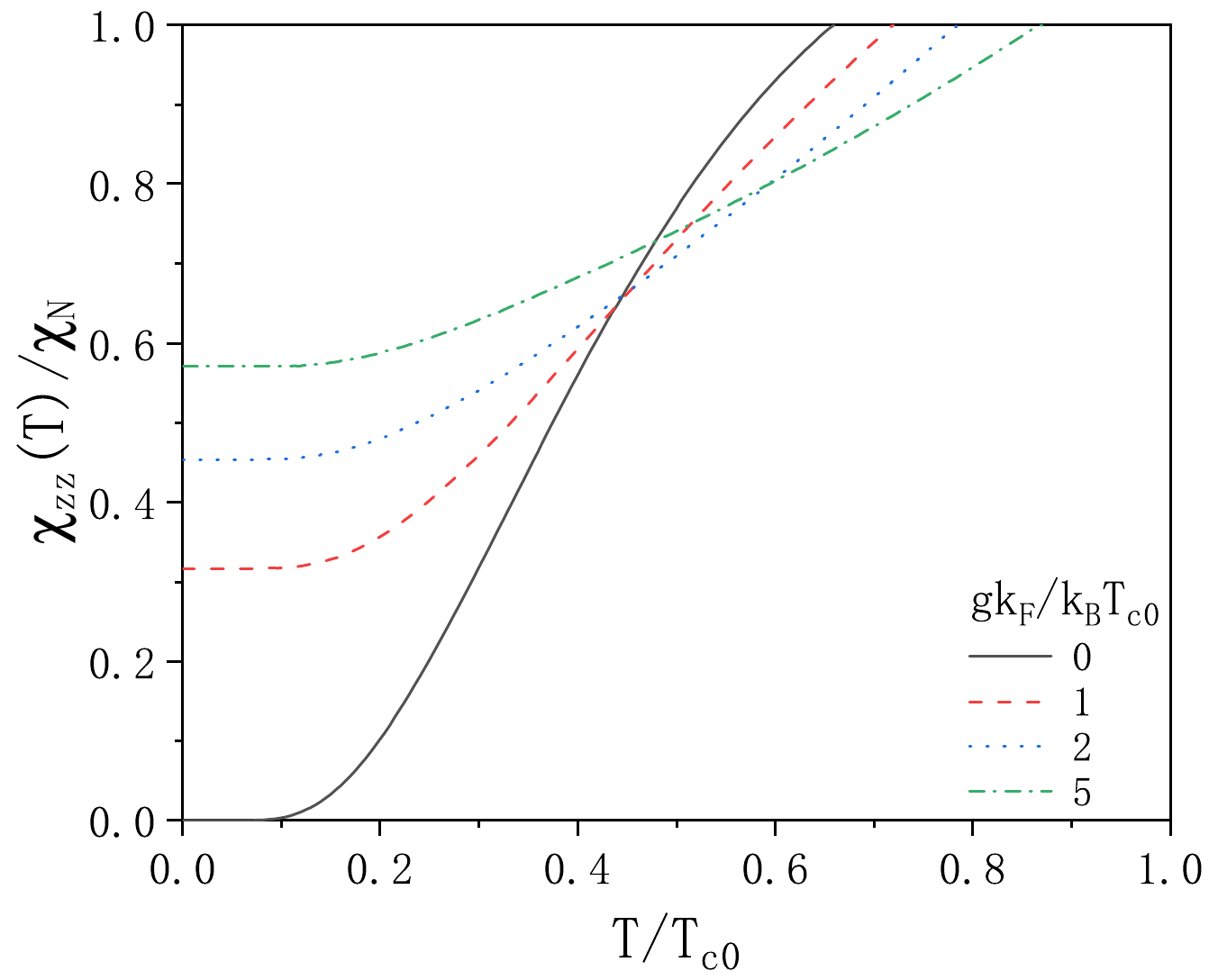} }
\caption{$s$-wave pairing state: Temperature-dependent spin susceptibility $\chi(T)$ with both Zeeman field $H_z$ and Rashba SOC $g$. (a) Rashba SOC fixed at $g k_F / k_B T_{c0} = 1$, with varying $H_z$; the transition becomes first-order for small $T_c / T_{c0}$, similar to $g = 0$~\cite{Maki1964}. (b) Zeeman field fixed at $H_z / H_0 = 0.8$, with varying Rashba SOC. Here the field scale $H_0 = H_P$ is provided at the end of Section~\ref{sec:s-wave-H}.}
\label{sus_magSOC}
\end{figure*}

\begin{figure*}[tb]
\centering
\subfigure[]{
\includegraphics[width=0.48\linewidth]{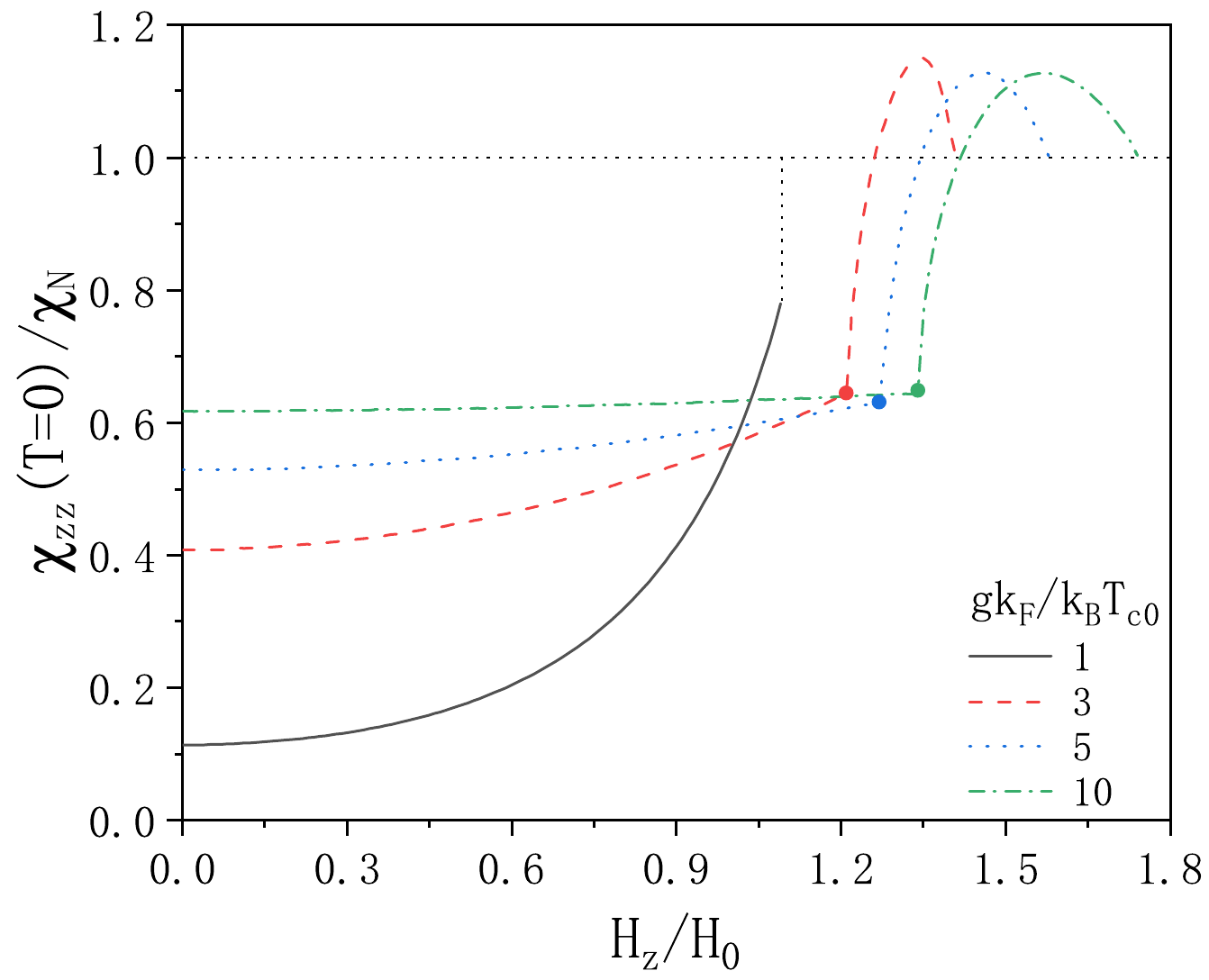}
}
\subfigure[]{
\includegraphics[width=0.48\linewidth]{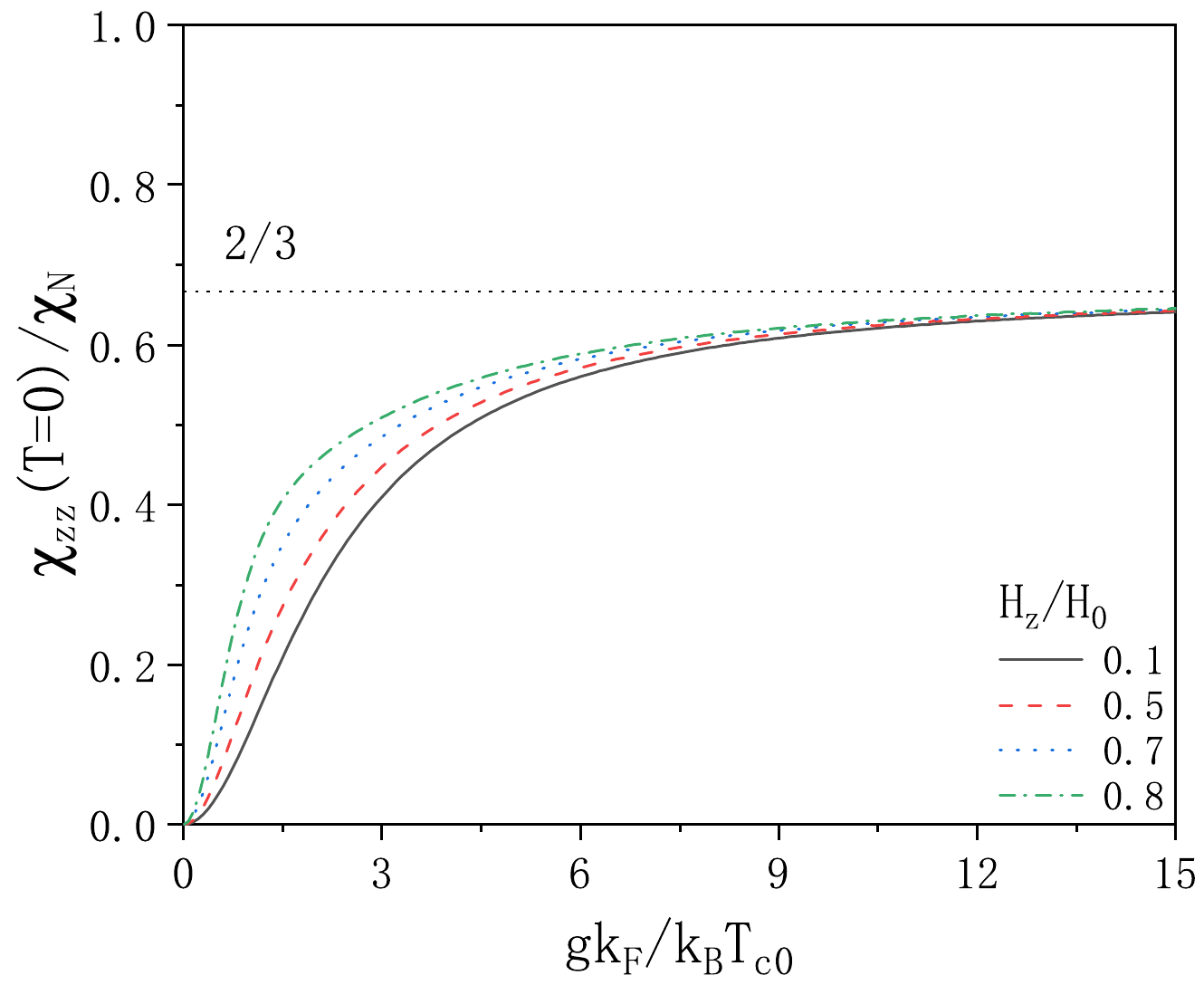}
}
\caption{$s$-wave pairing state: Zero-temperature spin susceptibility $\chi_{zz}(T=0)$ as a function of (a) Zeeman field $H_z$ (fixed Rashba SOC) and (b) Rashba SOC $g$ (fixed Zeeman field). In (a), curves end at critical field $H_c^Z$, with a kink at $H_{c2}^Z$ (solid circles) for $g > g_c$, indicating Bogoliubov Fermi surface formation~\cite{pang2025}.}
\label{sus_zero}
\end{figure*}

We now analyze the $s$-wave superconducting phase with both Zeeman field and Rashba SOC. The spin susceptibility $\chi_{zz}(T)$ is computed by solving the gap equation in Eq.~\eqref{eq:sgap}. Key results are presented in Figs.~\ref{sus_magSOC} and \ref{sus_zero}, leading to the following observations:
\begin{itemize}
\item{} In an $s$-wave superconductor, $\chi_{zz}(T)$ decreases as temperature drops below $T_c$, irrespective of $H_z$ and $g$.~\footnote{The only exception, specifically the upturn around $T_c$ in the curve with $H_z / H_0 = 1$ in Fig.~\ref{sus_magSOC}(a), may be attributed to the first-order phase transition and the proximity to regions hosting a Bogoliubov Fermi surface~\cite{pang2025}.}
\item{} Without Rashba SOC ($g = 0$), $\chi_{zz}(T=0) = 0$ for any $H_z$. With finite $g$, $\chi_{zz}(T=0)$ becomes positive and non-zero.
\item{} As shown in Fig.~\ref{sus_zero}(a), each curve ends at a critical Zeeman field $H_c^Z$, where superconductivity is suppressed. For $g < g_c$, $\chi_{zz}(T=0, g, H_z)$ increases monotonically with $H_z$. For $g > g_c$, it shows non-monotonic behavior with a kink at $H_z = H_{c2}^Z < H_c^Z$,  where $H_{c2}^Z$ is the field above which a Bogoliubov Fermi surface emerges~\cite{pang2025}, and $H_c^Z$ is the ultimate critical field where superconductivity is suppressed.

\item{} In Fig.~\ref{sus_zero}(b), for $H_z < H_P$, $\chi_{zz}(T=0, g, H_z)$ increases with $g$, approaching $2 \chi_N / 3$ as $gk_F/\Delta \to \infty$.
\end{itemize}

\section{$p$-wave pairing states}

\begin{table*}[tb]
\caption{Examples of $p$-wave pairing states and corresponding spin susceptibility without Zeeman field or Rashba SOC.~\footnote{Here $Y(T)$ is defined in Eq.~\eqref{Yosidap}.} }
\label{tab:pwave}
\setlength{\tabcolsep}{2.0ex}
\renewcommand\arraystretch{2.0}
\begin{tabular}{c|c|c|c|c|c|c}
\hline\hline
&\multicolumn{2}{c|}{Opposite-spin pairing} & \multicolumn{4}{c}{Equal-spin pairing}\\
\hline
Notation & $(k_x+ik_y)\hat{z}$ & $k_z\hat{z}$ & $k_x\hat{x}+k_y\hat{y}$ & $k_y\hat{x}-k_x\hat{y}$ & $k_x\hat{x}-k_y\hat{y}$ & $k_y\hat{x}+k_x\hat{y}$\\
\hline

\multirow{2}{*}{$\mathbf{d}(\mathbf{k})$} & \multirow{2}{*}{$\Delta\sin\theta_{\mathbf{k}}\mathrm{e}^{i\varphi_{\mathbf{k}}}\hat{z}$} & \multirow{2}{*}{$\Delta\cos\theta_{\mathbf{k}} \hat{z}  $   } & \multicolumn{4}{c}{$\Delta\sin\theta_{\mathbf{k}}\left(\cos\phi_{\mathbf{k}}\hat{x}+\sin\phi_{\mathbf{k}}\hat{y}\right)$}\\
\cline{4-7}

&  && $\phi_{\mathbf{k}}=\varphi_{\mathbf{k}}$ & $\phi_{\mathbf{k}}=\varphi_{\mathbf{k}}-\pi/2$ & $\phi_{\mathbf{k}}=-\varphi_{\mathbf{k}}$ & $\phi_{\mathbf{k}}=\pi/2-\varphi_{\mathbf{k}}$\\
\hline

$\chi_{zz}/\chi_{N}$ & \multicolumn{2}{c|}{$Y(T)$} & \multicolumn{4}{c}{1}\\
\hline
$\chi_{xx}/\chi_{N}$ & \multicolumn{2}{c|}{1}& \multicolumn{4}{c}{$\frac{1}{2}[1+Y(T)]$}\\
\hline\hline
\end{tabular}
\end{table*}

\begin{figure*}[tb]
\centering
\subfigure[]{
\includegraphics[width=0.48\linewidth]{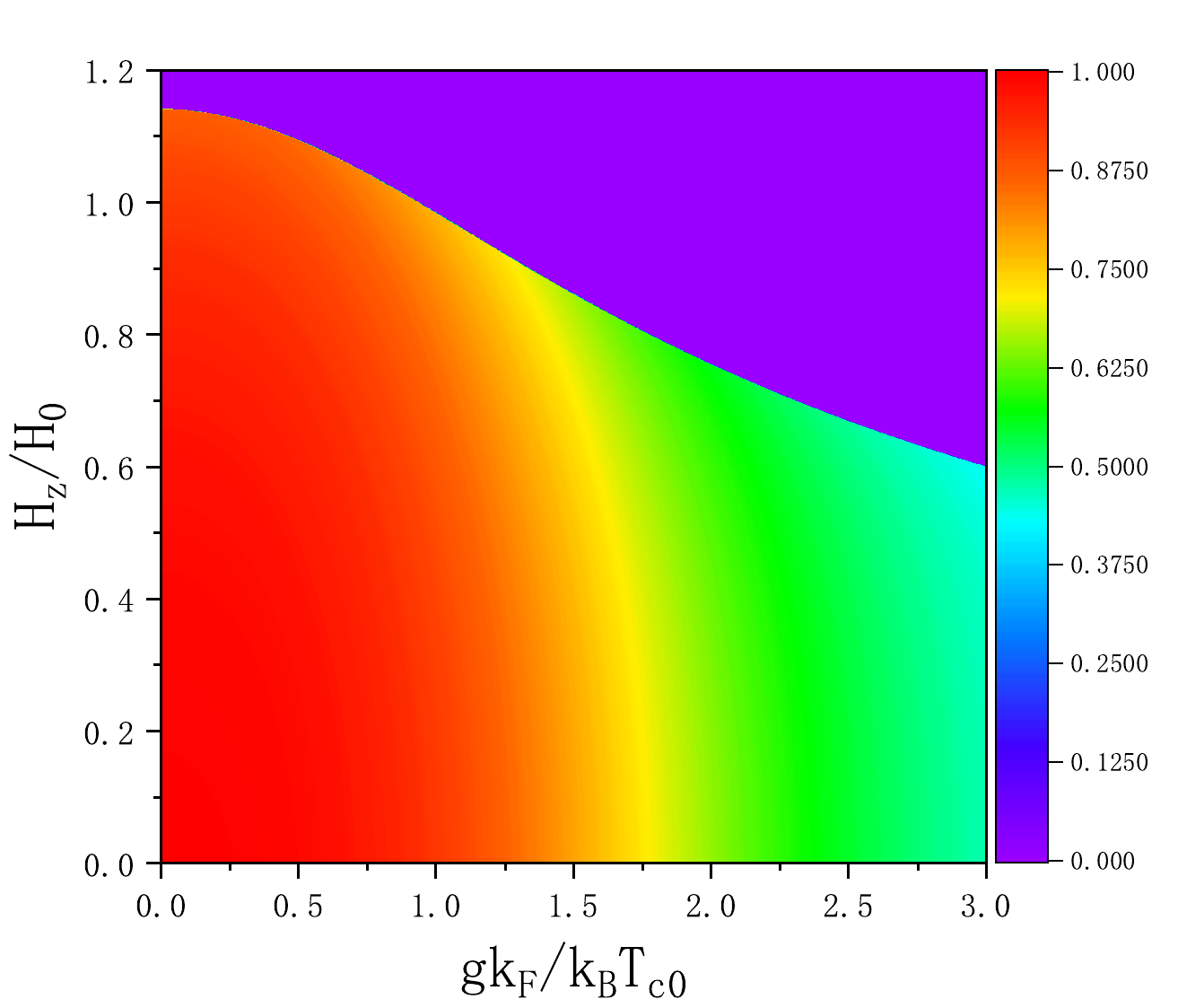}
}
\subfigure[]{
\includegraphics[width=0.48\linewidth]{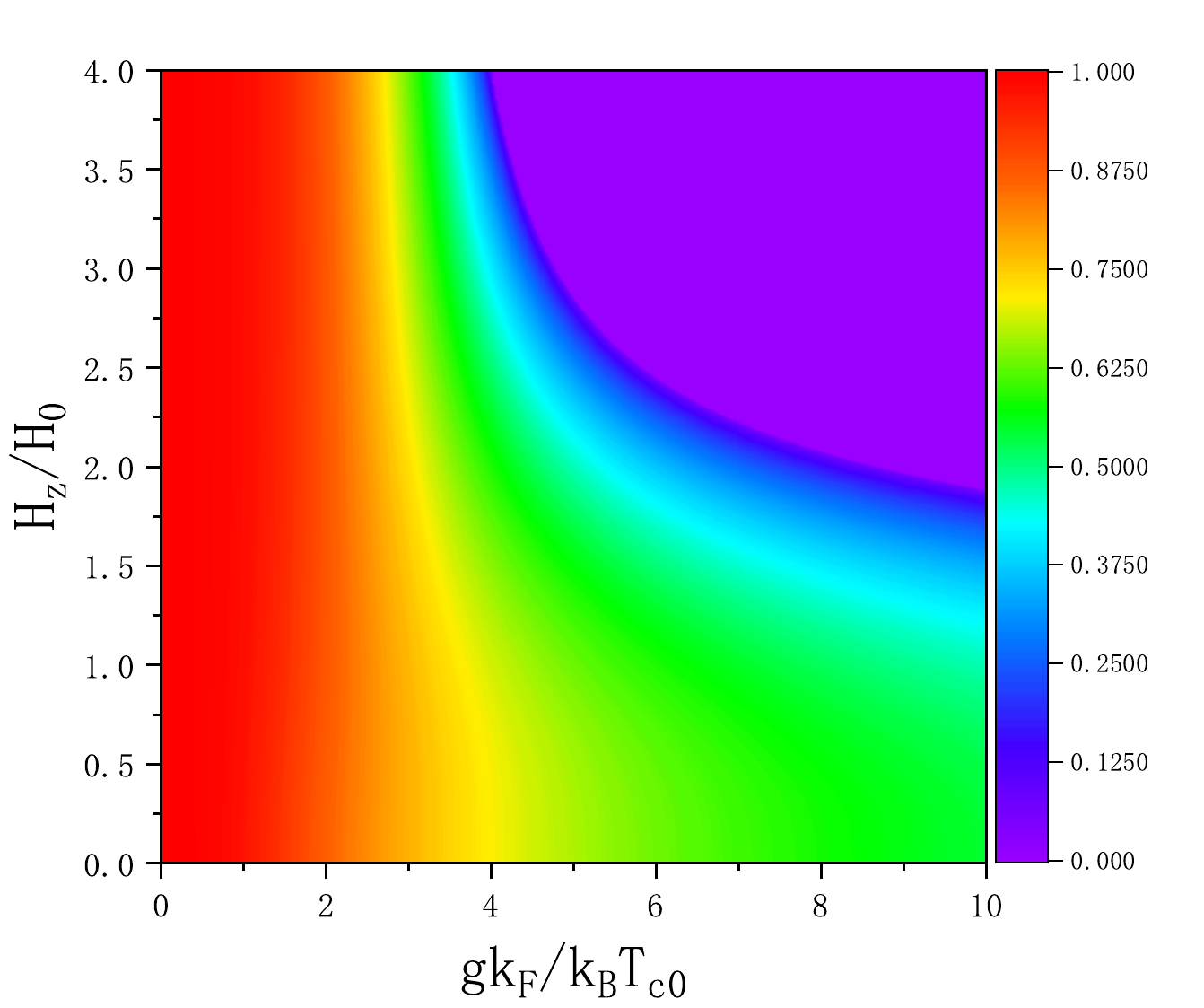}
}
\caption{ Dependence of the gap function on $H_z$ and Rashba SOC at zero temperature for (a) $k_z\hat{z}$ and (b) $k_x\hat{x}+k_y\hat{y}$ pairing states. The first-order phase transition for the $k_z\hat{z}$ pairing state and the continuous phase transition for the $k_x\hat{x}+k_y\hat{y}$ pairing state are in agreement with the corresponding results reported in Ref.~\cite{pang2025}. 
}
\label{fig:delta_zero_p}
\end{figure*}

\subsection{Opposite-spin pairing states}
In $p$-wave superconductors, the dominant pairing channel corresponds to $l=1$. We start by defining the simplest $p$-wave pairing model, based on a simplified matrix element from Eq.~\eqref{eq:Vl}, expressed as:
\begin{equation}\label{eq:Vp}
\begin{split}
V(\mathbf{k}, \mathbf{k}') &= 3 V_1 \cos \theta_{\mathbf{k}, \mathbf{k}'} \\
&= 4 \pi V_1 \sum_{m=0, \pm 1} Y_{1, m}(\theta_{\mathbf{k}}, \varphi_{\mathbf{k}}) Y_{1, m}^{*}(\theta_{\mathbf{k}'}, \varphi_{\mathbf{k}'}),
\end{split}
\end{equation}
where $Y_{l,m}(\theta, \varphi)$ are spherical harmonics.

To ensure a robust framework, we restrict our analysis to unitary states, where $\Delta^{\dagger}(\mathbf{k}) \Delta(\mathbf{k})$ is proportional to the identity matrix. This condition yields a quasiparticle energy dispersion of:
\begin{equation} \begin{aligned}
E_{\mathbf{k}} &= \sqrt{\xi_{\mathbf{k}}^2 + |\mathbf{d}(\mathbf{k})|^2}.
\label{eq:Ek-us}
\end{aligned} \end{equation}
For unitary states, $\mathbf{d}(\mathbf{k})$ is a real vector with three components, aside from an overall phase factor~\cite{Sigrist1991}.

When both a Zeeman field and Rashba SOC are present, deriving an analytical expression for the quasiparticle energy dispersion is highly challenging and often not feasible. In contrast, analytical solutions can be obtained when only one of these fields is applied~\cite{pang2025}. Despite this limitation, numerical methods allow us to solve the system effectively, analogous to the approach used for $s$-wave pairing states.

Moving to the classification of pairing states, $p$-wave and spin-triplet superconductors are divided into opposite-spin pairing (OSP) and equal-spin pairing (ESP) states~\cite{Leggett1975}. This categorization depends on the spin quantization axis and can vary accordingly. For clarity, details are provided in Ref.~\cite{pang2025}. In this work, unless otherwise stated, the spin quantization axis is set along the $z$-axis, either in the absence of an external magnetic field or aligned with the Zeeman field direction.

In OSP states, the superconducting condensate minimally responds to a Zeeman field along the quantization axis $\hat{\alpha}$ (e.g., $\hat{x}$, $\hat{y}$, or $\hat{z}$), avoiding Zeeman energy loss. Thus, only quasiparticle excitations contribute to magnetization within the Pauli limit, resulting in a reduced spin susceptibility $\chi_{\alpha\alpha}(T)$ below $T_c$. In ESP states, however, the condensate behaves similarly to the normal state under a Zeeman field along $\hat{\alpha}$, maintaining a constant spin susceptibility $\chi_{\alpha\alpha}(T) = \chi_N$ across $T_c$~\cite{Leggett1975}.

Finally, to contextualize these concepts, Table~\ref{tab:pwave} lists examples of $p$-wave pairing states and the corresponding temperature-dependent spin susceptibility without Zeeman field or Rashba SOC.
Note that the Yosida function $Y(T)$ in Table~\ref{tab:pwave} differs from the $s$-wave case and is computed using:
\begin{equation} \begin{aligned}
Y(T) &= \int \frac{d \Omega_{\mathbf{k}}}{4 \pi} \int_{-\Lambda}^{\Lambda} d \xi_{\mathbf{k}} \frac{\beta}{4} \left( \cosh \frac{\beta E_{\mathbf{k}}}{2} \right)^{-2},
\label{Yosidap}
\end{aligned} \end{equation}
where the integrand depends on $\theta_{\mathbf{k}}$. 

In the subsequent subsections, we will analyze the OSP and ESP states from Table~\ref{tab:pwave} in detail. By employing the pairing model in Eq.~\eqref{eq:Vp} and specifying the relevant $\mathbf{d}(\mathbf{k})$ vector, the gap equation from Eq.~\eqref{gapeqwhole} can be solved self-consistently, enabling a thorough exploration of the impacts of the Zeeman field and/or Rashba SOC on these states. Fig.~\ref{fig:delta_zero_p} shows the solutions of the gap equations for two representative pairing states.

In contrast to the $s$-wave pairing state, where the Zeeman field reduces the critical temperature $T_c$ while Rashba SOC preserves $T_c$ in the absence of a Zeeman field and can restore it otherwise, $p$-wave superconductors exhibit markedly different behaviors. As demonstrated in Ref.~\cite{pang2025}, for OSP states such as $(k_x + i k_y) \hat{z}$ and $k_z \hat{z}$, both Rashba SOC and a Zeeman field parallel to the spin quantization axis decrease $T_c$, whereas a perpendicular Zeeman field has no effect. For ESP states, Rashba SOC and a perpendicular Zeeman field lower $T_c$, while a parallel Zeeman field leaves it unchanged. Additionally, the combination of an external Zeeman field and Rashba SOC can significantly alter the nodal configuration of the quasiparticle energy gap. In this work, we focus on these effects to provide a comprehensive understanding of the temperature dependence of the spin susceptibility in $p$-wave superconductors.

Building on the previous classification in this section, we begin by examining the OSP states, characterized by the following d-vectors:
\begin{subequations}\label{eq:dvec1}
\begin{equation}
\mathbf{d}(\mathbf{k}) = \Delta \sin \theta_{\mathbf{k}} \mathrm{e}^{i \varphi_{\mathbf{k}}} \hat{z},
\end{equation}    
and
\begin{equation}
\mathbf{d}(\mathbf{k}) = \Delta \cos \theta_{\mathbf{k}}  \hat{z}  ,
\end{equation}
\end{subequations}
which correspond to the pairing states $(k_x + i k_y) \hat{z}$ and $k_z \hat{z}$ as listed in Table~\ref{tab:pwave}.

\subsubsection{Zeeman field effect}

\begin{figure*}[tb]
\centering
\subfigure[]{
\includegraphics[width=0.48\linewidth]{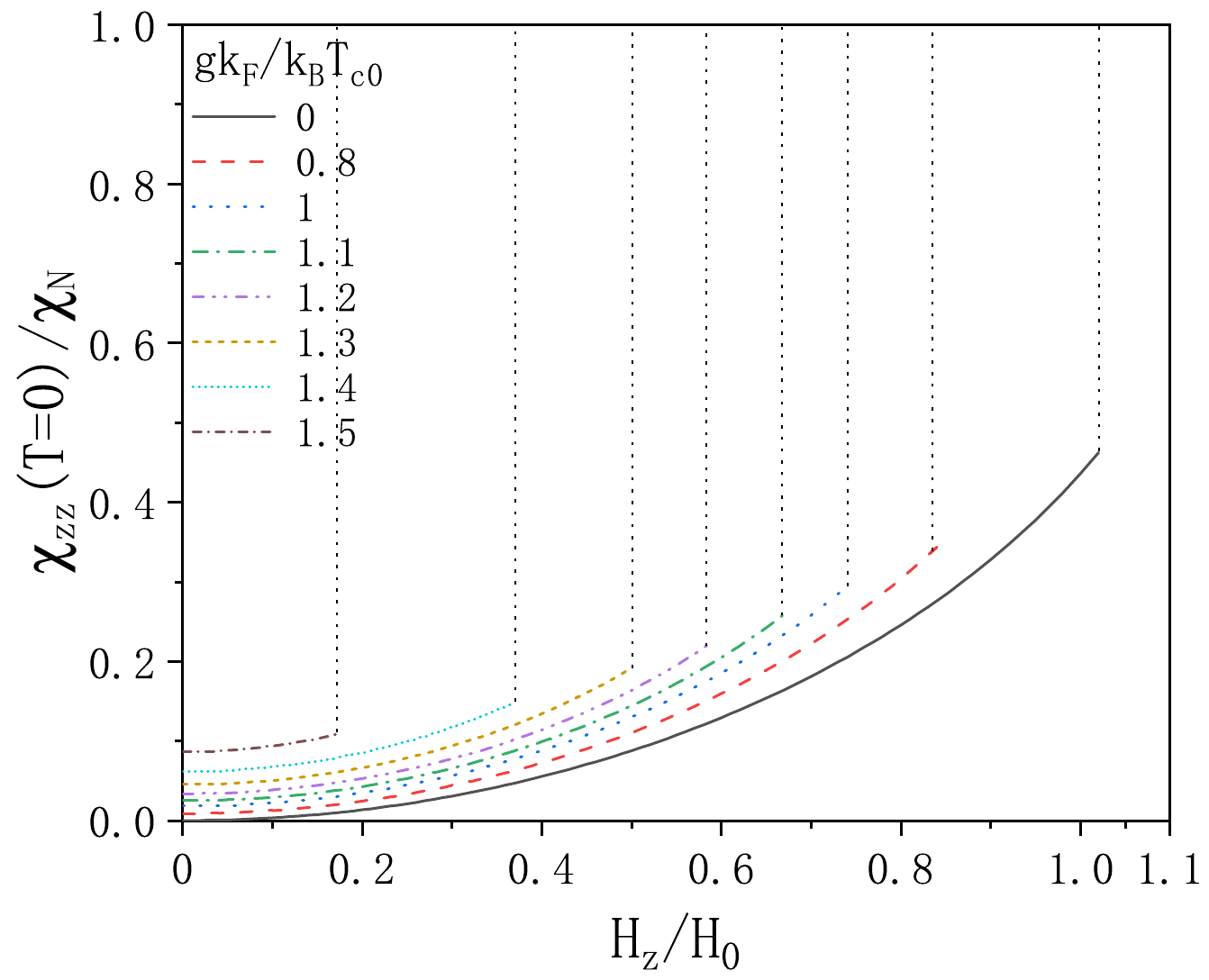}
}
\subfigure[]{
\includegraphics[width=0.48\linewidth]{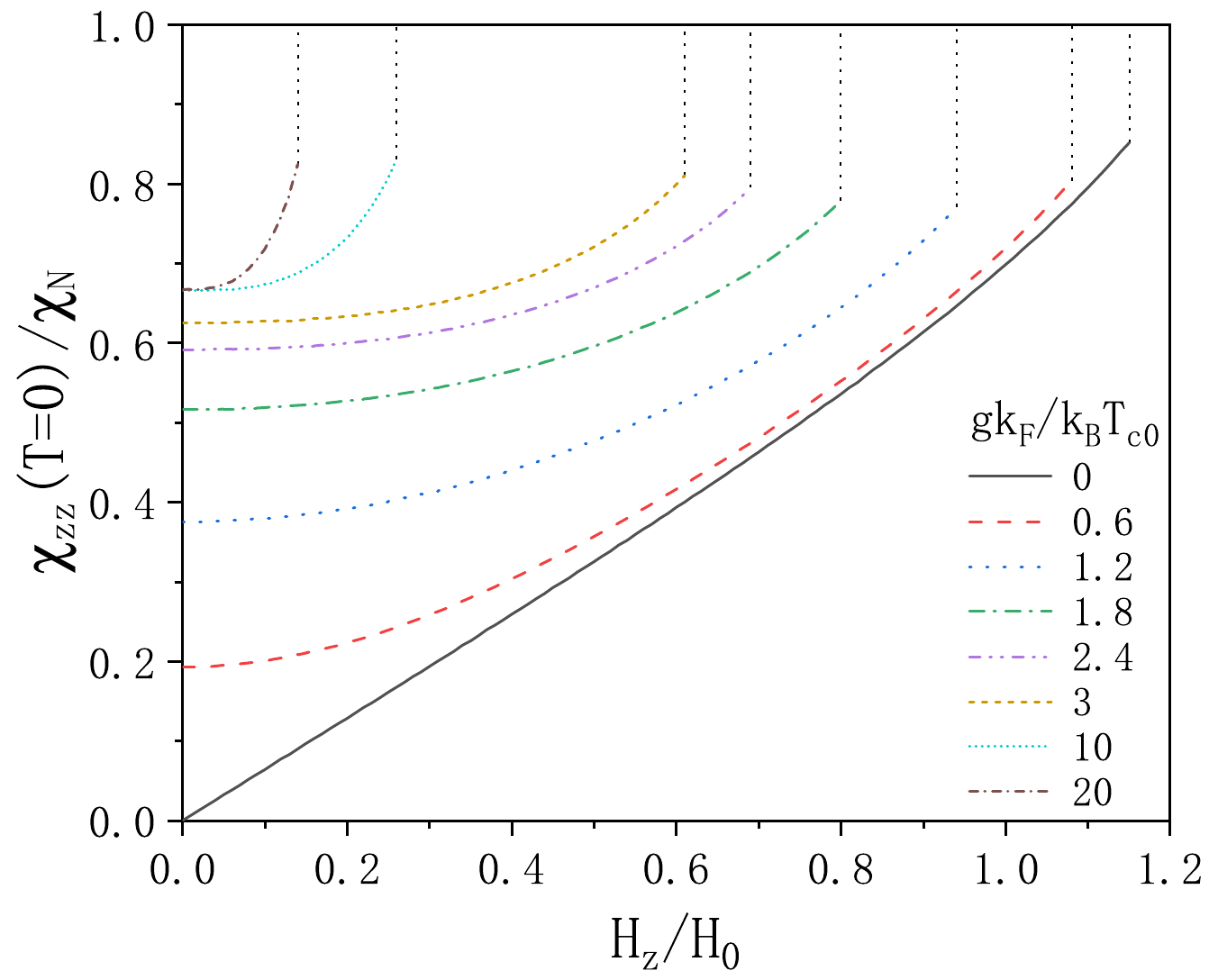}
}
\caption{ OSP states: Zero-temperature spin susceptibility $\chi_{zz}(0)$ as a function of the Zeeman field $H_{z}$ (with Rashba SOC held fixed) for (a) $(k_{x} + i k_{y})\hat{z}$ pairing and (b) $k_{z}\hat{z}$ pairing.
}
\label{fig:sus_zero_osp}
\end{figure*}

\begin{figure*}[tb]
\centering
\subfigure[]{
\includegraphics[width=0.48\linewidth]{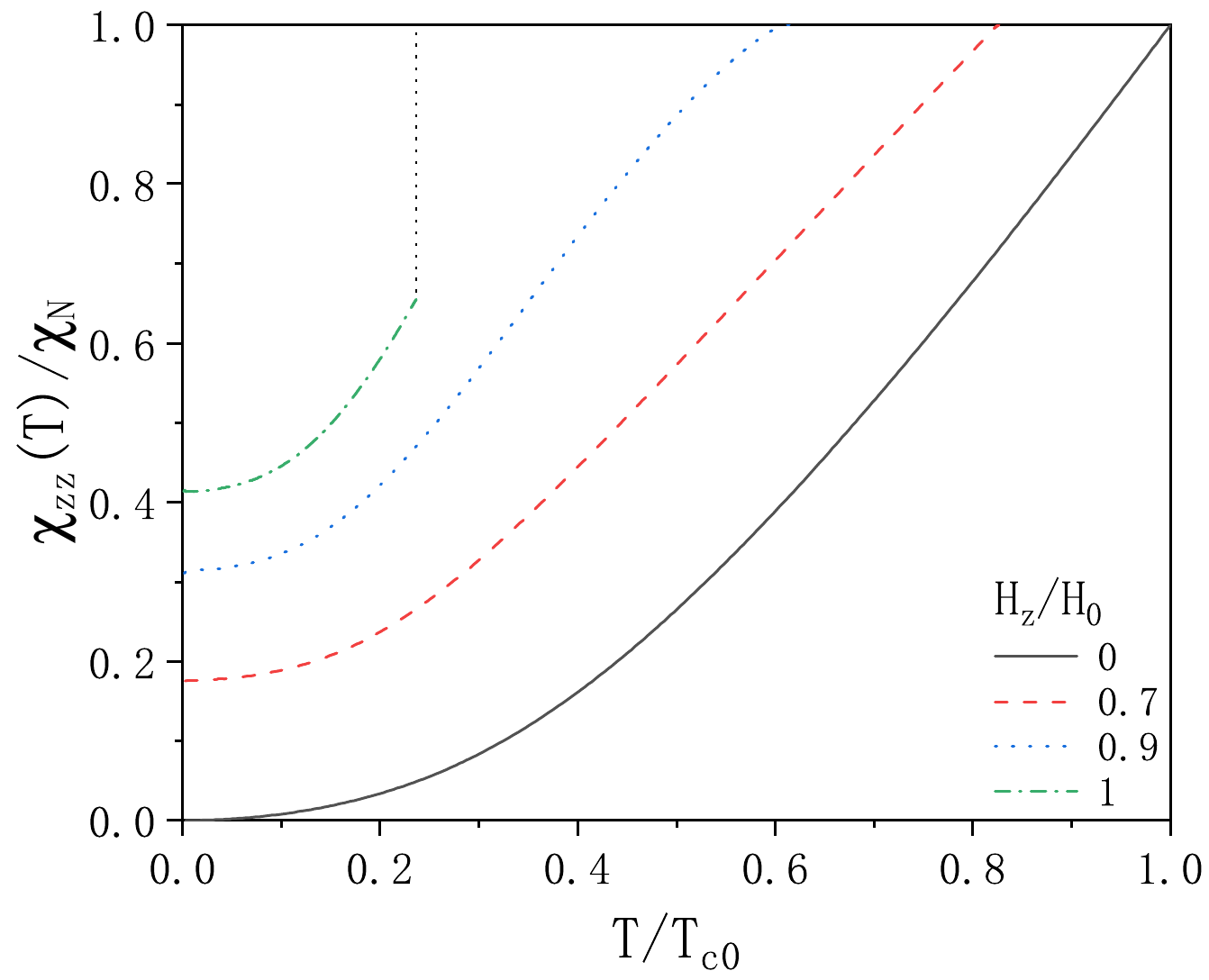}
}
\subfigure[]{
\includegraphics[width=0.48\linewidth]{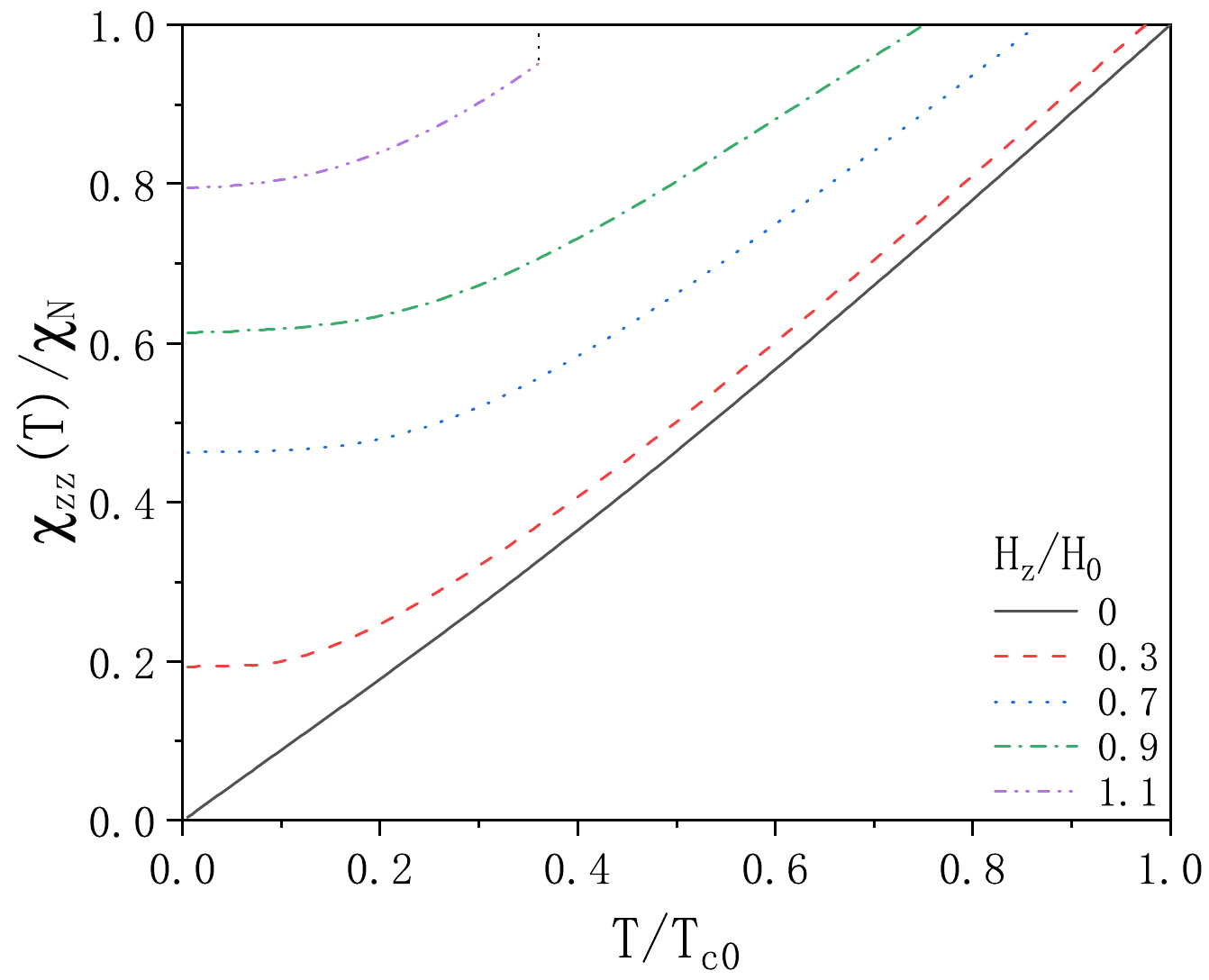}
}
\caption{Temperature-dependent spin susceptibility $\chi_{zz}(T)$ for two OSP states, (a) $(k_x + i k_y) \hat{z}$ and (b) $k_z \hat{z}$ , under a parallel Zeeman field $\mathbf{H} = H_z \hat{z}$, where Rashba SOC is absent. The definition of $H_0$ is provided at the end of Section~\ref{sec:s-wave-H}.}
\label{fig:OSP_parallel_mag}
\end{figure*}

Continuing from the general framework of OSP states, we first examine the impact of a Zeeman field aligned along the $z$-axis, i.e., $\mathbf{H} = H_z \hat{z}$. Using the Kubo formula from Eq.~\eqref{Kubo}, the static and uniform paramagnetic susceptibility $\chi_{zz}(\mathbf{q} = 0, \omega = 0)$ adopts the same form as in the $s$-wave pairing case, as given in Eq.~\eqref{chimag}, with the quasiparticle energy dispersion:
\begin{equation}
\label{eq:ospEkHz}
E_{\mathbf{k}\pm} = E_{\mathbf{k}} \pm \mu_B H_z.
\end{equation}
Substituting the specific pairing functions $\Delta(\mathbf{k})$ from Eq.~\eqref{eq:dvec1} into Eq.~\eqref{chimag} yields distinct spin susceptibilities $\chi_{zz}(T)$ for the $(k_x + i k_y) \hat{z}$ and $k_z \hat{z}$ states. At zero temperature, a finite $\chi_{zz}(T=0)$ emerges in a nonzero Zeeman field, resulting in:
\begin{subequations}\label{eq:chimagzOSP}
\begin{equation} 
\chi_{zz}(T=0, H_z) = \frac{1}{2} \chi_N \cdot \frac{\mu_B H_z}{|\Delta|} \ln \left| \frac{1 + \frac{\mu_B H_z}{|\Delta|}}{1 - \frac{\mu_B H_z}{|\Delta|}} \right|
\end{equation}
for the $(k_x + i k_y) \hat{z}$ pairing, and
\begin{equation}
\chi_{zz}(T=0, H_z) = \frac{\pi}{2} \cdot \frac{\mu_B H_z}{|\Delta|} \chi_N
\end{equation}
\end{subequations}
for the $k_z \hat{z}$ pairing, respectively. Here, $\Delta$ is the pairing strength solved self-consistently from Eqs.~\eqref{gapeqwhole}, \eqref{eq:Vp}, and \eqref{eq:dvec1}. This residual susceptibility arises from a finite Bogoliubov Fermi surface around nodal points at $\theta_{\mathbf{k}} = 0, \pi$ or a nodal line at $\theta_{\mathbf{k}} = \pi/2$ in Eq.~\eqref{eq:ospEkHz}, contrasting with the Rashba SOC-induced residual susceptibility in $s$-wave states, which stems from a fully gapped spectrum.

For comparison, Fig.~\ref{fig:sus_zero_osp} shows the zero-temperature spin susceptibility $\chi_{zz}(0)$ for the $(k_x + i k_y)\hat{z}$ and $k_z\hat{z}$ pairing states. In contrast to the $s$-wave case shown in Fig.~\ref{sus_zero}(a), Fig.~\ref{fig:sus_zero_osp} indicates that the superconducting states terminate at lower Zeeman fields as the Rashba SOC increases, and no kinks appear. As discussed in Ref.~\cite{pang2025}, Rashba SOC reduces the maximum Zeeman field that can be sustained by all $p$-wave pairing states considered below. Consequently, similar kinks are not expected in the remaining $p$-wave states, consistent with the behavior of the two OSP states. Moreover, Fig.~\ref{fig:sus_zero_osp}(b) shows that $\chi_{zz}(0)/\chi_{N} \to 2/3$ in the strong-SOC limit as the Zeeman field approaches zero.

At finite temperatures, $\chi_{zz}(T)$ is evaluated self-consistently, revealing a decrease with falling temperature, as illustrated in Fig.~\ref{fig:OSP_parallel_mag}. Similar to the $s$-wave case, the superconducting transition at $T_c$ can be first-order when $T_c$ falls below a threshold $T_0$~\cite{Maki1964}, though the ratio $T_0 / T_{c0}$ varies by pairing state.

Next, we consider a perpendicular Zeeman field along the $x$-axis, $\mathbf{H} = H_x \hat{x}$, without loss of generality, as a rotation in the $xy$-plane only shifts the phase for $(k_x + i k_y) \hat{z}$ and leaves $k_z \hat{z}$ unchanged. The quasiparticle energy dispersion becomes:
\begin{equation}\label{eq:ospEkHx}
E_{\mathbf{k}\pm} = \sqrt{(\xi_{\mathbf{k}} \pm \mu_B H_x)^2 + |d_z(\mathbf{k})|^2}.
\end{equation}
In this case, the spin susceptibility $\chi_{xx}(T)$ remains temperature-independent, with $\chi_{xx}(T) = \chi_N$ for $T < T_c$, as these states behave as ESP states when the quantization axis aligns with the field direction~\cite{pang2025}. This constancy is explained by the decomposition in Eq.~\eqref{eq:chikd}, where $\chi_{xx}^{ph-} = 0$ and $\chi_{xx}^{ph+} + \chi_{xx}^{pp} = \chi_N$ (detailed in Appendix~\ref{app:chixx_osp}).

To summarize, the anisotropic OSP states $(k_x + i k_y) \hat{z}$ and $k_z \hat{z}$ exhibit distinct responses to parallel ($\mathbf{H} = H_z \hat{z}$) and perpendicular ($\mathbf{H} = H_x \hat{x}$) magnetic fields:
\begin{itemize}
\item The energy dispersion $E_{\mathbf{k}\pm}$ varies differently with $H_x$ and $H_z$ [as shown in Eqs.~\eqref{eq:ospEkHx} and \eqref{eq:ospEkHz}].
\item Under a parallel field, $\chi_{zz}(T)$ decreases from $T_c$ to a value given in Eqs.~\eqref{eq:chimagzOSP}, while under a perpendicular field, $\chi_{xx}(T)$ remains constant at $\chi_N$ for $T < T_c$.
\end{itemize}
These results hold for any Zeeman field direction in the $xy$-plane and remain valid if the d-vector components in Eq.~\eqref{eq:dvec1} are generalized to arbitrary $d_z(\mathbf{k})$.

\subsubsection{Effect of Rashba SOC}

\begin{figure}[tb]
\centering
\subfigure[]{
\includegraphics[width=0.96\linewidth]{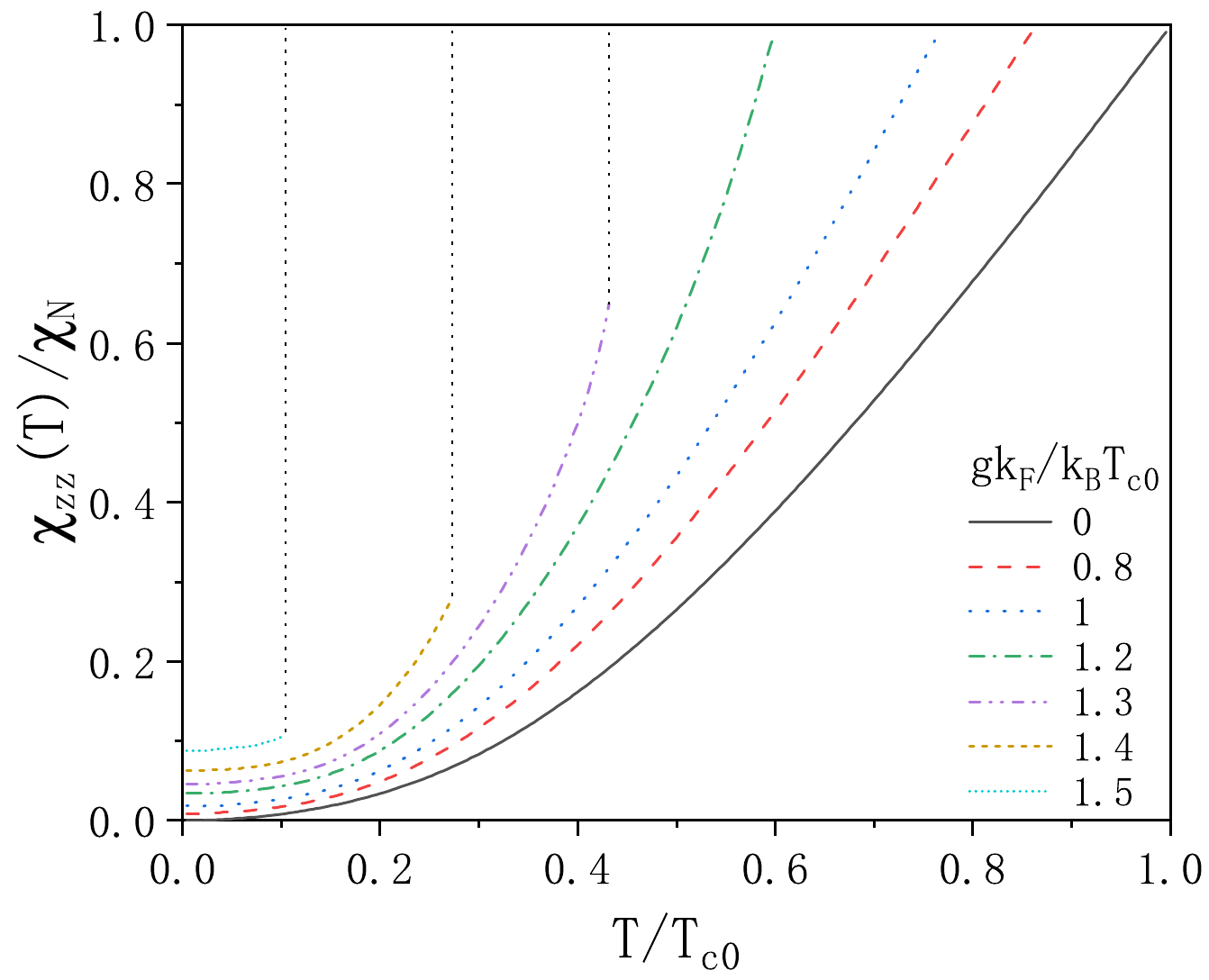}
}
\subfigure[]{
\includegraphics[width=0.96\linewidth]{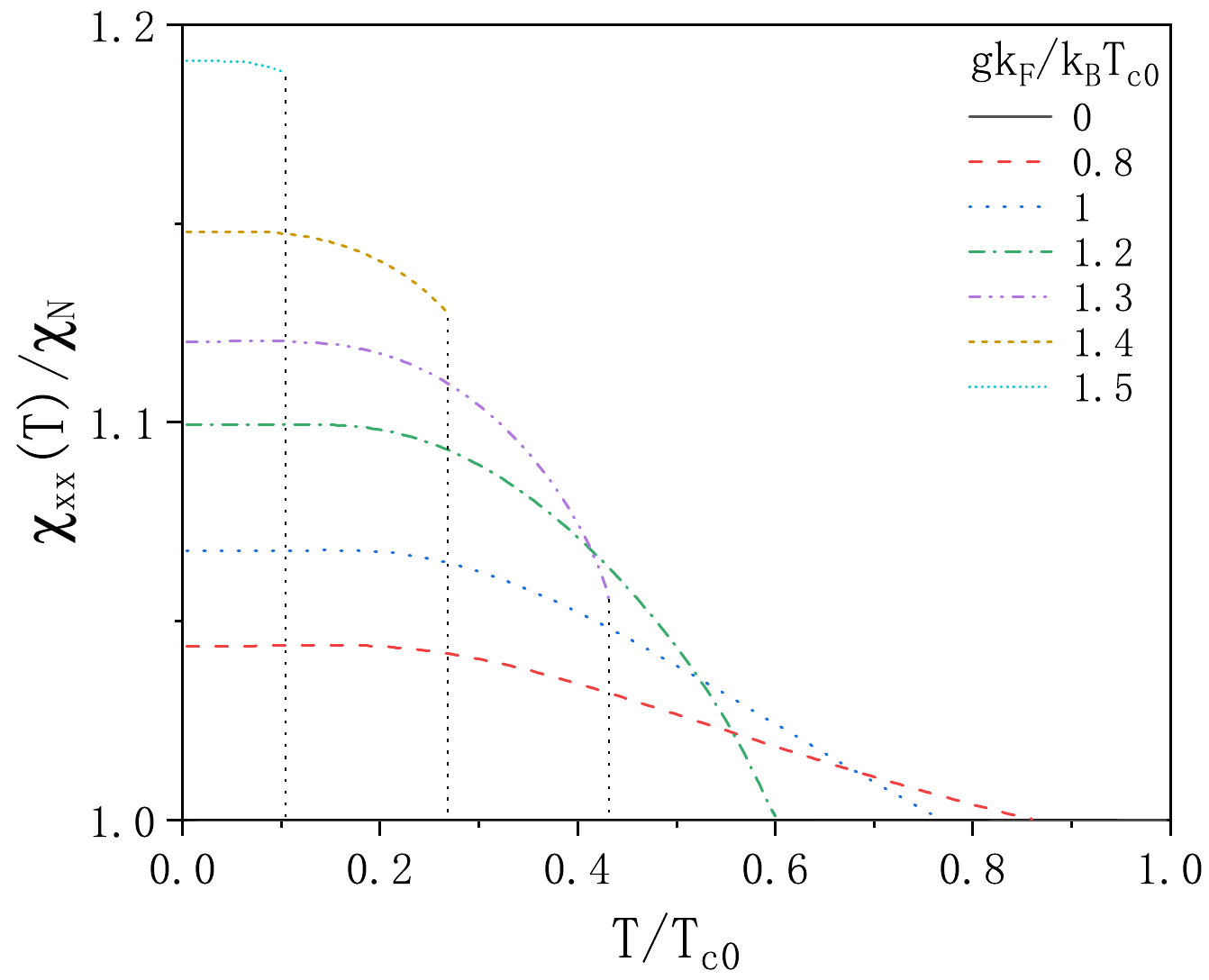}
}
\caption{The $(k_x + i k_y) \hat{z}$ pairing state: Effect of Rashba SOC on spin susceptibility. (a) $\chi_{zz}(T)$ and (b) $\chi_{xx}(T)$, with finite Rashba SOC and no Zeeman field.}
\label{sus_socpx}
\end{figure}

\begin{figure}[tb]
\centering
\subfigure[]{
\includegraphics[width=0.96\linewidth]{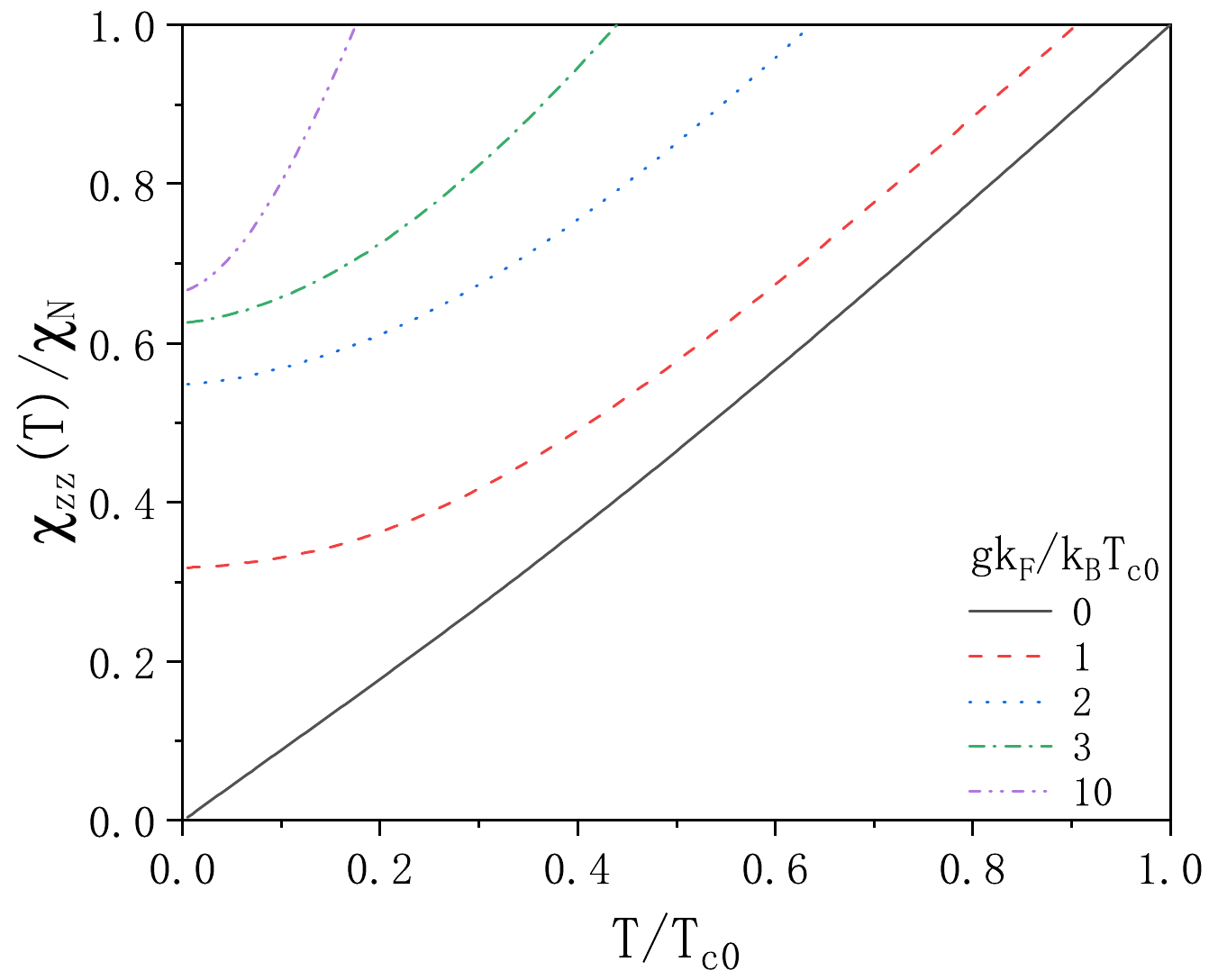}
}
\subfigure[]{
\includegraphics[width=0.96\linewidth]{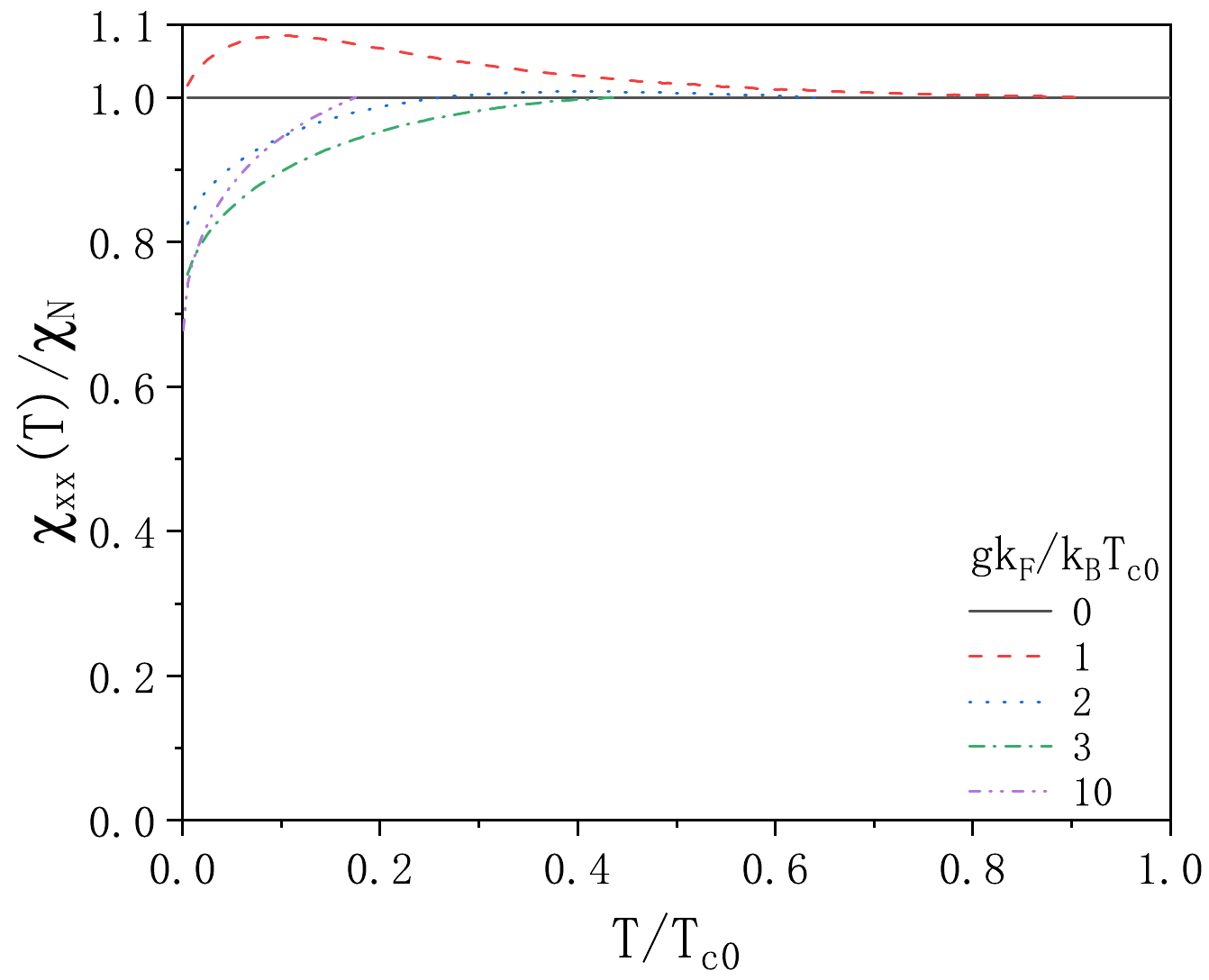}
}
\caption{The $k_z \hat{z}$ pairing state: Effect of Rashba SOC on spin susceptibility. (a) $\chi_{zz}(T)$ and (b) $\chi_{xx}(T)$, with finite Rashba SOC and no Zeeman field.}
\label{fig:chizzxx_kzz_soc}
\end{figure}

\begin{figure}[tb]
\centering
\includegraphics[width=0.90\linewidth]{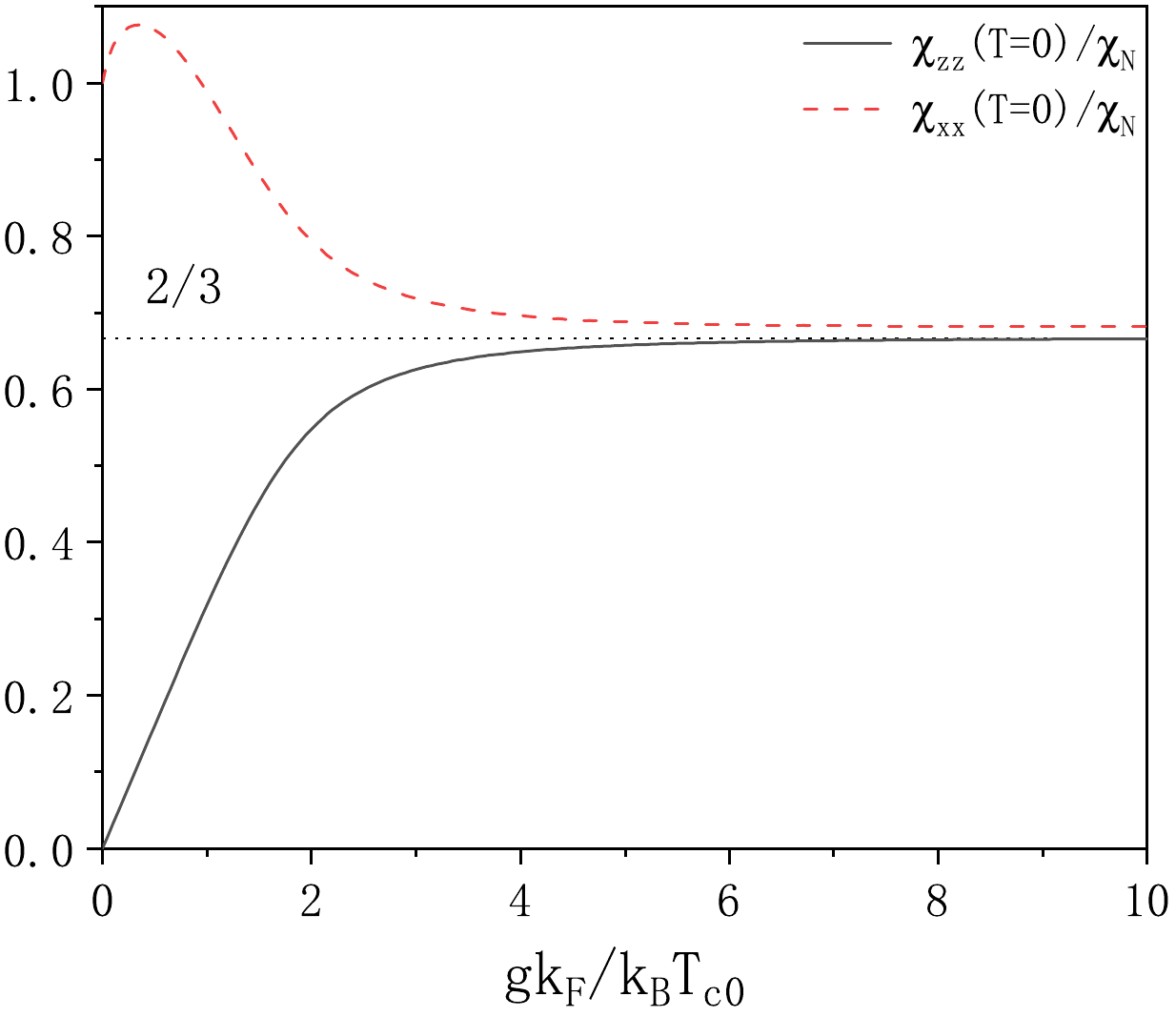}
\caption{The $k_z \hat{z}$ pairing state: Zero-temperature spin susceptibility $\chi_{zz}(T=0)/\chi_N$ and $\chi_{xx}(T=0)/\chi_N$ as a function of Rashba SOC, with no Zeeman field. Both approach $2/3$ in the strong SOC limit.}
\label{fig:chizzxx_kzz_zerotem_soc}
\end{figure}

Transitioning from the Zeeman field effects, we now analyze the influence of a finite Rashba SOC in the absence of any magnetic field. By computing the spin susceptibility components $\chi_{zz}(T)$ and $\chi_{xx}(T)$ for the OSP states $(k_x + i k_y) \hat{z}$ and $k_z \hat{z}$, as illustrated in Figs.~\ref{sus_socpx} and \ref{fig:chizzxx_kzz_soc}, we observe distinct behaviors.

Regarding temperature dependence, $\chi_{zz}(T)$ decreases with temperature below $T_c$ and retains a finite value at $T=0$ when Rashba SOC is nonzero, as shown in Figs.~\ref{sus_socpx}(a) and \ref{fig:chizzxx_kzz_soc}(a). In contrast, $\chi_{xx}(T)$ displays more intricate dynamics: for the $(k_x + i k_y) \hat{z}$ state, it increases monotonically below $T_c$, while for the $k_z \hat{z}$ state, it initially rises and then decreases, as depicted in Figs.~\ref{sus_socpx}(b) and \ref{fig:chizzxx_kzz_soc}(b).

Focusing on zero-temperature spin susceptibility as a function of Rashba SOC strength $g$, Fig.~\ref{sus_socpx} reveals that for $(k_x + i k_y) \hat{z}$, both $\chi_{zz}(T=0)$ and $\chi_{xx}(T=0)$ increase monotonically with $g$, until it completely suppresses superconductivity. For $k_z \hat{z}$, Fig.~\ref{fig:chizzxx_kzz_zerotem_soc} shows that $\chi_{zz}(T=0)$ also increases monotonically, but $\chi_{xx}(T=0)$ rises initially and then decreases, approaching $2 \chi_N /3 $ at large $g$. These patterns can be explained as follows:
\begin{itemize}
\item{} According to Eq.~\eqref{eq:xT0}, the zero-temperature spin susceptibility stems exclusively from particle-particle and hole-hole processes, with $\chi^{pp}(T=0)$ contributing, while $\chi^{ph+}(T=0) = \chi^{ph-}(T=0) = 0$.
\item For $\chi_{zz}(T=0)$, analysis using the Bogoliubov transformation matrices in Eqs.~\eqref{eq:chizzpp+_T0} and \eqref{eq:chizzpp-_T0} indicates that it arises solely from the inter-band term of $\chi_{zz}^{pp}(T=0)$, converging to $2 \chi_N /3 $ in the strong SOC limit, as outlined in Appendix~\ref{app:strongSOC_limit}.
\item{} For $\chi_{xx}(T=0)$, both intra- and inter-band terms, derived from Eqs.~\eqref{eq:chixxpp+_T0} and \eqref{eq:chixxpp-_T0}, play roles:
\begin{itemize}
\item{} In weak SOC, assuming a nearly constant gap $\Delta(T=0, g) \approx \Delta(T=0, g=0)$, the intra-band terms dominate and increase with $g$ due to the factor $1/E_{\mathbf{k}+} + 1/E_{\mathbf{k}-} > 2/E_{\mathbf{k}}$ [from Eq.~\eqref{eq:ekpekm}], accounting for $\chi_{xx}(T=0) > \chi_N$ at small $g$ (see Appendix~\ref{app:osp_chixx_soc}).
\item{} In strong SOC, gap suppression reduces the intra-band contribution, allowing inter-band terms to dominate and drive susceptibilities toward $2 \chi_N /3 $.
 \end{itemize}
\end{itemize}

In summary, the zero-temperature spin susceptibility is governed by particle-particle and hole-hole processes, with intra- and inter-band contributions. At low $g$, intra-band terms lead to an increase and potential exceedance of $\chi_N$ for $\chi_{xx}(T=0)$, but as $g$ grows, gap suppression shifts dominance to inter-band terms, resulting in convergence to $2 \chi_N /3 $. In this study,  the SOC strength is relative to the superconducting gap $\Delta_0$. The ``weak" and ``strong" regimes are defined by the ratio $g k_F / \Delta_0$. For candidate materials like K$_2$Cr$_3$As$_3$, where $\Delta_0 \sim 1$ meV, a strong SOC regime ($g k_F \gg \Delta_0$) is plausible if SOC is of the order 10 meV. For both OSP states, $\chi_{zz}(T=0)$ increases monotonically, while $\chi_{xx}(T=0)$ exhibits non-monotonicity in the $k_z \hat{z}$ case. Furthermore, the $(k_x + i k_y) \hat{z}$ state transitions to the normal state at a relatively small value of Rashba SOC, $g k_F / k_B T_{c0} = 1.52$~\cite{pang2025}, restricting $\chi_{xx}(T=0)$ to its increasing phase.

\subsubsection{Combination of Zeeman field and Rashba SOC}

\begin{figure}[tb]
\centering
\subfigure[]{
\includegraphics[width=0.96\linewidth]{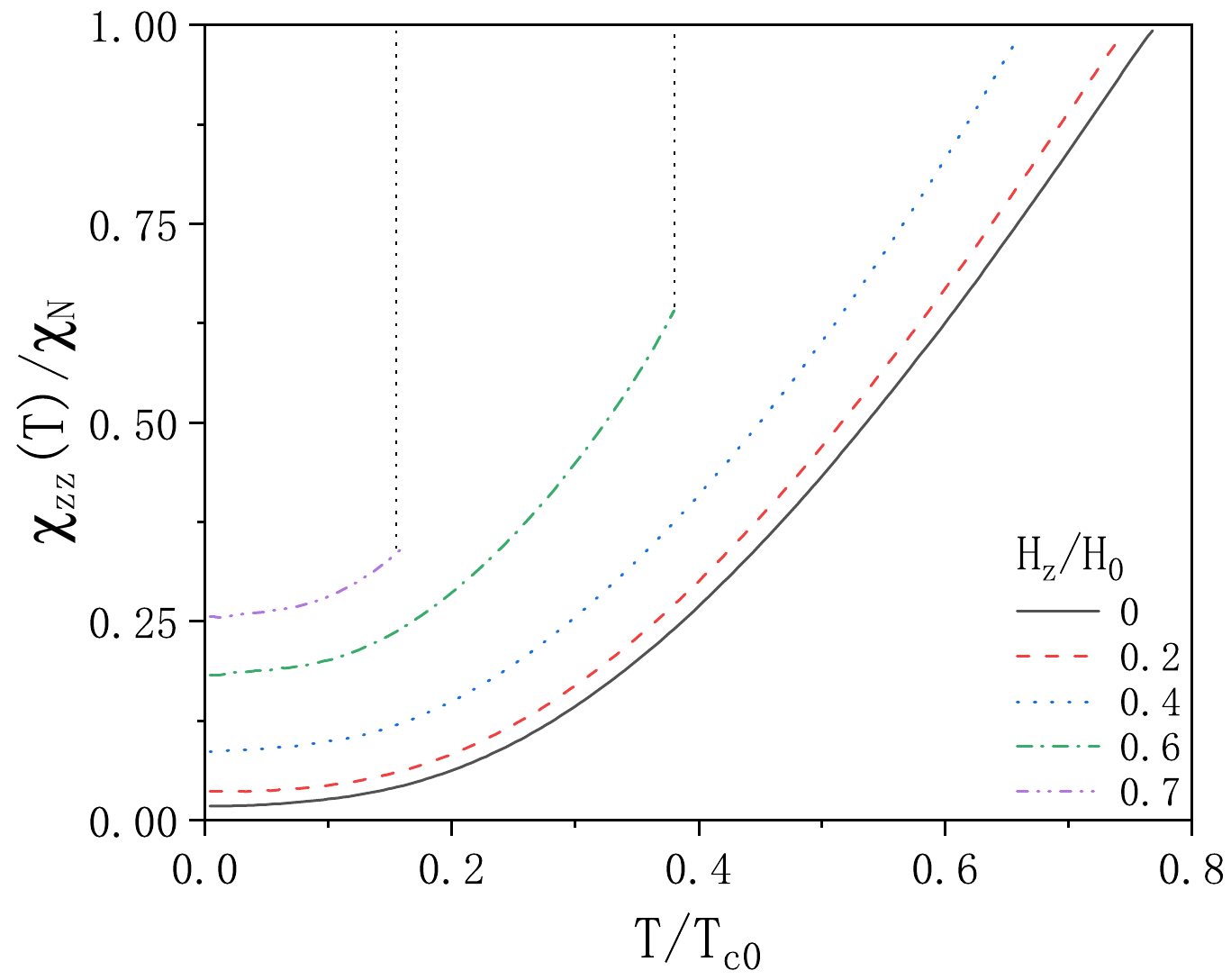}
}
\subfigure[]{
\includegraphics[width=0.96\linewidth]{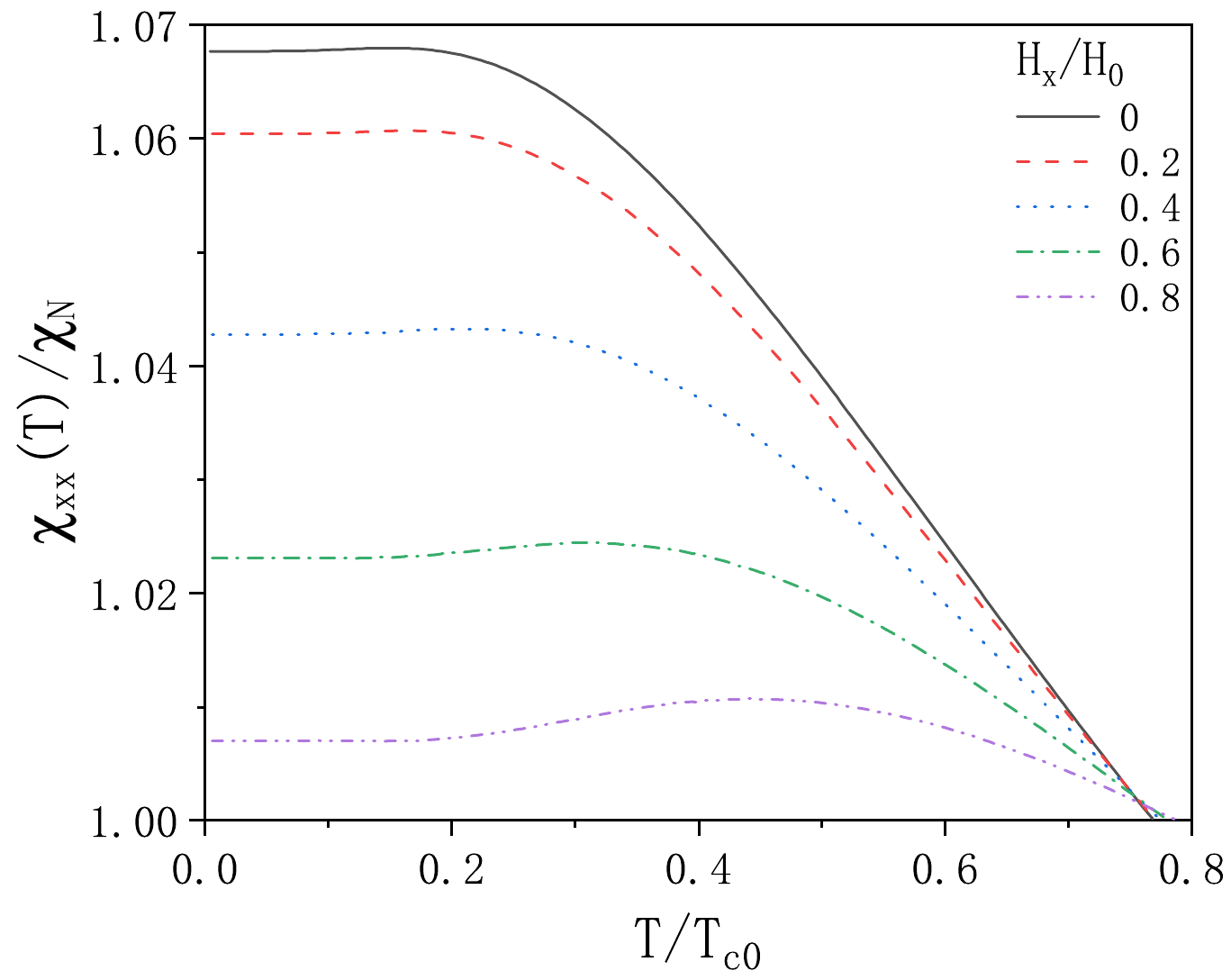}
}
\caption{The $(k_x + i k_y) \hat{z}$ pairing state: Spin susceptibility functions (a) $\chi_{zz}(T)$ and (b) $\chi_{xx}(T)$, with Rashba SOC strength fixed at $g k_F / k_B T_{c0} = 1$. The Zeeman field varies and is applied along the $z$- and $x$-directions, respectively. The definition of $H_0$ is provided at the end of Section~\ref{sec:s-wave-H}.}
\label{sus_socmagp1}
\end{figure}

\begin{figure}[tb]
\centering
\subfigure[]{
\includegraphics[width=0.96\linewidth]{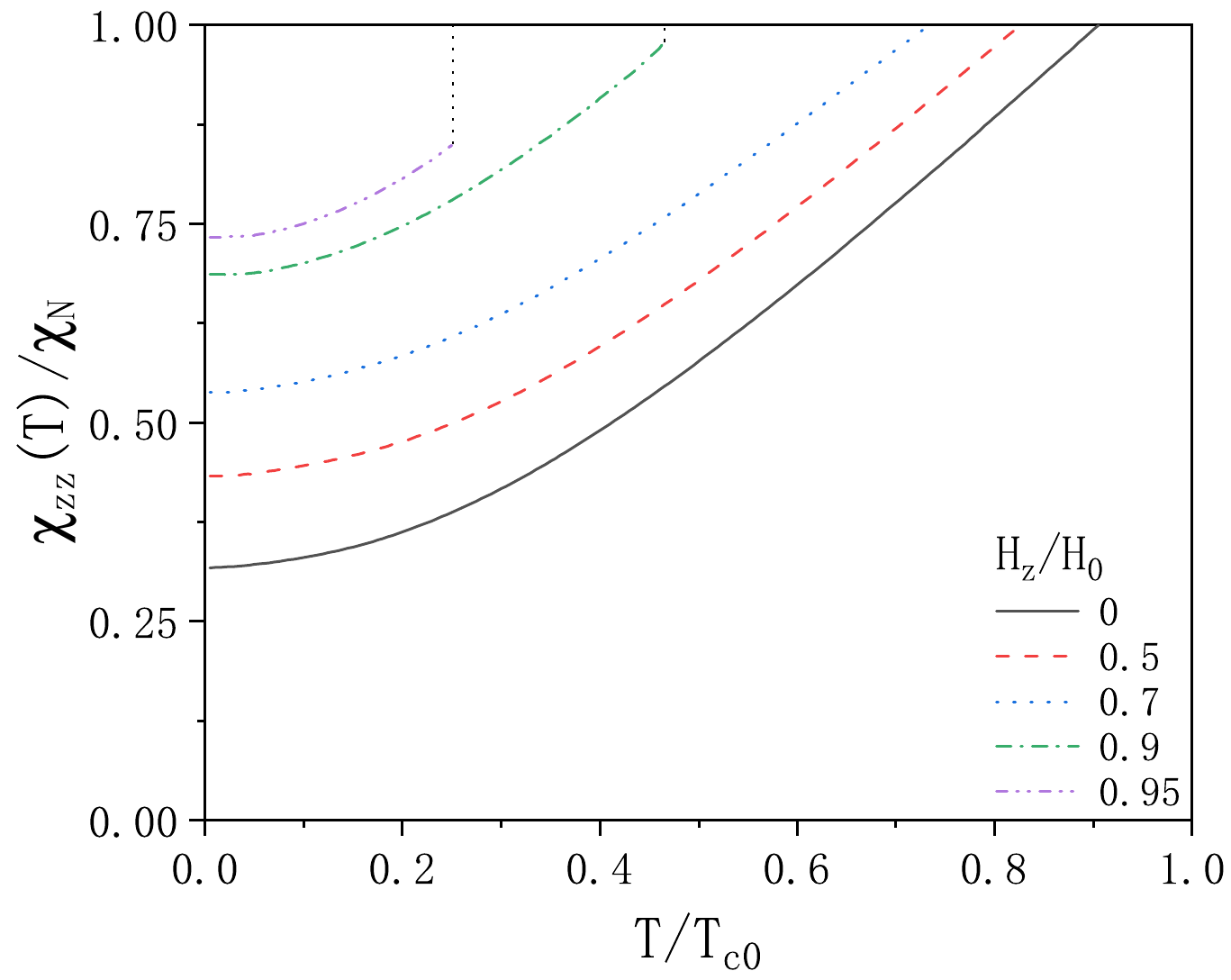}
}
\subfigure[]{
\includegraphics[width=0.96\linewidth]{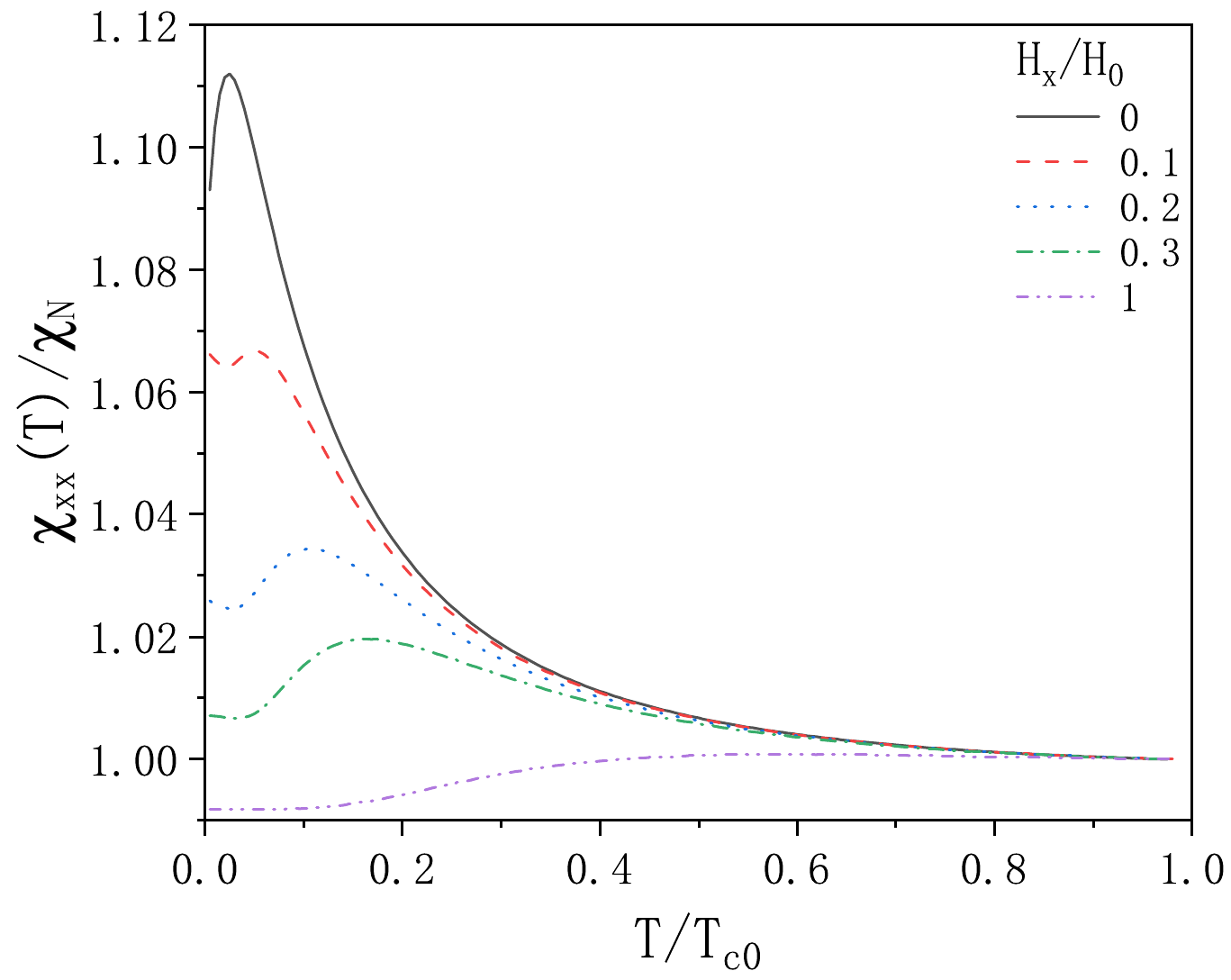}
}
\caption{The $k_z \hat{z}$ pairing state: Spin susceptibility functions (a) $\chi_{zz}(T)$ and (b) $\chi_{xx}(T)$, with Rashba SOC strength fixed at (a) $g k_F / k_B T_{c0} = 1$ and (b) $g k_F / k_B T_{c0} = 0.5$. The Zeeman field varies and is applied along the $z$- and $x$-directions, respectively. The definition of $H_0$ is provided at the end of Section~\ref{sec:s-wave-H}.}
\label{fig:suszzxx_kzz_magsoc}
\end{figure}

Building on the individual effects of the Zeeman field and Rashba SOC discussed earlier, we now examine their combined influence, where both factors interact to modify the spin susceptibility. The spin susceptibility components $\chi_{zz}(T)$ and $\chi_{xx}(T)$ are computed numerically, with results presented in Figs.~\ref{sus_socmagp1} and \ref{fig:suszzxx_kzz_magsoc}. Here, the Rashba SOC strength is fixed at $g k_F = k_B T_{c0}$ or $g k_F = 0.5 k_B T_{c0}$, while the Zeeman field varies in magnitude and direction.

The analysis reveals that $\chi_{zz}(T)$ decreases with temperature and retains a finite value at $T = 0$, while $\chi_{xx}(T)$ exceeds its normal-state value $\chi_N$ below the superconducting transition temperature $T_c$. Notably, a Zeeman field applied along the $x$-axis ($\mathbf{H} = H_x \hat{x}$) tends to suppress $\chi_{xx}(T)$ toward $\chi_N$ in the superconducting state ($T < T_c$), highlighting the interplay between the fields.

\subsection{Equal-spin pairing states}

Following the examination of OSP states, we now turn to the four ESP states listed in Table~\ref{tab:pwave}. These states can be unified through their d-vectors, expressed as:
\begin{equation}\label{eq:dvec4}
\mathbf{d}(\mathbf{k}) = \Delta \sin \theta_{\mathbf{k}} \left( \cos \phi_{\mathbf{k}} \hat{x} + \sin \phi_{\mathbf{k}} \hat{y} \right),
\end{equation}
where the angle $\phi_{\mathbf{k}}$ is specified in Table~\ref{tab:pwave}. As with the OSP states, detailed information on transition temperatures and quasiparticle excitation energies is available in Ref.~\cite{pang2025}.

\subsubsection{Zeeman field effect}

\begin{figure}[tb]
\centering
\includegraphics[width=1.0\linewidth]{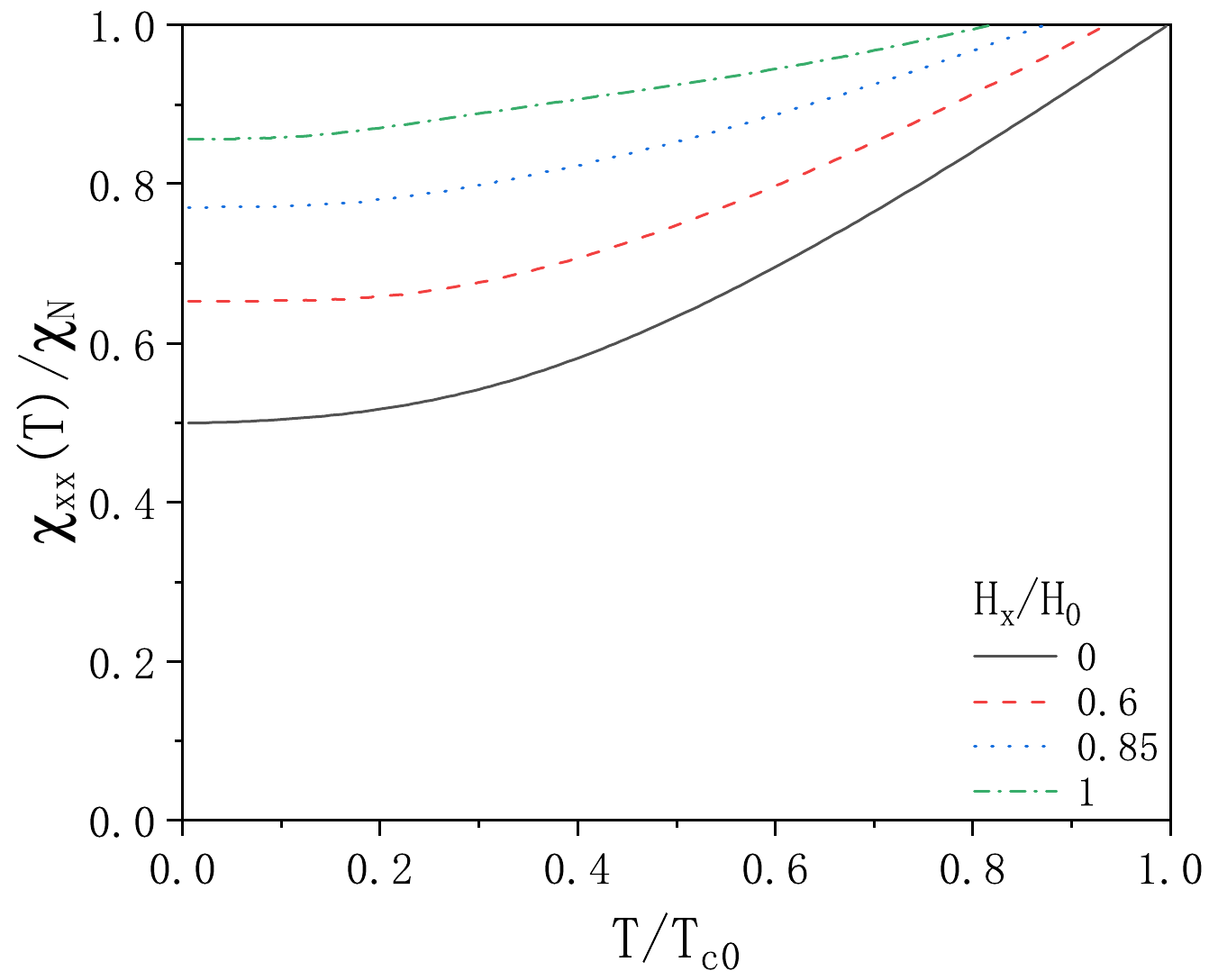} 
\caption{ESP states listed in Table~\ref{tab:pwave}: Spin susceptibility $\chi_{xx}(T)$ under a perpendicular Zeeman field $\mathbf{H} = H_x \hat{x}$. The definition of $H_0$ is provided at the end of Section~\ref{sec:s-wave-H}.}
\label{sus_socmagp}
\end{figure}

We begin by examining the effect of a parallel Zeeman field aligned with the $z$-axis, $\mathbf{H} = H_z \hat{z}$, in the absence of Rashba SOC. For all four ESP states, this field does not alter the superconducting transition temperature $T_c$~\cite{pang2025}, and the spin susceptibility remains temperature-independent, with $\chi_{zz}(T) = \chi_N$.

Next, we consider a perpendicular Zeeman field along the $x$-axis, $\mathbf{H} = H_x \hat{x}$. The results, illustrated in Fig.~\ref{sus_socmagp}, show that the spin susceptibility $\chi_{xx}(T)$ is identical across all four pairing states, as they share the same energy dispersion before and after the field is applied, as detailed in Ref.~\cite{pang2025}. As expected, the ratio $\chi_{xx}(T)/\chi_N$ decreases monotonically below $T_c$ and approaches $1/2$ at $T = 0$ and $H_x=0$, while increasing with the field strength $H_x$.

\subsubsection{Effect of Rashba SOC}

\begin{figure}[tb]
\centering
\subfigure[]{
\includegraphics[width=0.96\linewidth]{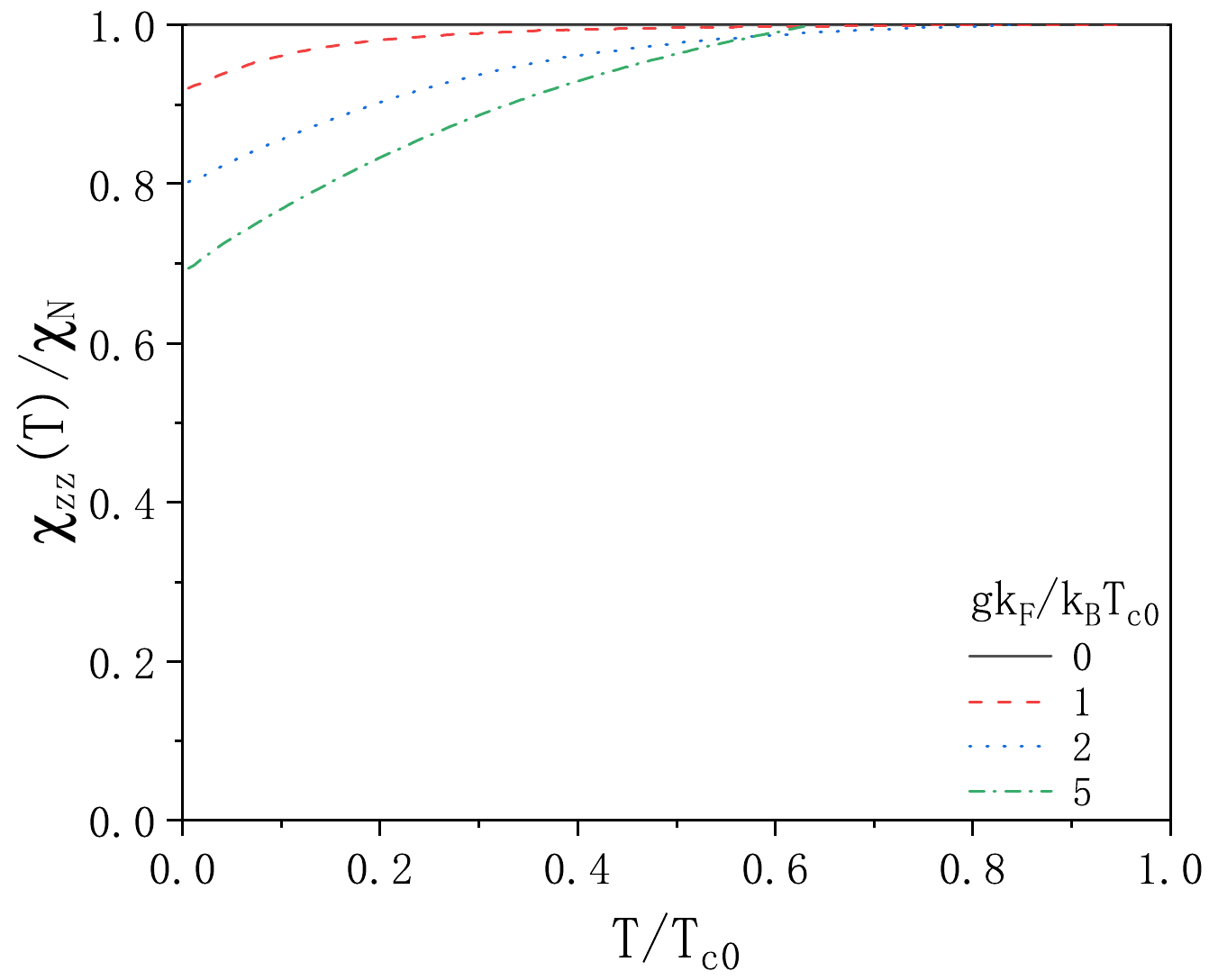} 
}
\subfigure[]{
\includegraphics[width=0.96\linewidth]{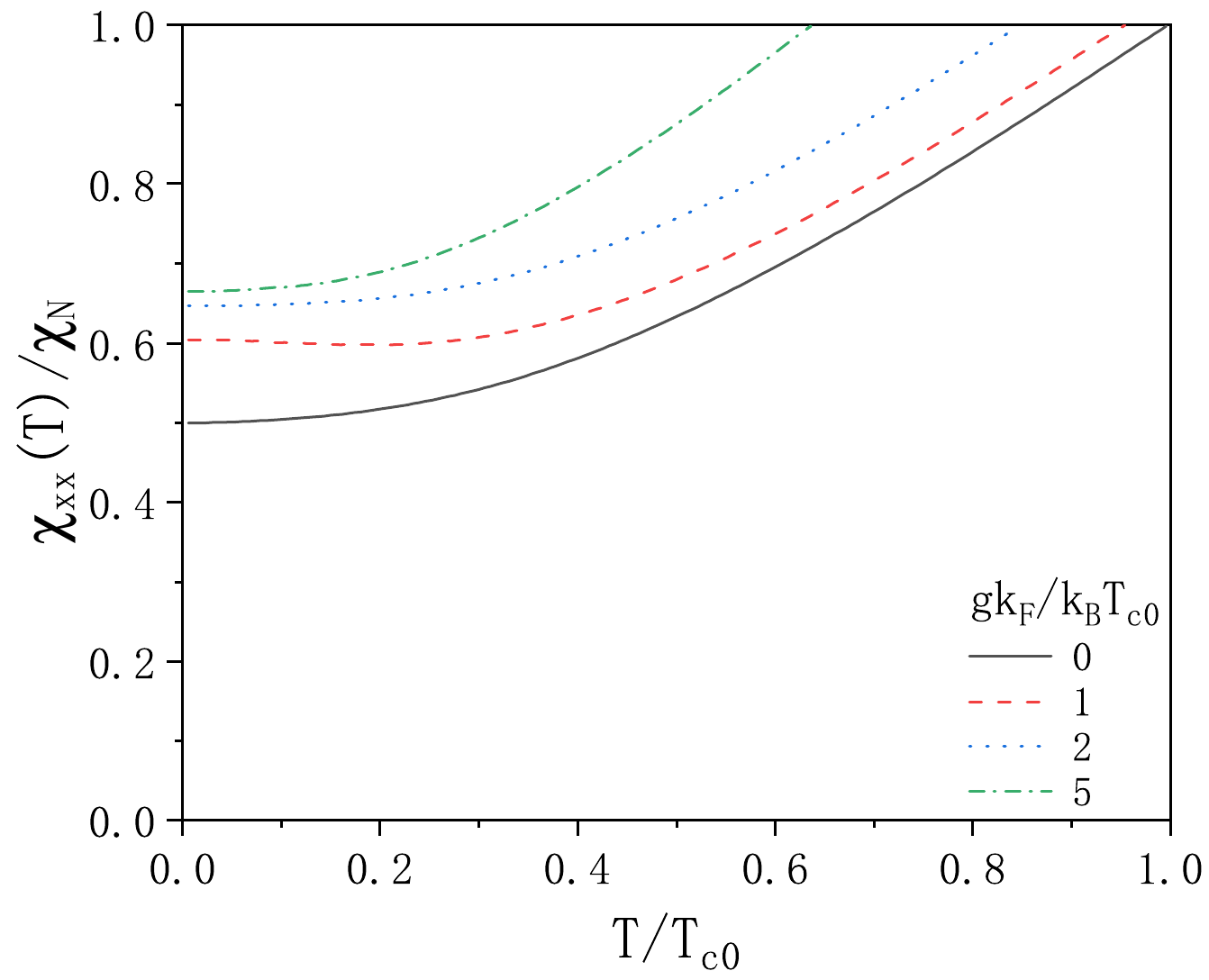}
}
\caption{The $k_x \hat{x} + k_y \hat{y}$ pairing state: Spin susceptibility functions (a) $\chi_{zz}(T)$ and (b) $\chi_{xx}(T)$, with finite Rashba SOC and no Zeeman field.}
\label{sus_socesp1}
\end{figure}

\begin{figure}[tb]
\centering
\subfigure[]{
\includegraphics[width=0.96\linewidth]{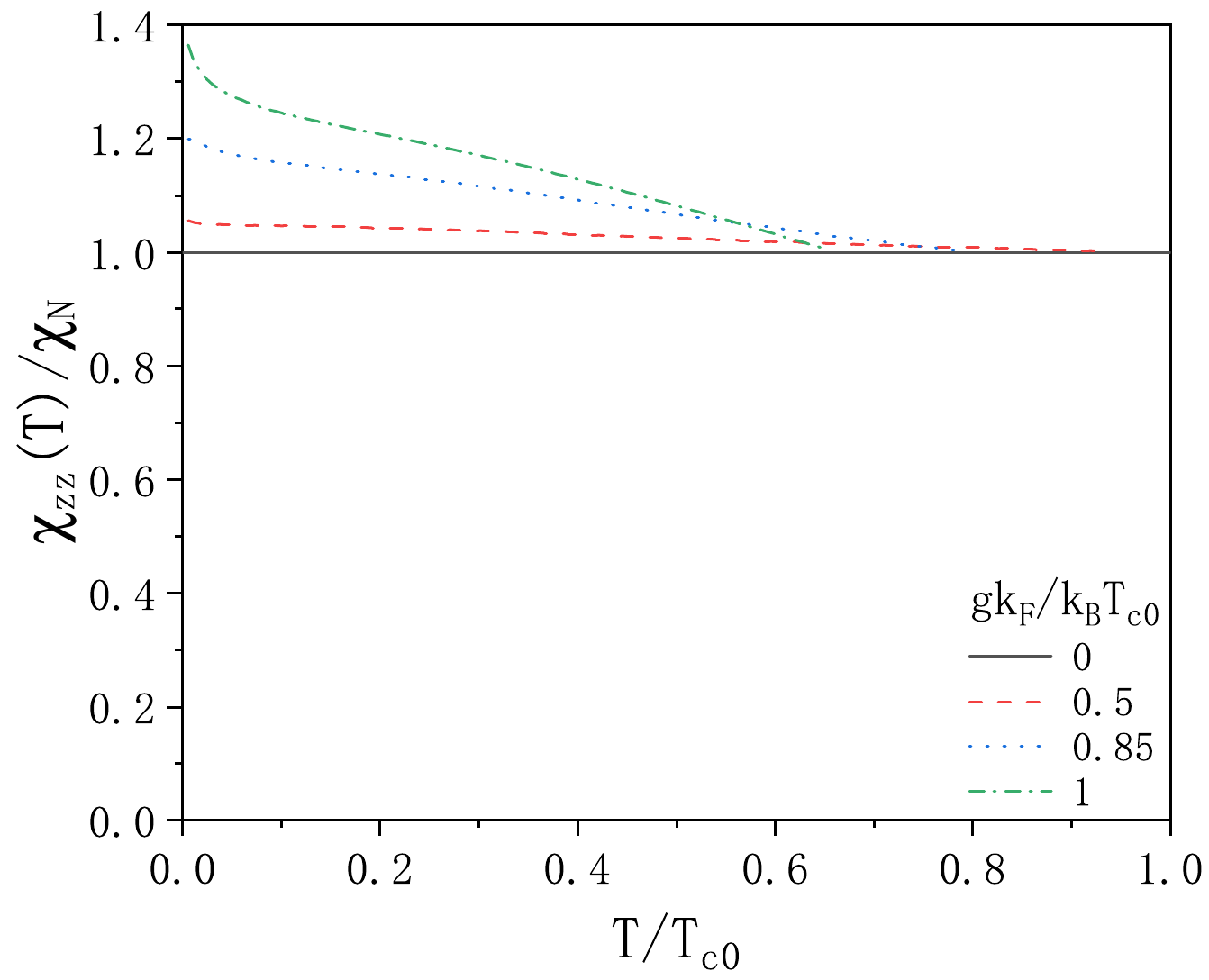}
}
\subfigure[]{
\includegraphics[width=0.96\linewidth]{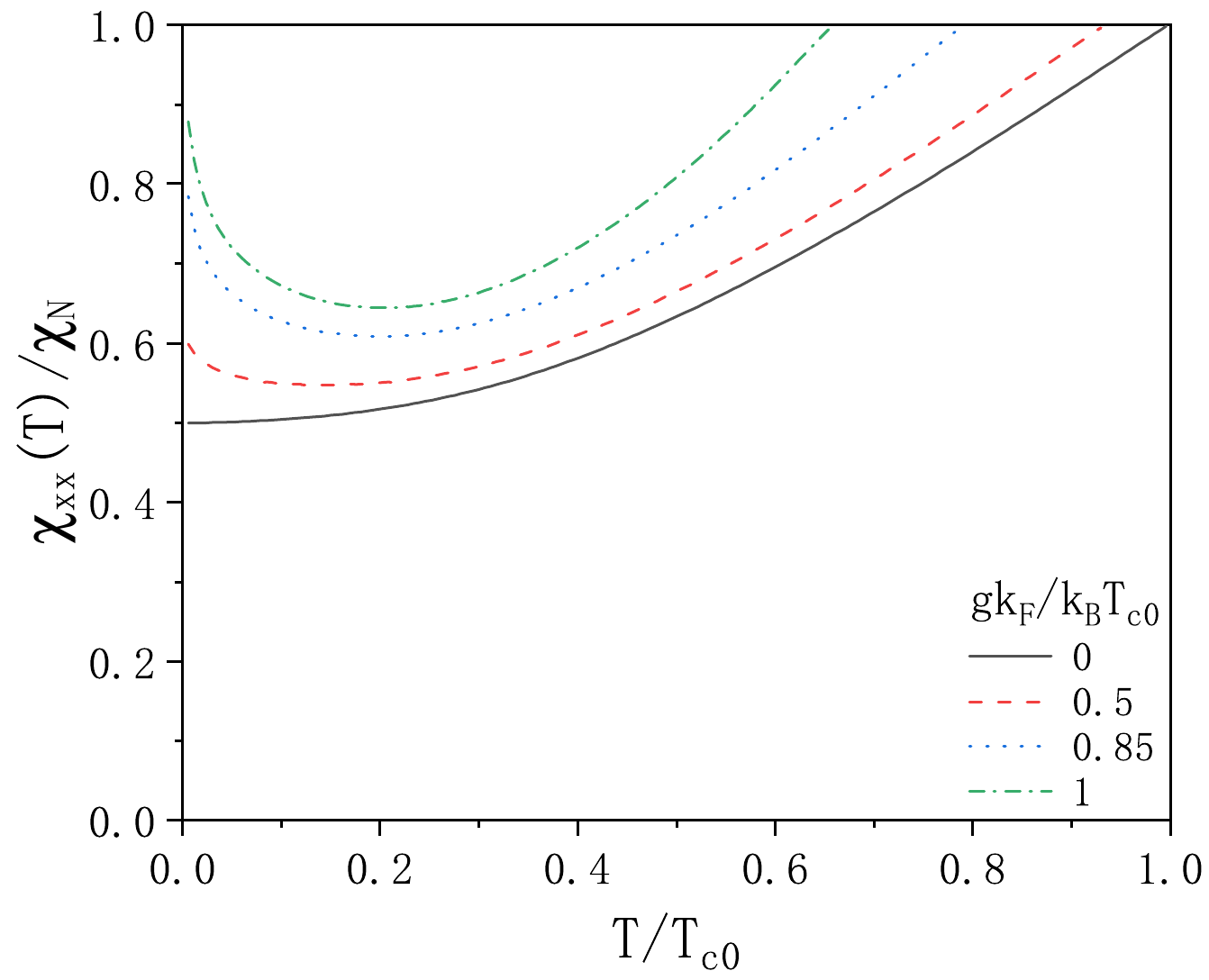}
}
\caption{The $k_y \hat{x} - k_x \hat{y}$ pairing state: Spin susceptibility functions (a) $\chi_{zz}(T)$ and (b) $\chi_{xx}(T)$, with finite Rashba SOC and no Zeeman field.}
\label{sus_socesp2}
\end{figure}

\begin{figure}[tb]
\centering
\subfigure[]{
\includegraphics[width=0.96\linewidth]{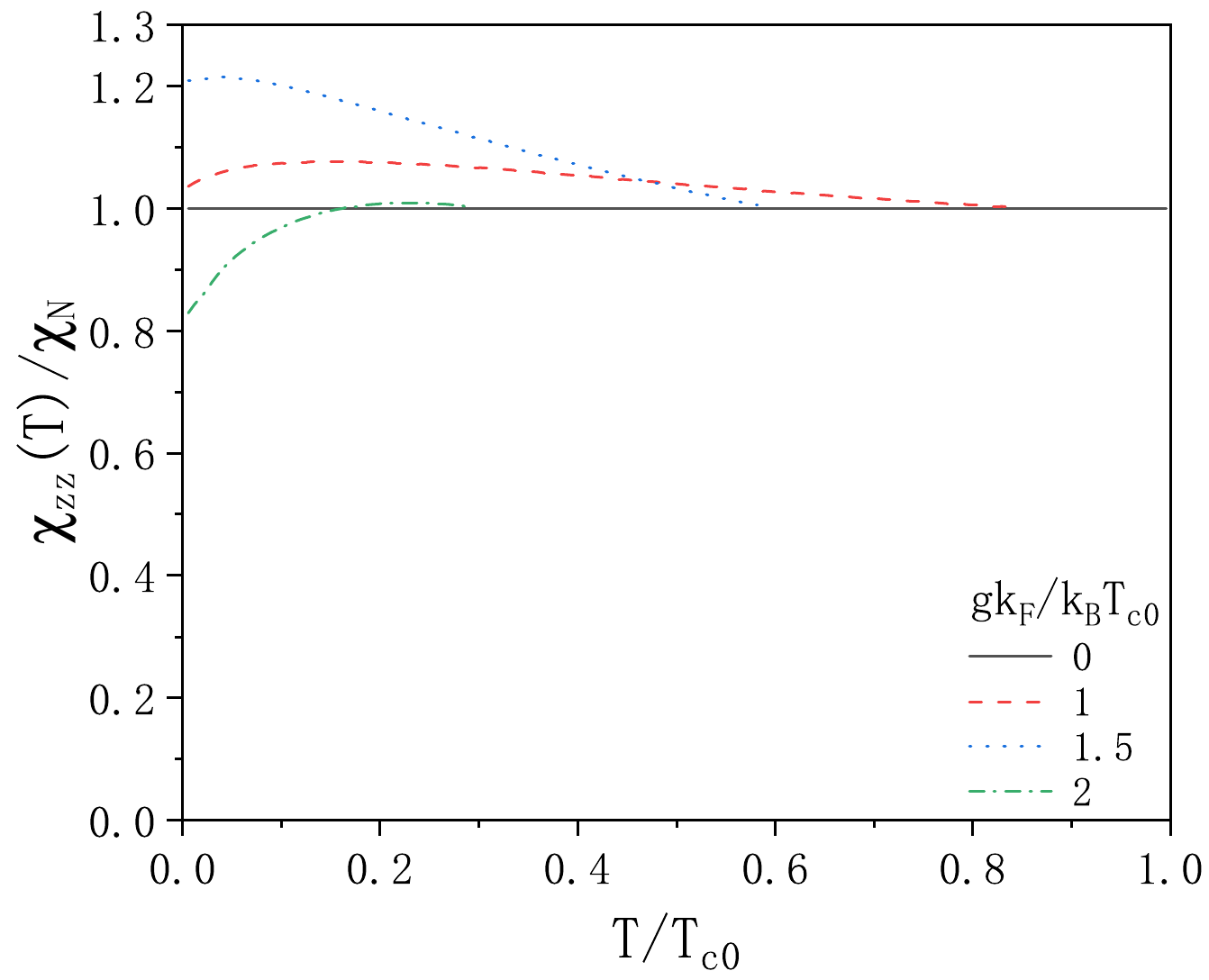} 
}
\subfigure[]{
\includegraphics[width=0.96\linewidth]{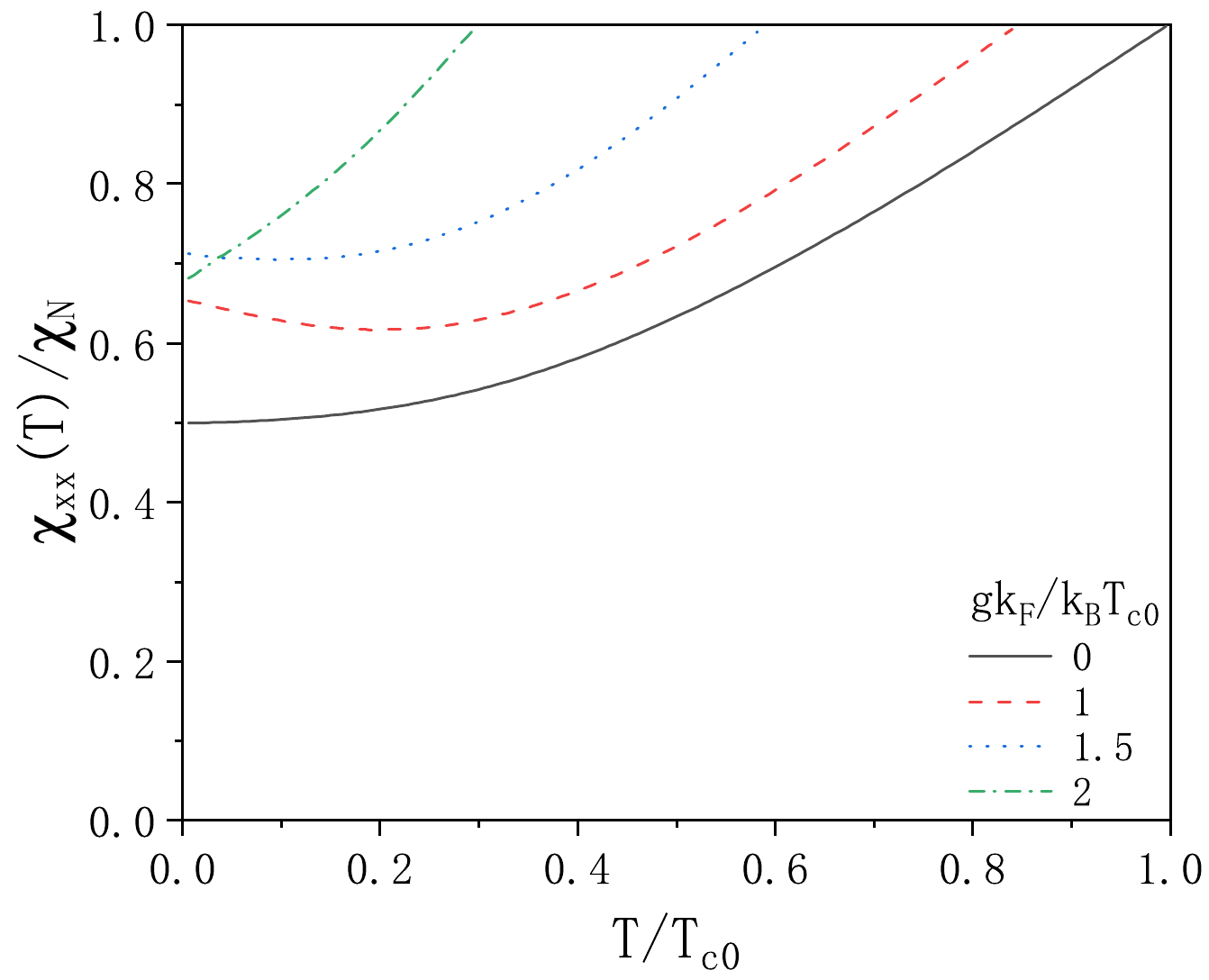}
}
\caption{The $k_y \hat{x} + k_x \hat{y}$ and $k_x \hat{x} - k_y \hat{y}$ pairing states: Spin susceptibility functions (a) $\chi_{zz}(T)$ and (b) $\chi_{xx}(T)$, with finite Rashba SOC and no Zeeman field.}
\label{sus_socesp3}
\end{figure}

We begin by examining the numerical results for the spin susceptibility components $\chi_{zz}(T)$ and $\chi_{xx}(T)$ in the four ESP states under the influence of Rashba SOC, as illustrated in Figs.~\ref{sus_socesp1}, \ref{sus_socesp2}, and \ref{sus_socesp3}. Consistent with Ref.~\cite{pang2025}, Rashba SOC reduces the superconducting transition temperature $T_c$ for all four ESP states. To understand these effects, we first analyze the temperature dependence of the spin susceptibility, followed by a detailed look at the zero-temperature behavior and its interpretations.

In terms of temperature dependence:
\begin{itemize}
\item{} For $\chi_{zz}(T)$, the absence of Rashba SOC ($g=0$) results in $\chi_{zz}(T) = \chi_N$ for all four ESP states.
However, with a finite $g$, the behavior varies: $\chi_{zz}(T)$ may decrease monotonically for the $k_y \hat{x} - k_x \hat{y}$ state, increase monotonically for the $k_x \hat{x} + k_y \hat{y}$ state, or increase and then decrease for the $k_x \hat{x} - k_y \hat{y}$ and $k_y \hat{x} + k_x \hat{y}$ states below $T_c$. 
\item{} For $\chi_{xx}(T)$, Rashba SOC generally induces a decreasing trend as temperature lowers, with values exceeding $\chi_N/2$ at zero temperature across all states.
\item{} A notable exception is the $k_y \hat{x} - k_x \hat{y}$ state in Fig.~\ref{sus_socesp2}(b), where $\chi_{xx}(T)$ exhibits non-monotonic behavior as temperature increases from zero, linked to nodal surfaces introduced by Rashba SOC~\cite{pang2025}. At zero temperature, wave vectors near these nodal surfaces contribute significantly to $\chi_{xx}(T=0)$ via the factor $[ 1 - 2f(E_{\mathbf{k}-}) ]/E_{\mathbf{k}-}$ in $\chi_{xx}^{pp+}$ from Eq.~\eqref{eq:chi_ppp}, which diminishes rapidly with rising temperature, driving the overall decrease.
\item{} Given that a continuous phase transition maintains $\chi_{\mu\mu}(T_c) = \chi_N$, the observed temperature dependence necessitates a closer examination of the zero-temperature susceptibility to uncover the underlying mechanisms.
\end{itemize}

\begin{figure}[tb]
\centering
\subfigure[]{
\includegraphics[width=0.90\linewidth]{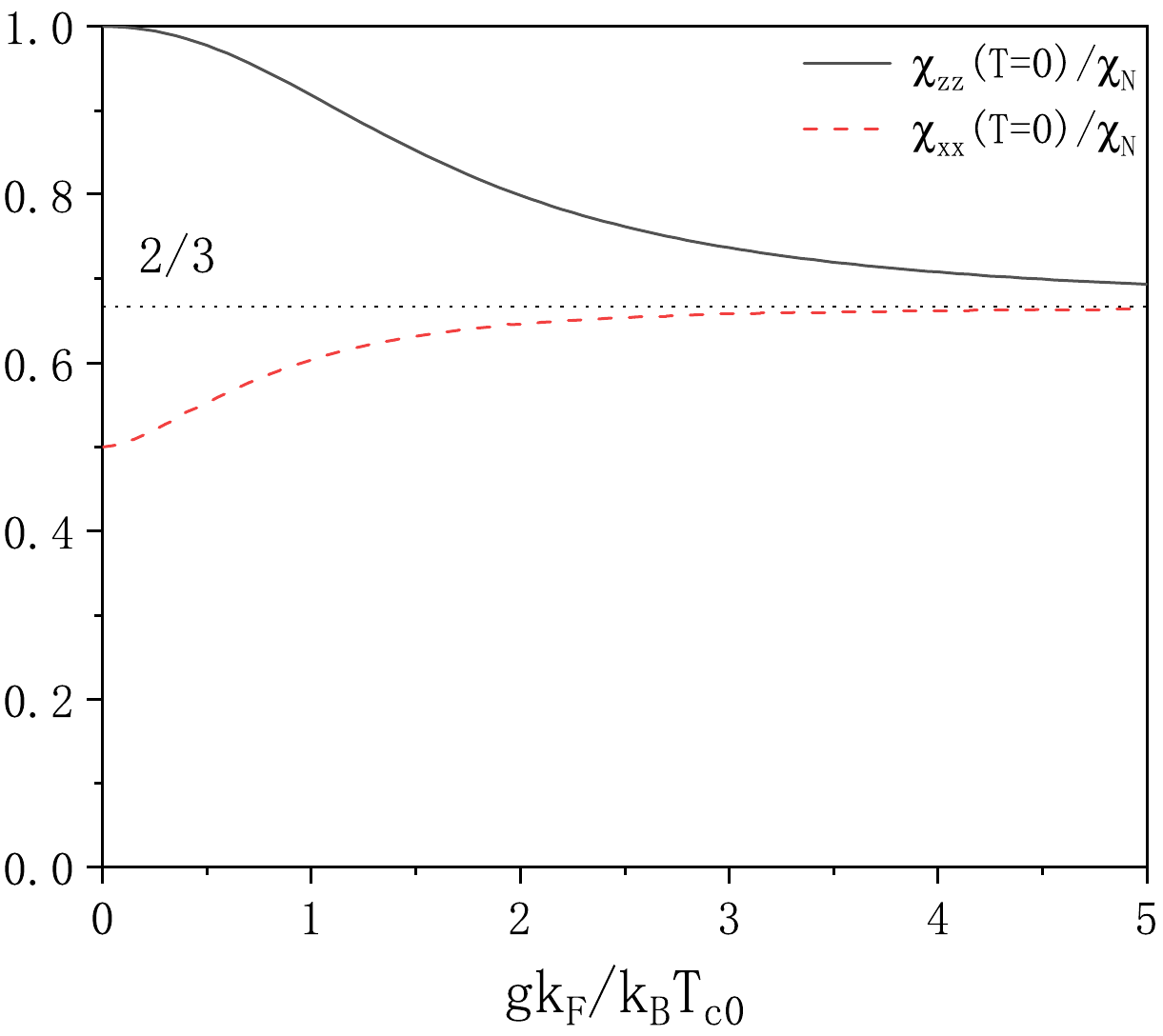}
}
\subfigure[]{
\includegraphics[width=0.90\linewidth]{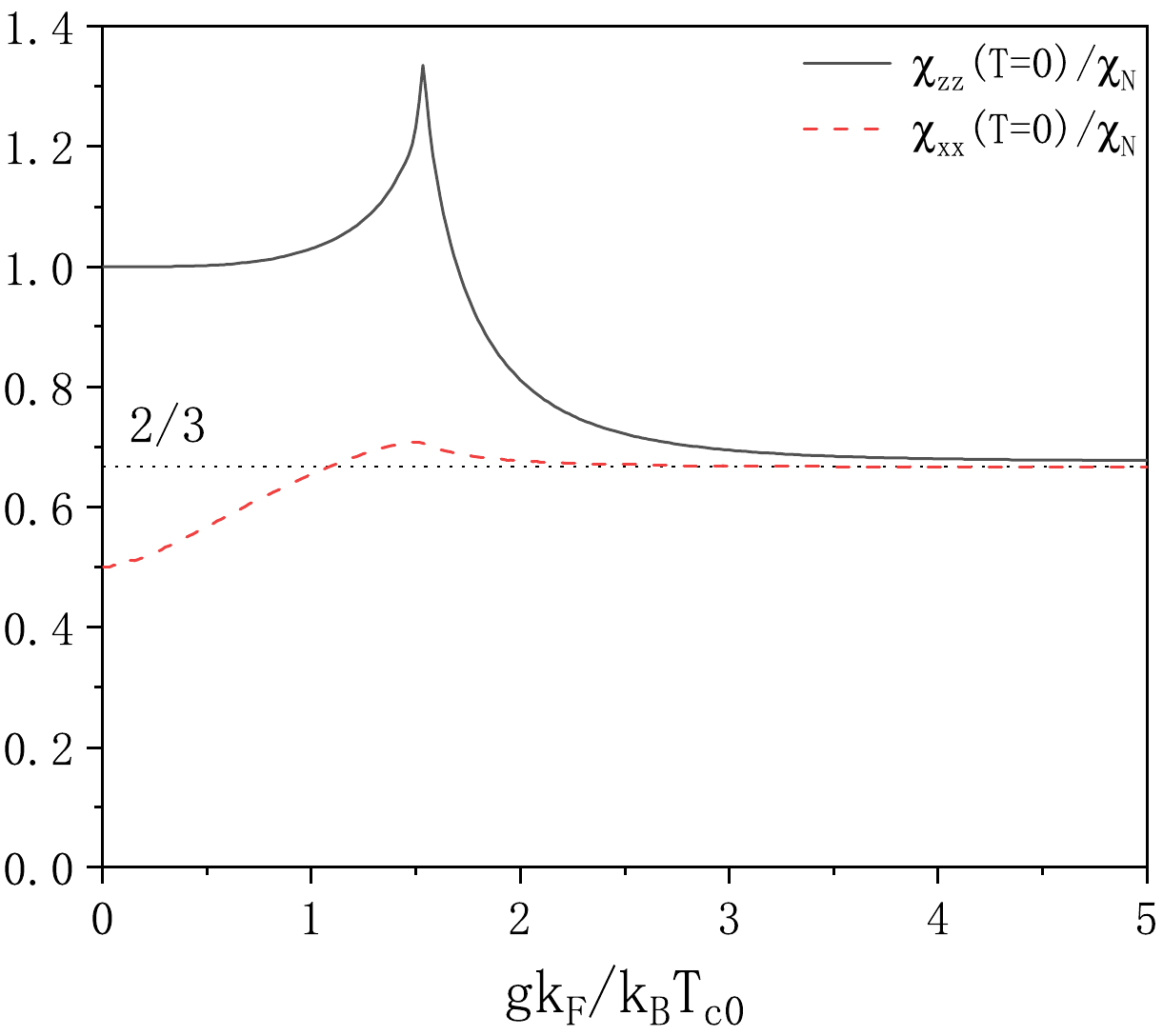}
}
\caption{Effects of Rashba SOC on zero-temperature spin susceptibility with components $\chi_{zz}(T=0)$ and $\chi_{xx}(T=0)$ for pairing states (a) $k_x \hat{x} + k_y \hat{y}$, (b) $k_x \hat{x} - k_y \hat{y}$ and $k_y \hat{x} + k_x \hat{y}$. Both components approach $2 \chi_N /3  $ under strong Rashba SOC. For (b) pairing states $k_x \hat{x} - k_y \hat{y}$ and $k_y \hat{x} + k_x \hat{y}$, the component $\chi_{zz}(T=0)$ diverges when $g k_F = \Delta(T=0, g)$.}
\label{fig:chizzxx_kxx-kyy_zerotem_soc}
\end{figure}

Shifting focus to the zero-temperature limit, the Rashba SOC strength $g$ influences the susceptibility in distinct ways, as summarized in the following key characteristics derived from Figs.~\ref{sus_socesp1}, \ref{sus_socesp2}, \ref{sus_socesp3}, and \ref{fig:chizzxx_kxx-kyy_zerotem_soc}:
\begin{itemize}
\item For the $k_x \hat{x} + k_y \hat{y}$ pairing state, both $\chi_{zz}(T=0)$ and $\chi_{xx}(T=0)$ approach $2 \chi_N /3 $ monotonically as $g$ increases.
\item For the $k_y \hat{x} - k_x \hat{y}$ pairing state, both $\chi_{zz}(T=0)$ and $\chi_{xx}(T=0)$ increase monotonically with $g$ until a critical value is reached, after which a quantum phase transition to the normal state occurs~\cite{pang2025}.
\item For the $k_x \hat{x} - k_y \hat{y}$ and $k_y \hat{x} + k_x \hat{y}$ pairing states, $\chi_{zz}(T=0)$ and $\chi_{xx}(T=0)$ initially increase and then decrease to $2 \chi_N /3 $ as $g$ grows, with $\chi_{zz}(T=0)$ diverging at $g k_F = \Delta(T=0,g)$~as illustrated in Fig.~\ref{fig:chizzxx_kxx-kyy_zerotem_soc}(b).
\end{itemize}

These observations can be interpreted through the theoretical expressions for spin susceptibility, which reveal the roles of intra- and inter-band terms:
\begin{itemize}
\item As demonstrated in opposite-spin pairing (OSP) states, the intra- and inter-band components of $\chi_{\mu\mu}^{pp}(T=0)$ dictate the susceptibility behavior in weak and strong SOC limits, accounting for the trends in the $k_x \hat{x} - k_y \hat{y}$ and $k_y \hat{x} + k_x \hat{y}$ states.
\item For the $k_x \hat{x} + k_y \hat{y}$ pairing, $\sin \Omega_{\mathbf{k}} = 0$ indicates that $\chi_{zz}(T=0)$ depends exclusively on the inter-band term of $\chi_{zz}^{pp}(T=0)$, resulting in a monotonic decrease with increasing $g$ rather than non-monotonicity.
\item For the $k_y \hat{x} - k_x \hat{y}$ pairing, strong Rashba SOC triggers a phase transition to the normal state, restricting susceptibility to its increasing phase, akin to the $(k_x + i k_y) \hat{z}$ OSP state. Rashba SOC also introduces nodal surfaces, enhancing $\chi_{\mu\mu}(T=0)$ contributions near these regions through the factor $1/E_{\mathbf{k}-}$ in $\chi_{\mu\mu}^{pp+}(T=0)$~\cite{pang2025}.
\item The divergence of $\chi_{zz}(T=0)$ at $g k_F = \Delta(T=0, g)$ for the $k_x \hat{x} - k_y \hat{y}$ and $k_y \hat{x} + k_x \hat{y}$ states is due to nodal lines induced by Rashba SOC. Wave vectors near these lines significantly boost susceptibility, and as outlined in Appendix~\ref{app_diverge}, the integral converges for $g k_F \neq \Delta$ but diverges at the nodal line-equator intersection when $g k_F = \Delta$. In contrast, $\chi_{xx}(T=0)$ avoids divergence because of the vanishing prefactor $\cos^2 \theta_{\mathbf{k}}$ in Eq.~\eqref{eq:chixxpp+_T0} at $\theta = \pi/2$.
\end{itemize}

\section{Summary and Discussions}
\label{sec:summary}

In this work, we have carried out a comprehensive study of the static spin susceptibility of superconductors that experience both a Zeeman magnetic field and Rashba-type SOC. The analysis, valid for arbitrary field strength and SOC magnitude, covers isotropic $s$-wave spin-singlet as well as several representative $p$-wave spin-triplet pairing states, and may be viewed as a non-perturbative extension of classical linear-response treatments.

\paragraph{General Framework.}
Starting from a single-band Bogoliubov-de Gennes Hamiltonian [Eq.~(\ref{eq:H})] that contains the Zeeman term $\mu_{B}\mathbf{H}\cdot\hat\sigma$ and the Rashba term $g\,\mathbf{k}\cdot\hat\sigma$, we derived (i)~self-consistent gap equations for arbitrary pairing channels and (ii)~a compact Kubo expression for uniform susceptibility, decomposed into intra-/inter-band particle-hole and particle-particle contributions [Eqs.~(\ref{eq:chikd})]. Two universal results follow:
(i)~only particle-particle channels survive at $T=0$, and (ii)~only particle-hole channels survive at $T=T_{c}$, so that $\chi(T_{c}^{-})=\chi(T_{c}^{+})=\chi_{N}$ for a continuous superconducting transition. These identities are indifferent to the pairing symmetry and constitute useful benchmarks for future numerical or experimental work.

\paragraph{$s$-wave and Spin-singlet Pairing.}
For conventional BCS pairing, we recover the Yosida temperature suppression when both Zeeman field and Rashba SOC are absent. A pure Zeeman field lowers $T_{c}$ towards the Pauli limit and produces a first-order transition when $T_{c}/T_{c0}<0.556$ [Fig.~\ref{fig:sus_mag}], whereas pure SOC leaves $T_{c}$ untouched but generates a residual susceptibility $\chi(0) \to 2\chi_{N}/3$ in the strong SOC limit [Fig.~\ref{sus_SOC}]. When both terms are present, the competition is highly non-trivial:
(i)~$\chi_{zz}(T)$ always decreases below $T_{c}$ but never vanishes; (ii)~for sufficiently large $g$ a Bogoliubov Fermi surface appears between two Zeeman critical fields, producing kinks in $\chi_{zz}(0)$ [Fig.~\ref{sus_zero}].

\paragraph{$p$-wave and Spin-triplet Pairing.}
The spin-triplet pairing states divide naturally into OSP and ESP classes (Table~\ref{tab:pwave}), each displaying a characteristic anisotropy.

\begin{itemize}
\item \textbf{OSP states} [$(k_{x}+ik_{y})\hat{z}$, $k_{z}\hat{z}$] respond to a parallel field like spin-singlet pairing and to a transverse field like ESP: $\chi_{zz}(T)$ is reduced below $T_{c}$ for $\mathbf H\parallel\hat z$~[Fig.~\ref{fig:OSP_parallel_mag}], whereas $\chi_{xx}(T)=\chi_{N}$ for $\mathbf H\perp\hat z$. Rashba SOC gaps the nodal quasiparticles, drives $\chi_{\mu\mu}(T=0) \to 2\chi_{N}/3$, and can convert the Zeeman-induced transition into one involving a Bogoliubov Fermi surface
[Figs.~\ref{sus_socpx} -- \ref{fig:suszzxx_kzz_magsoc}].

\item \textbf{ESP states} [$k_{x}\hat{x}\pm k_{y}\hat{y}$, $k_{y}\hat{x}\pm k_{x}\hat{y}$]       keep $\chi_{zz}(T)=\chi_{N}$ for $\mathbf H\parallel\hat z$, while a transverse field or SOC lowers $T_{c}$ and reduces the in-plane susceptibility $\chi_{xx}$(T) toward $\chi_{N}/2$ at low temperature [Figs.~\ref{sus_socesp1} -- \ref{sus_socesp3}].
In two of the ESP states ($k_{x}\hat{x}-k_{y}\hat{y}$ and $k_{y}\hat{x} + k_{x}\hat{y}$) SOC introduces line nodes~\cite{pang2025} that cause a divergent $\chi_{zz}(T=0)$ when $gk_{F}=\Delta$ [Fig.~\ref{fig:chizzxx_kxx-kyy_zerotem_soc}(b)], signalling a topological change of the nodal structure.
\end{itemize}

It is noteworthy that finite Rashba SOC generally enhance spin susceptibility below the superconducting $T_c$, causing specific components of them to exceed the normal-state value $\chi_N = \chi(T=T_c^{+})$. For instance, this enhancement is evident in the in-plane component $\chi_{xx}(T)$ for OSP states, as demonstrated in Figs.~\ref{sus_socpx}(b), \ref{fig:chizzxx_kzz_soc}(b), \ref{fig:chizzxx_kzz_zerotem_soc}, \ref{sus_socmagp1}(b), and \ref{fig:suszzxx_kzz_magsoc}. Similarly, the parallel component $\chi_{zz}(T)$ is affected in ESP states, except for the $k_x\hat{x} + k_y\hat{y}$ state, as shown in Figs.~\ref{sus_socesp2}(a), \ref{sus_socesp3}(a), and \ref{fig:chizzxx_kxx-kyy_zerotem_soc}. This discrepancy arises because Rashba SOC introduces additional nodal surfaces in the $k_y\hat{x} - k_x\hat{y}$ state and nodal lines in the $k_x\hat{x} - k_y\hat{y}$ and $k_y\hat{x} + k_x\hat{y}$ states, but does not generate any extra gap nodes in the $k_x\hat{x} + k_y\hat{y}$ state~\cite{pang2025}, thereby weakening similar enhancements.

\paragraph{Implications for Experiment}

Our results provide concrete predictions for the Knight shift in non-centrosymmetric and strongly anisotropic superconductors, particularly in regimes where Zeeman energies and Rashba SOC splittings are comparable to the superconducting gap. These predictions offer guidance for future experimental and theoretical studies in the following aspects: (i) the appearance of a residual plateau near $2\chi_{N}/3$ as SOC is tuned (e.g., via chemical pressure) indicates that the Van Vleck contribution associated with band splitting dominates the magnetic response; (ii) the field-direction dependence of the spin susceptibility $\chi(T)$ provides a clear diagnostic of the underlying pairing symmetry. For OSP states, a magnetic field applied parallel to the spin-quantization axis suppresses $\chi_{\parallel}(T)$, while a perpendicular field leaves $\chi_{\perp}(T)$ unchanged at $\chi_{N}$. ESP states show the opposite trend. This strong anisotropy directly reflects how the condensate responds to a magnetic field depending on its spin structure; and (iii)  a key prediction of this work is that tuning the Rashba SOC can drive certain spin-triplet (ESP) states into a regime where SOC induces line nodes in the quasiparticle spectrum. In this regime, the zero-temperature longitudinal susceptibility $\chi_{zz}(0)$ diverges because the particle-particle term in Eq.~\eqref{eq:xT0} becomes singular along the SOC-induced nodal line, as illustrated in Fig.~\ref{fig:chizzxx_kxx-kyy_zerotem_soc}(b). Thus, in systems where Rashba SOC is tunable -- such as gated thin films -- a divergence in $\chi_{zz}(0)$ serves as a direct experimental signature of SOC-driven nodal-line formation. Since gate-controlled SOC has already been realized in a variety of two-dimensional materials and heterostructures~\cite{Nitta97,Miller03,Caviglia10,Manchon15}, similar control should also be achievable in thin films whose bulk electronic structure is three-dimensional.

To contextualize these implications, we focus on A$_2$Cr$_3$As$_3$ (A = Na, K, Rb, or Cs), a recently discovered family of superconductors with $T_c$ up to 8 K~\cite{Bao15,Tang15_1,Tang15_2,Mu18}. Density functional theory calculations reveal that Cr-3d orbitals dominate the electronic states near the Fermi level, featuring three bands: quasi-1D $\alpha$ and $\beta$ bands with flat Fermi surfaces and a 3D $\gamma$ band~\cite{Jiang15,Wu15}. The significant contribution of the 3D $\gamma$ band to the density of states justifies our theoretical model's assumption of a spherical Fermi surface. Moreover, the $P6m2$ space group of these compounds lacks spatial inversion symmetry~\cite{Bao15}, supporting the inclusion of Rashba SOC in our analysis.

A specific NMR experiment on K$_2$Cr$_3$As$_3$ single crystals~\cite{Triplet2021} provides strong evidence for spin-triplet pairing, with the $^{75}$As Knight shift remaining unchanged below $T_c$ for magnetic fields in the $ab$ plane but approaching zero along the $c$ axis at low temperatures. This behavior aligns with an  OSP state and a d-vector oriented along the $c$ axis, consistent with prior theoretical predictions~\cite{ZHOU2017208}. In this work, we have examined relevant OSP and ESP states, including $(k_x + i k_y) \hat{z}$ [as shown in Figs.~\ref{fig:OSP_parallel_mag}(a), \ref{sus_socpx}, and \ref{sus_socmagp1}] and $k_z \hat{z}$ [as depicted in Figs.~\ref{fig:OSP_parallel_mag}(b), \ref{fig:chizzxx_kzz_soc}, \ref{fig:chizzxx_kzz_zerotem_soc}, and \ref{fig:suszzxx_kzz_magsoc}].

Several key features from our analysis can act as diagnostic tools to distinguish between the $(k_x + i k_y) \hat{z}$ and $k_z \hat{z}$ OSP states in future experiments: (1) Starting with the baseline of no external fields, the spin susceptibility $\chi_{zz}(T)$ exhibits distinct low-temperature power-law decays: $\chi_{zz}(T) \propto T^3$ for $(k_x + i k_y) \hat{z}$ due to a nodal point, versus $\chi_{zz}(T) \propto T$ for $k_z \hat{z}$ due to a nodal line. (2) Introducing a finite parallel Zeeman field $\mathbf{H} = H_z \hat{z}$ induces a non-zero zero-temperature susceptibility, scaling as $\chi_{zz}(T=0) \propto (\mu_B H_z / \Delta)^2$ for $(k_x + i k_y) \hat{z}$ and $\chi_{zz}(T=0) \propto \mu_B H_z / \Delta$ for $k_z \hat{z}$ in the low-field limit ($\mu_B H_z \ll \Delta$), as given in Eqs.~\eqref{eq:chimagzOSP}.  (3) In the presence of Rashba SOC, superconductivity is fully suppressed in $(k_x + i k_y) \hat{z}$ for $g k_F / k_B T_{c0} \gtrsim 1.5$ without a Zeeman field, while it persists in $k_z \hat{z}$~\cite{pang2025}.  (4) Furthermore, finite Rashba SOC produces a residual $\chi_{zz}(T=0)$ that is more significant in $k_z \hat{z}$ [Fig.~\ref{sus_socpx}(a)] than in $(k_x + i k_y) \hat{z}$ [Fig.~\ref{fig:chizzxx_kzz_soc}(a)], and it enhances $\chi_{xx}(T)$ above $\chi_N$ below $T_c$ for $(k_x + i k_y) \hat{z}$ until suppression, whereas $\chi_{xx}(T)$ can drop below $\chi_N$ in $k_z \hat{z}$ under strong SOC.

\paragraph{Outlook.}
The theoretical framework developed in this work provides a robust foundation for exploring the spin response of superconductors under strong spin-orbit coupling and magnetic fields. Our results, particularly the universal constraints on spin susceptibility and the detailed predictions for various pairing symmetries, open several avenues for future research that naturally extend from our findings.

A key unresolved question concerns the role of \emph{multi-orbital physics} and the potential for \emph{mixed-parity pairing} in non-centrosymmetric superconductors. Our single-band, unitary-state analysis provides clear benchmarks; however, real materials often feature multiple electronic bands. Extending this theory to multi-orbital systems would allow us to investigate how inter-band interactions and orbital-dependent pairing modify the susceptibility signatures we have identified. This is particularly relevant for materials like A$_2$Cr$_3$As$_3$, where several bands cross the Fermi level. Furthermore, the lack of inversion symmetry can intrinsically mix even- and odd-parity components in the order parameter. A natural and important extension of this work would be to quantify how such parity mixing influences the spin susceptibility, potentially leading to new, hybrid behaviors that could serve as additional experimental diagnostics.
Another promising direction involves exploring \emph{non-unitary} triplet states, where the superconducting state spontaneously breaks time-reversal symmetry. Our current framework for unitary states could be generalized to investigate how the intrinsic magnetization in non-unitary phases interacts with an external Zeeman field and SOC, potentially leading to novel magnetic phenomena and unique signatures in the susceptibility.

From a methodological perspective, incorporating realistic Fermi surface anisotropies -- for example, modeling the quasi-one-dimensional bands in A$_2$Cr$_3$As$_3$ --would bridge the gap between our simplified model and specific material properties. This would enable more quantitative comparisons with experiments and help identify which features are robust versus model-dependent. Finally, generalizing the theory to finite wave-vector susceptibility would pave the way for a first-principles understanding of the spin-lattice relaxation rate $1/T_1$ under strong SOC and Zeeman fields, offering a complementary and powerful experimental probe.

In conclusion, the findings presented here not only advance our understanding of the spin response in superconductors but also set the stage for a deeper investigation into the interplay of symmetry, topology, and correlation effects. The future directions outlined above will further elucidate the rich physics of unconventional superconductors, with our work serving as a foundational reference.

\section*{Acknowledgement}
This work is partially supported by National Key Research and Development Program of China (No. 2022YFA1403403), and National Natural Science Foundation of China (No. 12274441, 12034004).

\appendix	

\section{Bogoliubov Transformation and the Sum Rule in Eq.~\eqref{eq:sum}}\label{app:BTM}

The Hamiltonian for a superconducting system can be expressed as:
\begin{equation}
H = \frac{1}{2} \sum_{\mathbf{k}} C^{\dagger}_{\mathbf{k}} H(\mathbf{k}) C_{\mathbf{k}},
\end{equation}
where $C_{\mathbf{k}} = (c_{\mathbf{k}\uparrow}, c_{\mathbf{k}\downarrow}, c_{-\mathbf{k}\uparrow}^{\dagger}, c_{-\mathbf{k}\downarrow}^{\dagger})^{T}$, and
\begin{equation}
H(\mathbf{k}) = \begin{pmatrix}
H_0(\mathbf{k}) & \Delta(\mathbf{k}) \\
\Delta^{\dagger}(\mathbf{k}) & -H_0(-\mathbf{k})^{T}
\end{pmatrix}.
\end{equation}

To diagonalize this Hamiltonian, we introduce the Bogoliubov-transformed operators $\Psi_{\mathbf{k}} = (\psi_{\mathbf{k}+}, \psi_{\mathbf{k}-}, \psi_{-\mathbf{k}+}^{\dagger}, \psi_{-\mathbf{k}-}^{\dagger})^{T}$, related to the original operators by $C_{\mathbf{k}} = U_{\mathbf{k}} \Psi_{\mathbf{k}}$, with the transformation matrix
\begin{equation}
U_{\mathbf{k}} = \begin{pmatrix}
u_{\mathbf{k}} & v_{\mathbf{k}} \\
v_{-\mathbf{k}}^* & u_{-\mathbf{k}}^*
\end{pmatrix}.
\end{equation}
This transformation diagonalizes the Hamiltonian, yielding
\begin{equation}
U_{\mathbf{k}}^{\dagger} H(\mathbf{k}) U_{\mathbf{k}} = \mbox{diag}(E_{\mathbf{k}+}, E_{\mathbf{k}-}, -E_{-\mathbf{k}+}, -E_{-\mathbf{k}-}).
\end{equation}

Since $U_{\mathbf{k}}$ is unitary, it satisfies $U_{\mathbf{k}} U_{\mathbf{k}}^{\dagger} = U_{\mathbf{k}}^{\dagger} U_{\mathbf{k}} = \mathbb{I}$, which implies the following conditions:
\begin{equation}
\begin{aligned}
u_{\mathbf{k}} u_{\mathbf{k}}^{\dagger} + v_{\mathbf{k}} v_{\mathbf{k}}^{\dagger} &= \mathbb{I}, \\
u_{\mathbf{k}}^{\dagger} u_{\mathbf{k}} + v_{-\mathbf{k}}^{T} v_{-\mathbf{k}}^* &= \mathbb{I}, \\
u_{\mathbf{k}} v_{-\mathbf{k}}^{T} + v_{\mathbf{k}} u_{-\mathbf{k}}^{T} &= 0.
\end{aligned}
\end{equation}

Using these properties, the sum rule in Eq.~\eqref{eq:sum} can be verified as follows:
\begin{widetext}
\begin{equation}
\begin{aligned}
&\quad \sum_{s_1, s_2} \left| (u_{\mathbf{k}}^{\dagger} \sigma_\mu u_{\mathbf{k}})^{s_1 s_2} - (v_{-\mathbf{k}}^{\dagger} \sigma_{\mu} v_{-\mathbf{k}})^{s_2 s_1} \right|^2 + \left| (u_{\mathbf{k}}^{\dagger} \sigma_{\mu} v_{\mathbf{k}})^{s_1 s_2} - (u_{-\mathbf{k}}^{\dagger} \sigma_{\mu} v_{-\mathbf{k}})^{s_2 s_1} \right|^2 \\
&= \sum_{s_1, s_2} \left[ u_{\mathbf{k}}^{\dagger} \sigma_\mu u_{\mathbf{k}} - (v_{-\mathbf{k}}^{\dagger} \sigma_{\mu} v_{-\mathbf{k}})^{T} \right]^{s_1 s_2} \left[ u_{\mathbf{k}}^{\dagger} \sigma_\mu u_{\mathbf{k}} - (v_{-\mathbf{k}}^{\dagger} \sigma_{\mu} v_{-\mathbf{k}})^{T} \right]^{s_2 s_1} + \left[ u_{\mathbf{k}}^{\dagger} \sigma_{\mu} v_{\mathbf{k}} - (u_{-\mathbf{k}}^{\dagger} \sigma_{\mu} v_{-\mathbf{k}})^{T} \right]^{s_1 s_2} \left[ v_{\mathbf{k}}^{\dagger} \sigma_{\mu} u_{\mathbf{k}} - (v_{-\mathbf{k}}^{\dagger} \sigma_{\mu} u_{-\mathbf{k}})^{T} \right]^{s_2 s_1} \\
&= \mbox{tr}(u_{\mathbf{k}}^{\dagger} \sigma_\mu u_{\mathbf{k}} u_{\mathbf{k}}^{\dagger} \sigma_\mu u_{\mathbf{k}}) + \mbox{tr}(v_{-\mathbf{k}}^{\dagger} \sigma_{\mu} v_{-\mathbf{k}} v_{-\mathbf{k}}^{\dagger} \sigma_{\mu} v_{-\mathbf{k}}) - 2 \mbox{tr}(u_{\mathbf{k}}^{\dagger} \sigma_\mu u_{\mathbf{k}} v_{-\mathbf{k}}^{T} \sigma_\mu^* v_{-\mathbf{k}}^*) + \mbox{tr}(u_{\mathbf{k}}^{\dagger} \sigma_\mu v_{\mathbf{k}} v_{\mathbf{k}}^{\dagger} \sigma_\mu u_{\mathbf{k}}) + \mbox{tr}(u_{-\mathbf{k}}^{\dagger} \sigma_\mu v_{-\mathbf{k}} v_{-\mathbf{k}}^{\dagger} \sigma_\mu u_{-\mathbf{k}}) \\
&\quad- 2 \mbox{tr}(u_{\mathbf{k}}^{\dagger} \sigma_\mu v_{\mathbf{k}} u_{-\mathbf{k}}^{T} \sigma_\mu^* v_{-\mathbf{k}}^*) \\
&= \mbox{tr}[u_{\mathbf{k}}^{\dagger} \sigma_\mu (u_{\mathbf{k}} u_{\mathbf{k}}^{\dagger} + v_{\mathbf{k}} v_{\mathbf{k}}^{\dagger}) \sigma_\mu u_{\mathbf{k}}] - 2 \mbox{tr}[u_{\mathbf{k}}^{\dagger} \sigma_\mu (u_{\mathbf{k}} v_{-\mathbf{k}}^{T} + v_{\mathbf{k}} u_{-\mathbf{k}}^{T}) \sigma_\mu^* v_{-\mathbf{k}}^*]  + \mbox{tr}[v_{-\mathbf{k}}^{\dagger} \sigma_\mu (\mathbb{I} - u_{-\mathbf{k}} u_{-\mathbf{k}}^{\dagger}) \sigma_\mu v_{-\mathbf{k}}] + \mbox{tr}(u_{-\mathbf{k}}^{\dagger} \sigma_\mu v_{-\mathbf{k}} v_{-\mathbf{k}}^{\dagger} \sigma_\mu u_{-\mathbf{k}}) \\
&= \mbox{tr}(u_{\mathbf{k}}^{\dagger} u_{\mathbf{k}}) + \mbox{tr}(v_{-\mathbf{k}}^{\dagger} v_{-\mathbf{k}}) \\
&= \mbox{tr}(u_{\mathbf{k}}^{\dagger} u_{\mathbf{k}} + v_{-\mathbf{k}}^{T} v_{-\mathbf{k}}^*) = 2.
\end{aligned}
\end{equation}
Here, we utilize the fact that $v_{-\mathbf{k}}^{\dagger} v_{-\mathbf{k}}$ is Hermitian, so its diagonal elements are real, and $\mbox{tr}(v_{-\mathbf{k}}^{\dagger} v_{-\mathbf{k}}) = [\mbox{tr}(v_{-\mathbf{k}}^{\dagger} v_{-\mathbf{k}})]^* = \mbox{tr}(v_{-\mathbf{k}}^{T} v_{-\mathbf{k}}^*)$.

\end{widetext}

\section{Kubo Formula for Spin Susceptibility}\label{app:Kubo}

By solving the gap equation in Eq.~\eqref{gapeqwhole}, the pairing function $\Delta(\mathbf{k})$ is determined. The spin susceptibility, measurable via the NMR Knight shift, is then calculated using the Kubo formula:
\begin{equation}\label{Kubo}
\begin{split}
\chi(\mathbf{q}, i\omega_l) &= \int_{0}^{\beta} d\tau \, e^{i \omega_l \tau} \langle T_\tau \mathbf{M}(\mathbf{q}, \tau) \mathbf{M}(-\mathbf{q}, 0) \rangle, \\
\chi(\mathbf{q}, \omega) &= \chi(\mathbf{q}, i \omega_l \to \omega + i 0^+),
\end{split}
\end{equation}
where the magnetization operator is given by
\begin{equation}\label{magnetization}
\mathbf{M}(\mathbf{q}) = -\mu_B \sum_{\mathbf{k}, \alpha, \beta} \hat{\sigma}^{\alpha \beta} c_{\mathbf{k}-\mathbf{q}, \alpha}^{\dagger} c_{\mathbf{k}, \beta}.
\end{equation}

In tensor form, the spin susceptibility is expressed as
\begin{equation}\label{chi}
\chi_{\mu\nu}(\mathbf{q}, i\omega_l) = \frac{1}{Z} \sum_{n, m} \frac{\langle n | M_{\mu}(\mathbf{q}) | m \rangle \langle m | M_{\nu}(-\mathbf{q}) | n \rangle (\mathrm{e}^{-\beta E_m} - \mathrm{e}^{-\beta E_n})}{i \omega_l + E_n - E_m},
\end{equation}
with $Z$ as the partition function, $\beta = 1 / k_B T$ the inverse temperature, $\mu, \nu = x, y, z$, and $|n\rangle$, $|m\rangle$ the many-body eigenstates with energies $E_n$ and $E_m$.

Substituting Eq.~\eqref{magnetization} into Eq.~\eqref{chi} and performing the analytic continuation $i \omega_l \to \omega + i 0^+$, the static uniform susceptibility $\chi_{\mu\nu} \equiv \chi_{\mu\nu}(\mathbf{q} \to 0, \omega = 0)$ becomes:
\begin{equation}\label{chi0}
\begin{split}
\chi_{\mu\nu} &= \frac{\mu_B^2}{Z} \lim_{\mathbf{q} \to 0} \sum_{\alpha, \beta, \alpha', \beta'} \sigma_{\mu}^{\alpha \beta} \sigma_{\nu}^{\alpha' \beta'} \sum_{\mathbf{k}, \mathbf{k}'} \sum_{n, m} \frac{\mathrm{e}^{-\beta E_m} - \mathrm{e}^{-\beta E_n}}{E_n - E_m + i 0^+} \\
&\quad \times \langle n | c_{\mathbf{k}-\mathbf{q}, \alpha}^{\dagger} c_{\mathbf{k}, \beta} | m \rangle \langle m | c_{\mathbf{k}'+\mathbf{q}, \alpha'}^{\dagger} c_{\mathbf{k}', \beta'} | n \rangle.
\end{split}
\end{equation}

For a uniform superconducting state with zero center-of-mass momentum for Cooper pairs, the matrix element $\langle n | c_{\mathbf{k}-\mathbf{q}, \alpha}^{\dagger} c_{\mathbf{k}, \beta} | m \rangle \langle m | c_{\mathbf{k}'+\mathbf{q}, \alpha'}^{\dagger} c_{\mathbf{k}', \beta'} | n \rangle$ is nonzero only if $\mathbf{k}' = \mathbf{k} - \mathbf{q}$ or $\mathbf{k}' = -\mathbf{k}$. The eigenstates $|n\rangle$ and $|m\rangle$ are quasiparticle states over the superconducting ground state, enforcing energy conservation as $E_n - E_m = \pm E_{\mathbf{k} \pm} \pm E_{\mathbf{k}-\mathbf{q} \pm}$. Thus, $\chi_{\mu\nu}$ decomposes as:
\begin{subequations}\label{chi0_k}
\begin{equation}
\chi_{\mu\nu} = \sum_{\mathbf{k}} \chi_{\mu\nu}(\mathbf{k}),
\end{equation}
with the $\mathbf{k}$-dependent component:
\begin{equation}
\begin{split}
\chi_{\mu\nu}(\mathbf{k}) & = \frac{\mu_B^2}{Z} \lim_{\mathbf{q} \to 0} \sum_{\alpha, \beta, \alpha', \beta'} \sigma_{\mu}^{\alpha \beta} \sigma_{\nu}^{\alpha' \beta'} \sum_{n, m} \frac{\mathrm{e}^{-\beta E_m} - \mathrm{e}^{-\beta E_n}}{E_n - E_m + i 0^+} \\
&\quad \times \langle n | c_{\mathbf{k}-\mathbf{q}, \alpha}^{\dagger} c_{\mathbf{k}, \beta} | m \rangle \langle m | (c_{\mathbf{k}, \alpha'}^{\dagger} c_{\mathbf{k}-\mathbf{q}, \beta'} + c_{-\mathbf{k}+\mathbf{q}, \alpha'}^{\dagger} c_{-\mathbf{k}, \beta'}) | n \rangle.
\end{split}
\end{equation}
\end{subequations}

Using the Bogoliubov transformation from Eq.~\eqref{eq:BT}, $\chi_{\mu\nu}(\mathbf{k})$ is expressed as:
\begin{subequations}\label{eq:chi_k}
\begin{equation}
\chi_{\mu\nu}(\mathbf{k}) = \chi_{\mu\nu}^{ph}(\mathbf{k}) + \chi_{\mu\nu}^{pp}(\mathbf{k}),
\end{equation}
where $\chi_{\mu\nu}^{ph}$ and $\chi_{\mu\nu}^{pp}$ correspond to particle-hole and particle-particle/hole-hole processes, respectively. The particle-hole component further splits into:
\begin{equation}
\chi_{\mu\nu}^{ph}(\mathbf{k}) = \chi_{\mu\nu}^{ph+}(\mathbf{k}) + \chi_{\mu\nu}^{ph-}(\mathbf{k}),
\end{equation}
with $\chi_{\mu\nu}^{ph+}$ and $\chi_{\mu\nu}^{ph-}$ representing intraband and interband contributions.
\end{subequations}

This work focuses on the diagonal components ($\mu = \nu$), analyzing $\chi_{\mu\mu}^{ph+}$, $\chi_{\mu\mu}^{ph-}$, and $\chi_{\mu\mu}^{pp}$ separately. The matrix elements in Eq.~\eqref{chi0_k} are rewritten using the Bogoliubov transformation:
\begin{widetext}
\begin{equation}
\begin{aligned}
\langle n | c_{\mathbf{k}-\mathbf{q}, \alpha}^{\dagger} c_{\mathbf{k}, \beta} | m \rangle &= \sum_{s_1, s_2} \langle n | (u_{\mathbf{k}-\mathbf{q}}^{\alpha s_1 *} \psi_{\mathbf{k}-\mathbf{q}, s_1}^{\dagger} + v_{\mathbf{k}-\mathbf{q}}^{\alpha s_1 *} \psi_{\mathbf{q}-\mathbf{k}, s_1}) (u_{\mathbf{k}}^{\beta s_2} \psi_{\mathbf{k}, s_2} + v_{\mathbf{k}}^{\beta s_2} \psi_{-\mathbf{k}, s_2}^{\dagger}) | m \rangle, \\
\langle m | c_{\mathbf{k}, \alpha'}^{\dagger} c_{\mathbf{k}-\mathbf{q}, \beta'} | n \rangle &= \sum_{s_3, s_4} \langle m | (u_{\mathbf{k}}^{\alpha' s_3 *} \psi_{\mathbf{k}, s_3}^{\dagger} + v_{\mathbf{k}}^{\alpha' s_3 *} \psi_{-\mathbf{k}, s_3}) (u_{\mathbf{k}-\mathbf{q}}^{\beta' s_4} \psi_{\mathbf{k}-\mathbf{q}, s_4} + v_{\mathbf{k}-\mathbf{q}}^{\beta' s_4} \psi_{\mathbf{q}-\mathbf{k}, s_4}^{\dagger}) | n \rangle, \\
\langle m | c_{-\mathbf{k}+\mathbf{q}, \alpha'}^{\dagger} c_{-\mathbf{k}, \beta'} | n \rangle &= \sum_{s_3, s_4} \langle m | (u_{-\mathbf{k}+\mathbf{q}}^{\alpha' s_3 *} \psi_{-\mathbf{k}+\mathbf{q}, s_3}^{\dagger} + v_{-\mathbf{k}+\mathbf{q}}^{\alpha' s_3 *} \psi_{\mathbf{k}-\mathbf{q}, s_3}) (u_{-\mathbf{k}}^{\beta' s_4} \psi_{-\mathbf{k}, s_4} + v_{-\mathbf{k}}^{\beta' s_4} \psi_{\mathbf{k}, s_4}^{\dagger}) | n \rangle.
\end{aligned}
\end{equation}
\end{widetext}

Nonzero contributions to $\chi_{\mu\mu}^{ph+}(\mathbf{k})$ and $\chi_{\mu\mu}^{ph-}(\mathbf{k})$ arise from states like $|m\rangle = \psi^{\dagger} \psi' |n\rangle$ and $|m\rangle = \psi \psi'^{\dagger} |n\rangle$, while $\chi_{\mu\mu}^{pp}(\mathbf{k})$ involves $|m\rangle = \psi^{\dagger} \psi'^{\dagger} |n\rangle$ and $|m\rangle = \psi \psi' |n\rangle$. Configurations with $s_1 = s_2$ and $s_3 = s_4$ correspond to $\chi_{\mu\mu}^{ph+}(\mathbf{k})$, and $s_1 = \bar{s}_2$, $s_3 = \bar{s}_4$ to $\chi_{\mu\mu}^{ph-}(\mathbf{k})$, with $\bar{s}$ denoting the opposite band index.

For $\chi_{\mu\mu}^{ph\pm}(\mathbf{k})$, the relevant factor is evaluated as:
\begin{widetext}
\begin{equation}
\begin{aligned}
&\sum_{s_1, s_2, s_3, s_4} \left[ u_{\mathbf{k}-\mathbf{q}}^{\alpha s_1 *} u_{\mathbf{k}}^{\beta s_2} \langle n | \psi_{\mathbf{k}-\mathbf{q}, s_1}^{\dagger} \psi_{\mathbf{k}, s_2} | m \rangle \left( u_{\mathbf{k}}^{\alpha' s_3 *} u_{\mathbf{k}-\mathbf{q}}^{\beta' s_4} \langle m | \psi_{\mathbf{k}, s_3}^{\dagger} \psi_{\mathbf{k}-\mathbf{q}, s_4} | n \rangle + v_{-\mathbf{k}+\mathbf{q}}^{\alpha' s_3 *} v_{-\mathbf{k}}^{\beta' s_4} \langle m | \psi_{\mathbf{k}-\mathbf{q}, s_3} \psi_{\mathbf{k}, s_4}^{\dagger} | n \rangle \right) \right. \\
&\quad + \left. v_{\mathbf{k}-\mathbf{q}}^{\alpha s_1 *} v_{\mathbf{k}}^{\beta s_2} \langle n | \psi_{\mathbf{q}-\mathbf{k}, s_1} \psi_{-\mathbf{k}, s_2}^{\dagger} | m \rangle \left( v_{\mathbf{k}}^{\alpha' s_3 *} v_{\mathbf{k}-\mathbf{q}}^{\beta' s_4} \langle m | \psi_{-\mathbf{k}, s_3} \psi_{\mathbf{q}-\mathbf{k}, s_4}^{\dagger} | n \rangle + u_{-\mathbf{k}+\mathbf{q}}^{\alpha' s_3 *} u_{-\mathbf{k}}^{\beta' s_4} \langle m | \psi_{-\mathbf{k}+\mathbf{q}, s_3}^{\dagger} \psi_{-\mathbf{k}, s_4} | n \rangle \right) \right].
\end{aligned}
\end{equation}
Substituting into Eq.~\eqref{eq:chi_k} and taking the limit $\mathbf{q} \to 0$ yields:
\begin{equation}
\begin{aligned}
\chi_{\mu\mu}^{ph}(\mathbf{k}) &= \mu_B^2 \lim_{\mathbf{q} \to 0} \sum_{\alpha, \beta, \alpha', \beta'} \sigma_{\mu}^{\alpha \beta} \sigma_{\mu}^{\alpha' \beta'} \sum_{s_1, s_2} \left\{ u_{\mathbf{k}-\mathbf{q}}^{\alpha s_1 *} u_{\mathbf{k}}^{\beta s_2} u_{\mathbf{k}}^{\alpha' s_2 *} u_{\mathbf{k}-\mathbf{q}}^{\beta' s_1} \frac{f(E_{\mathbf{k}-\mathbf{q}, s_1}) [1 - f(E_{\mathbf{k}, s_2})] (\mathrm{e}^{\beta (E_{\mathbf{k}-\mathbf{q}, s_1} - E_{\mathbf{k}, s_2})} - 1)}{E_{\mathbf{k}-\mathbf{q}, s_1} - E_{\mathbf{k}, s_2}} \right. \\
&\quad - u_{\mathbf{k}-\mathbf{q}}^{\alpha s_1 *} u_{\mathbf{k}}^{\beta s_2} v_{-\mathbf{k}+\mathbf{q}}^{\alpha' s_1 *} v_{-\mathbf{k}}^{\beta' s_2} \frac{f(E_{\mathbf{k}-\mathbf{q}, s_1}) [1 - f(E_{\mathbf{k}, s_2})] (\mathrm{e}^{\beta (E_{\mathbf{k}-\mathbf{q}, s_1} - E_{\mathbf{k}, s_2})} - 1)}{E_{\mathbf{k}-\mathbf{q}, s_1} - E_{\mathbf{k}, s_2}} \\
&\quad + v_{\mathbf{k}-\mathbf{q}}^{\alpha s_1 *} v_{\mathbf{k}}^{\beta s_2} v_{\mathbf{k}}^{\alpha' s_2 *} v_{\mathbf{k}-\mathbf{q}}^{\beta' s_1} \frac{[1 - f(E_{\mathbf{q}-\mathbf{k}, s_1})] f(E_{-\mathbf{k}, s_2}) (\mathrm{e}^{\beta (E_{-\mathbf{k}, s_2} - E_{\mathbf{q}-\mathbf{k}, s_1})} - 1)}{E_{-\mathbf{k}, s_2} - E_{\mathbf{q}-\mathbf{k}, s_1}} \\
&\left.\quad - v_{\mathbf{k}-\mathbf{q}}^{\alpha s_1 *} v_{\mathbf{k}}^{\beta s_2} u_{-\mathbf{k}+\mathbf{q}}^{\alpha' s_1 *} u_{-\mathbf{k}}^{\beta' s_2} \frac{[1 - f(E_{\mathbf{q}-\mathbf{k}, s_1})] f(E_{-\mathbf{k}, s_2}) (\mathrm{e}^{\beta (E_{-\mathbf{k}, s_2} - E_{\mathbf{q}-\mathbf{k}, s_1})} - 1)}{E_{-\mathbf{k}, s_2} - E_{\mathbf{q}-\mathbf{k}, s_1}} \right\} \\
&= -\mu_B^2 \lim_{\mathbf{q}\to 0} \sum_{s_1, s_2} \left\{ \left[ (u_{\mathbf{k}-\mathbf{q}}^{\dagger} \sigma_\mu u_{\mathbf{k}})^{s_1 s_2} (u_{\mathbf{k}}^{\dagger} \sigma_\mu u_{\mathbf{k}-\mathbf{q}})^{s_2 s_1} - (u_{\mathbf{k}-\mathbf{q}}^{\dagger} \sigma_\mu u_{\mathbf{k}})^{s_1 s_2} (v_{-\mathbf{k}+\mathbf{q}}^{\dagger} \sigma_\mu v_{-\mathbf{k}})^{s_1 s_2} \right] \frac{f(E_{\mathbf{k}-\mathbf{q}, s_1}) - f(E_{\mathbf{k}, s_2})}{E_{\mathbf{k}-\mathbf{q}, s_1} - E_{\mathbf{k}, s_2}} \right. \\
&\left.\quad + \left[ (v_{\mathbf{k}-\mathbf{q}}^{\dagger} \sigma_\mu v_{\mathbf{k}})^{s_1 s_2} (v_{\mathbf{k}}^{\dagger} \sigma_\mu v_{\mathbf{k}-\mathbf{q}})^{s_2 s_1} - (v_{\mathbf{k}-\mathbf{q}}^{\dagger} \sigma_\mu v_{\mathbf{k}})^{s_1 s_2} (u_{-\mathbf{k}+\mathbf{q}}^{\dagger} \sigma_\mu u_{-\mathbf{k}})^{s_1 s_2} \right] \frac{f(E_{\mathbf{q}-\mathbf{k}, s_1}) - f(E_{-\mathbf{k}, s_2})}{E_{\mathbf{q}-\mathbf{k}, s_1} - E_{-\mathbf{k}, s_2}} \right\}.
\end{aligned}
\end{equation}
Thus, the components are:
\begin{equation}\label{eq:chiph_pm}
\begin{aligned}
\chi_{\mu\mu}^{ph+}(\mathbf{k}) &= -\mu_B^2 \sum_{s = \pm} \left[ (u_{\mathbf{k}}^{\dagger} \sigma_\mu u_{\mathbf{k}})^{s s} - (v_{-\mathbf{k}}^{\dagger} \sigma_\mu v_{-\mathbf{k}})^{s s} \right]^2 \left. \frac{d f(E)}{d E} \right|_{E = E_{\mathbf{k}, s}}, \\
\chi_{\mu\mu}^{ph-}(\mathbf{k}) &= -2 \mu_B^2 \left| (u_{\mathbf{k}}^{\dagger} \sigma_\mu u_{\mathbf{k}})^{+-} - (v_{-\mathbf{k}}^{\dagger} \sigma_\mu v_{-\mathbf{k}})^{-+} \right|^2 \frac{f(E_{\mathbf{k}, +}) - f(E_{\mathbf{k}, -})}{E_{\mathbf{k}, +} - E_{\mathbf{k}, -}} ,\\
\end{aligned}
\end{equation}

where the symmetrization by summing over $\mathbf{k}$ and $-\mathbf{k}$ has been used, which gives:
\begin{equation}\label{eq:chiph_pmk}
\begin{aligned}
\chi_{\mu\mu}^{ph+}(\mathbf{k}) + \chi_{\mu\mu}^{ph+}(-\mathbf{k}) &= -\mu_B^2 \sum_{s = \pm} \left[ (u_{\mathbf{k}}^{\dagger} \sigma_\mu u_{\mathbf{k}})^{s s} - (v_{-\mathbf{k}}^{\dagger} \sigma_\mu v_{-\mathbf{k}})^{s s} \right]^2 \left. \frac{d f(E)}{d E} \right|_{E = E_{\mathbf{k}, s}} \\
&\quad + \left[ (u_{-\mathbf{k}}^{\dagger} \sigma_\mu u_{-\mathbf{k}})^{s s} - (v_{\mathbf{k}}^{\dagger} \sigma_\mu v_{\mathbf{k}})^{s s} \right]^2 \left. \frac{d f(E)}{d E} \right|_{E = E_{-\mathbf{k}, s}}, \\
\chi_{\mu\mu}^{ph-}(\mathbf{k}) + \chi_{\mu\mu}^{ph-}(-\mathbf{k}) &= -2 \mu_B^2 \left\{\left| (u_{\mathbf{k}}^{\dagger} \sigma_\mu u_{\mathbf{k}})^{+-} - (v_{-\mathbf{k}}^{\dagger} \sigma_\mu v_{-\mathbf{k}})^{-+} \right|^2 \frac{f(E_{\mathbf{k}, +}) - f(E_{\mathbf{k}, -})}{E_{\mathbf{k}, +} - E_{\mathbf{k}, -}} \right.\\
&\quad \left.+ \left| (u_{-\mathbf{k}}^{\dagger} \sigma_\mu u_{-\mathbf{k}})^{+-} - (v_{\mathbf{k}}^{\dagger} \sigma_\mu v_{\mathbf{k}})^{-+} \right|^2 \frac{f(E_{-\mathbf{k}, +}) - f(E_{-\mathbf{k}, -})}{E_{-\mathbf{k}, +} - E_{-\mathbf{k}, -}}\right\}.
\end{aligned}
\end{equation}

For $\chi_{\mu\mu}^{pp}(\mathbf{k})$, the matrix elements are:
\begin{equation}
\begin{aligned}
&\sum_{s_1, s_2, s_3, s_4} \left[ u_{\mathbf{k}-\mathbf{q}}^{\alpha s_1 *} v_{\mathbf{k}}^{\beta s_2} \langle n | \psi_{\mathbf{k}-\mathbf{q}, s_1}^{\dagger} \psi_{-\mathbf{k}, s_2}^{\dagger} | m \rangle \left( v_{\mathbf{k}}^{\alpha' s_3 *} u_{\mathbf{k}-\mathbf{q}}^{\beta' s_4} \langle m | \psi_{-\mathbf{k}, s_3} \psi_{\mathbf{k}-\mathbf{q}, s_4} | n \rangle + v_{-\mathbf{k}+\mathbf{q}}^{\alpha' s_3 *} u_{-\mathbf{k}}^{\beta' s_4} \langle m | \psi_{\mathbf{k}-\mathbf{q}, s_3} \psi_{-\mathbf{k}, s_4} | n \rangle \right) \right. \\
&\quad + \left. v_{\mathbf{k}-\mathbf{q}}^{\alpha s_1 *} u_{\mathbf{k}}^{\beta s_2} \langle n | \psi_{\mathbf{q}-\mathbf{k}, s_1} \psi_{\mathbf{k}, s_2} | m \rangle \left( u_{\mathbf{k}}^{\alpha' s_3 *} v_{\mathbf{k}-\mathbf{q}}^{\beta' s_4} \langle m | \psi_{\mathbf{k}, s_3}^{\dagger} \psi_{\mathbf{q}-\mathbf{k}, s_4}^{\dagger} | n \rangle + u_{-\mathbf{k}+\mathbf{q}}^{\alpha' s_3 *} v_{-\mathbf{k}}^{\beta' s_4} \langle m | \psi_{-\mathbf{k}+\mathbf{q}, s_3}^{\dagger} \psi_{\mathbf{k}, s_4}^{\dagger} | n \rangle \right) \right].
\end{aligned}
\end{equation}
Taking the limit $\mathbf{q} \to 0$ and substituting into Eq.~\eqref{eq:chi_k} gives:
\begin{equation}
\begin{aligned}
\chi_{\mu\mu}^{pp}(\mathbf{k}) &= \mu_B^2 \lim_{\mathbf{q} \to 0} \sum_{\alpha, \beta, \alpha', \beta'} \sigma_{\mu}^{\alpha \beta} \sigma_{\mu}^{\alpha' \beta'} \sum_{s_1, s_2} \left\{ u_{\mathbf{k}-\mathbf{q}}^{\alpha s_1 *} v_{\mathbf{k}}^{\beta s_2} v_{\mathbf{k}}^{\alpha' s_2 *} u_{\mathbf{k}-\mathbf{q}}^{\beta' s_1} \frac{f(E_{\mathbf{k}-\mathbf{q}, s_1}) f(E_{-\mathbf{k}, s_2}) (\mathrm{e}^{\beta (E_{\mathbf{k}-\mathbf{q}, s_1} + E_{-\mathbf{k}, s_2})} - 1)}{E_{\mathbf{k}-\mathbf{q}, s_1} + E_{-\mathbf{k}, s_2}} \right. \\
&\quad - u_{\mathbf{k}-\mathbf{q}}^{\alpha s_1 *} v_{\mathbf{k}}^{\beta s_2} v_{-\mathbf{k}+\mathbf{q}}^{\alpha' s_1 *} u_{-\mathbf{k}}^{\beta' s_2} \frac{f(E_{\mathbf{k}-\mathbf{q}, s_1}) f(E_{-\mathbf{k}, s_2}) (\mathrm{e}^{\beta (E_{\mathbf{k}-\mathbf{q}, s_1} + E_{-\mathbf{k}, s_2})} - 1)}{E_{\mathbf{k}-\mathbf{q}, s_1} + E_{-\mathbf{k}, s_2}} \\
&\quad + v_{\mathbf{k}-\mathbf{q}}^{\alpha s_1 *} u_{\mathbf{k}}^{\beta s_2} u_{\mathbf{k}}^{\alpha' s_2 *} v_{\mathbf{k}-\mathbf{q}}^{\beta' s_1} \frac{[1 - f(E_{\mathbf{q}-\mathbf{k}, s_1})] [1 - f(E_{\mathbf{k}, s_2})] (\mathrm{e}^{-\beta (E_{\mathbf{q}-\mathbf{k}, s_1} + E_{\mathbf{k}, s_2})} - 1)}{-E_{\mathbf{q}-\mathbf{k}, s_1} - E_{\mathbf{k}, s_2}} \\
&\left.\quad - v_{\mathbf{k}-\mathbf{q}}^{\alpha s_1 *} u_{\mathbf{k}}^{\beta s_2} u_{-\mathbf{k}+\mathbf{q}}^{\alpha' s_1 *} v_{-\mathbf{k}}^{\beta' s_2} \frac{[1 - f(E_{\mathbf{q}-\mathbf{k}, s_1})] [1 - f(E_{\mathbf{k}, s_2})] (\mathrm{e}^{-\beta (E_{\mathbf{q}-\mathbf{k}, s_1} + E_{\mathbf{k}, s_2})} - 1)}{-E_{\mathbf{q}-\mathbf{k}, s_1} - E_{\mathbf{k}, s_2}} \right\} \\
&= -\mu_B^2 \sum_{s_1, s_2} \left\{ \left[ (u_{\mathbf{k}}^{\dagger} \sigma_\mu v_{\mathbf{k}})^{s_1 s_2} (v_{\mathbf{k}}^{\dagger} \sigma_\mu u_{\mathbf{k}})^{s_2 s_1} - (u_{\mathbf{k}}^{\dagger} \sigma_\mu v_{\mathbf{k}})^{s_1 s_2} (v_{-\mathbf{k}}^{\dagger} \sigma_\mu u_{-\mathbf{k}})^{s_1 s_2} \right] \frac{f(E_{\mathbf{k}, s_1}) + f(E_{-\mathbf{k}, s_2}) - 1}{E_{\mathbf{k}, s_1} + E_{-\mathbf{k}, s_2}} \right. \\
&\quad + \left. \left[ (v_{\mathbf{k}}^{\dagger} \sigma_\mu u_{\mathbf{k}})^{s_1 s_2} (u_{\mathbf{k}}^{\dagger} \sigma_\mu v_{\mathbf{k}})^{s_2 s_1} - (v_{\mathbf{k}}^{\dagger} \sigma_\mu u_{\mathbf{k}})^{s_1 s_2} (u_{-\mathbf{k}}^{\dagger} \sigma_\mu v_{-\mathbf{k}})^{s_1 s_2} \right] \frac{f(E_{-\mathbf{k}, s_1}) + f(E_{\mathbf{k}, s_2}) - 1}{E_{-\mathbf{k}, s_1} + E_{\mathbf{k}, s_2}} \right\}.
\end{aligned}
\end{equation}
Summing over $\mathbf{k}$ and $-\mathbf{k}$ yields:
\begin{equation}
\begin{aligned}
\chi_{\mu\mu}^{pp}(\mathbf{k}) + \chi_{\mu\mu}^{pp}(-\mathbf{k}) &= -\mu_B^2 \sum_{s_1, s_2} \left[ \left| (u_{\mathbf{k}}^{\dagger} \sigma_\mu v_{\mathbf{k}})^{s_1 s_2} - (u_{-\mathbf{k}}^{\dagger} \sigma_\mu v_{-\mathbf{k}})^{s_2 s_1} \right|^2 \frac{f(E_{\mathbf{k}, s_1}) + f(E_{-\mathbf{k}, s_2}) - 1}{E_{\mathbf{k}, s_1} + E_{-\mathbf{k}, s_2}} \right. \\
&\quad + \left. \left| (u_{-\mathbf{k}}^{\dagger} \sigma_\mu v_{-\mathbf{k}})^{s_1 s_2} - (u_{\mathbf{k}}^{\dagger} \sigma_\mu v_{\mathbf{k}})^{s_2 s_1} \right|^2 \frac{f(E_{-\mathbf{k}, s_1}) + f(E_{\mathbf{k}, s_2}) - 1}{E_{-\mathbf{k}, s_1} + E_{\mathbf{k}, s_2}} \right].
\end{aligned}
\end{equation}
Thus, $\chi_{\mu\mu}^{pp}(\mathbf{k})$ can be rewritten as:
\begin{equation}
\chi_{\mu\mu}^{pp}(\mathbf{k}) = -\mu_B^2 \sum_{s_1, s_2} \left| (u_{\mathbf{k}}^{\dagger} \sigma_\mu v_{\mathbf{k}})^{s_1 s_2} - (u_{-\mathbf{k}}^{\dagger} \sigma_\mu v_{-\mathbf{k}})^{s_2 s_1} \right|^2 \frac{f(E_{\mathbf{k}, s_1}) + f(E_{-\mathbf{k}, s_2}) - 1}{E_{\mathbf{k}, s_1} + E_{-\mathbf{k}, s_2}}.
\end{equation}

Summing over all $\mathbf{k}$ gives the expressions in Eq.~\eqref{eq:chikd} from the main text:
\begin{equation*}
\begin{split}
\chi_{\mu\mu}^{ph+} &= -\mu_B^2 \sum_{\mathbf{k}} \sum_{s = \pm} \left[ (u_{\mathbf{k}}^{\dagger} \sigma_\mu u_{\mathbf{k}})^{s s} - (v_{-\mathbf{k}}^{\dagger} \sigma_\mu v_{-\mathbf{k}})^{s s} \right]^2 \left. \frac{d f(E)}{d E} \right|_{E = E_{\mathbf{k}, s}}, \\
\chi_{\mu\mu}^{ph-} &= -2 \mu_B^2 \sum_{\mathbf{k}} \left| (u_{\mathbf{k}}^{\dagger} \sigma_\mu u_{\mathbf{k}})^{+-} - (v_{-\mathbf{k}}^{\dagger} \sigma_\mu v_{-\mathbf{k}})^{-+} \right|^2 \frac{f(E_{\mathbf{k}, +}) - f(E_{\mathbf{k}, -})}{E_{\mathbf{k}, +} - E_{\mathbf{k}, -}}, \\
\chi_{\mu\mu}^{pp} &= -\mu_B^2 \sum_{\mathbf{k}} \sum_{s_1, s_2} \left| (u_{\mathbf{k}}^{\dagger} \sigma_\mu v_{\mathbf{k}})^{s_1 s_2} - (u_{-\mathbf{k}}^{\dagger} \sigma_\mu v_{-\mathbf{k}})^{s_2 s_1} \right|^2 \frac{f(E_{\mathbf{k}, s_1}) + f(E_{-\mathbf{k}, s_2}) - 1}{E_{\mathbf{k}, s_1} + E_{-\mathbf{k}, s_2}}.
\end{split}
\end{equation*}
\end{widetext}

\section{$s$-wave Pairing State with Only Zeeman Field or Rashba SOC}\label{app:BT}

\subsection{A Single Zeeman Field}

For an $s$-wave superconductor subjected to a single Zeeman field, the Hamiltonian is given by:
\begin{equation}
H(\mathbf{k}) = \begin{pmatrix}
\xi_{\mathbf{k}+} & 0 & 0 & \Delta \\
0 & \xi_{\mathbf{k}-} & -\Delta & 0 \\
0 & -\Delta^* & -\xi_{\mathbf{k}+} & 0 \\
\Delta^* & 0 & 0 & -\xi_{\mathbf{k}-}
\end{pmatrix},
\end{equation}
where $\xi_{\mathbf{k}\pm} = \xi_{\mathbf{k}} \pm \mu_B H_z$. The corresponding $4 \times 4$ Bogoliubov transformation matrix $U_{\mathbf{k}}$ is:
\begin{equation}
\begin{pmatrix}
\frac{E_{\mathbf{k}} + \xi_{\mathbf{k}}}{[2 E_{\mathbf{k}} (E_{\mathbf{k}} + \xi_{\mathbf{k}})]^{1/2}} & 0 & 0 & \frac{-\Delta}{[2 E_{\mathbf{k}} (E_{\mathbf{k}} + \xi_{\mathbf{k}})]^{1/2}} \\
0 & \frac{E_{\mathbf{k}} + \xi_{\mathbf{k}}}{[2 E_{\mathbf{k}} (E_{\mathbf{k}} + \xi_{\mathbf{k}})]^{1/2}} & \frac{\Delta}{[2 E_{\mathbf{k}} (E_{\mathbf{k}} + \xi_{\mathbf{k}})]^{1/2}} & 0 \\
0 & \frac{-\Delta^*}{[2 E_{\mathbf{k}} (E_{\mathbf{k}} + \xi_{\mathbf{k}})]^{1/2}} & \frac{E_{\mathbf{k}} + \xi_{\mathbf{k}}}{[2 E_{\mathbf{k}} (E_{\mathbf{k}} + \xi_{\mathbf{k}})]^{1/2}} & 0 \\
\frac{\Delta^*}{[2 E_{\mathbf{k}} (E_{\mathbf{k}} + \xi_{\mathbf{k}})]^{1/2}} & 0 & 0 & \frac{E_{\mathbf{k}} + \xi_{\mathbf{k}}}{[2 E_{\mathbf{k}} (E_{\mathbf{k}} + \xi_{\mathbf{k}})]^{1/2}}
\end{pmatrix},
\end{equation}
which remains identical to the case without a Zeeman field. Consequently, the factorial matrices in Eq.~\eqref{eq:chikd} are:
\begin{equation}
\begin{aligned}
u_{\mathbf{k}}^{\dagger} \sigma_z u_{\mathbf{k}} &= \frac{1}{2} \left( 1 + \frac{\xi_{\mathbf{k}}}{E_{\mathbf{k}}} \right) \sigma_z, \\
v_{-\mathbf{k}}^{\dagger} \sigma_z v_{-\mathbf{k}} &= -\frac{1}{2} \left( 1 - \frac{\xi_{\mathbf{k}}}{E_{\mathbf{k}}} \right) \sigma_z, \\
u_{\mathbf{k}}^{\dagger} \sigma_z v_{\mathbf{k}} &= -\frac{\Delta}{2 E_{\mathbf{k}}} \sigma_x, \\
u_{-\mathbf{k}}^{\dagger} \sigma_z v_{-\mathbf{k}} &= -\frac{\Delta}{2 E_{\mathbf{k}}} \sigma_x.
\end{aligned}
\end{equation}
This leads to the spin susceptibility components:
\begin{equation}\label{eq:chizz_swave_mag}
\begin{aligned}
\chi_{zz} &= \sum_{\mathbf{k}} \left[ \chi_{zz}^{ph+}(\mathbf{k}) + \chi_{zz}^{ph-}(\mathbf{k}) + \chi_{zz}^{pp}(\mathbf{k}) \right], \\
\chi_{zz}^{ph+}(\mathbf{k}) &= -\mu_B^2 \left[ \left. \frac{df(E)}{dE} \right|_{E = E_{\mathbf{k}+}} + \left. \frac{df(E)}{dE} \right|_{E = E_{\mathbf{k}-}} \right], \\
\chi_{zz}^{ph-}(\mathbf{k}) &= \chi_{zz}^{pp}(\mathbf{k}) = 0.
\end{aligned}
\end{equation}

\subsection{Single Rashba SOC}

Transitioning to the case with only a finite Rashba SOC strength $g$ and no Zeeman field, the gap equation remains identical to that of a conventional $s$-wave superconductor for all $g$ values. For dominant wave vectors, the Bogoliubov transformation matrices are given by~\cite{pang2025}:
\begin{equation}
\begin{split}
u_{\mathbf{k}} &= \begin{pmatrix}
a_{\mathbf{k}+} \cos \frac{\theta_{\mathbf{k}}}{2} & a_{\mathbf{k}-} \sin \frac{\theta_{\mathbf{k}}}{2} \\
a_{\mathbf{k}+} \sin \frac{\theta_{\mathbf{k}}}{2} \mathrm{e}^{i \varphi_{\mathbf{k}}} & -a_{\mathbf{k}-} \cos \frac{\theta_{\mathbf{k}}}{2} \mathrm{e}^{i \varphi_{\mathbf{k}}}
\end{pmatrix}, \\
v_{\mathbf{k}} &= \begin{pmatrix}
b_{\mathbf{k}+} \cos \frac{\theta_{\mathbf{k}}}{2} \mathrm{e}^{-i \varphi_{\mathbf{k}}} & -b_{\mathbf{k}-} \sin \frac{\theta_{\mathbf{k}}}{2} \mathrm{e}^{-i \varphi_{\mathbf{k}}} \\
b_{\mathbf{k}+} \sin \frac{\theta_{\mathbf{k}}}{2} & b_{\mathbf{k}-} \cos \frac{\theta_{\mathbf{k}}}{2}
\end{pmatrix},
\end{split}
\label{eq:bt_s_soc}
\end{equation}
with the coefficients defined as:
\begin{equation}
\begin{aligned}
a_{\mathbf{k}\pm} &= \frac{\xi_{\mathbf{k}\pm} + E_{\mathbf{k}\pm}}{[2 E_{\mathbf{k}\pm} (\xi_{\mathbf{k}\pm} + E_{\mathbf{k}\pm})]^{1/2}}, \\
b_{\mathbf{k}\pm} &= \frac{\Delta}{[2 E_{\mathbf{k}\pm} (\xi_{\mathbf{k}\pm} + E_{\mathbf{k}\pm})]^{1/2}}.
\end{aligned}
\end{equation}

\subsection{Derivation of Eq.~\eqref{eq:chizz_s_soc}}

We derive Eq.~\eqref{eq:chizz_s_soc} for the case where $g k_F \ll \Lambda$. Using Eq.~\eqref{eq:bt_s_soc}, the matrices for $-\mathbf{k}$ are:
\begin{equation}
\begin{aligned}
u_{-\mathbf{k}} &= \begin{pmatrix}
a_{\mathbf{k}+} \sin \frac{\theta_{\mathbf{k}}}{2} & a_{\mathbf{k}-} \cos \frac{\theta_{\mathbf{k}}}{2} \\
-a_{\mathbf{k}+} \cos \frac{\theta_{\mathbf{k}}}{2} \mathrm{e}^{i \varphi_{\mathbf{k}}} & a_{\mathbf{k}-} \sin \frac{\theta_{\mathbf{k}}}{2} \mathrm{e}^{i \varphi_{\mathbf{k}}}
\end{pmatrix}, \\
v_{-\mathbf{k}} &= \begin{pmatrix}
-b_{\mathbf{k}+} \sin \frac{\theta_{\mathbf{k}}}{2} \mathrm{e}^{-i \varphi_{\mathbf{k}}} & b_{\mathbf{k}-} \cos \frac{\theta_{\mathbf{k}}}{2} \mathrm{e}^{-i \varphi_{\mathbf{k}}} \\
b_{\mathbf{k}+} \cos \frac{\theta_{\mathbf{k}}}{2} & b_{\mathbf{k}-} \sin \frac{\theta_{\mathbf{k}}}{2}
\end{pmatrix}.
\end{aligned}
\end{equation}
The factorial matrices from Eq.~\eqref{eq:chikd} are then:
\begin{equation}
\begin{aligned}
u_{\mathbf{k}}^{\dagger} \sigma_z u_{\mathbf{k}} &= \begin{pmatrix}
a_{\mathbf{k}+}^2 \cos \theta_{\mathbf{k}} & a_{\mathbf{k}+} a_{\mathbf{k}-} \sin \theta_{\mathbf{k}} \\
a_{\mathbf{k}+} a_{\mathbf{k}-} \sin \theta_{\mathbf{k}} & -a_{\mathbf{k}-}^2 \cos \theta_{\mathbf{k}}
\end{pmatrix}, \\
v_{-\mathbf{k}}^{\dagger} \sigma_z v_{-\mathbf{k}} &= \begin{pmatrix}
-|b_{\mathbf{k}+}|^2 \cos \theta_{\mathbf{k}} & -b_{\mathbf{k}+}^* b_{\mathbf{k}-} \sin \theta_{\mathbf{k}} \\
-b_{\mathbf{k}+} b_{\mathbf{k}-}^* \sin \theta_{\mathbf{k}} & |b_{\mathbf{k}-}|^2 \cos \theta_{\mathbf{k}}
\end{pmatrix}, \\
u_{\mathbf{k}}^{\dagger} \sigma_z v_{\mathbf{k}} &= \mathrm{e}^{-i \varphi_{\mathbf{k}}} \begin{pmatrix}
a_{\mathbf{k}+} b_{\mathbf{k}+} \cos \theta_{\mathbf{k}} & -a_{\mathbf{k}+} b_{\mathbf{k}-} \sin \theta_{\mathbf{k}} \\
a_{\mathbf{k}-} b_{\mathbf{k}+} \sin \theta_{\mathbf{k}} & a_{\mathbf{k}-} b_{\mathbf{k}-} \cos \theta_{\mathbf{k}}
\end{pmatrix}, \\
u_{-\mathbf{k}}^{\dagger} \sigma_z v_{-\mathbf{k}} &= \mathrm{e}^{-i \varphi_{\mathbf{k}}} \begin{pmatrix}
a_{\mathbf{k}+} b_{\mathbf{k}+} \cos \theta_{\mathbf{k}} & a_{\mathbf{k}+} b_{\mathbf{k}-} \sin \theta_{\mathbf{k}} \\
-a_{\mathbf{k}-} b_{\mathbf{k}+} \sin \theta_{\mathbf{k}} & a_{\mathbf{k}-} b_{\mathbf{k}-} \cos \theta_{\mathbf{k}}
\end{pmatrix}.
\end{aligned}
\end{equation}
\begin{widetext}
This results in the spin susceptibility components:
\begin{equation}
\begin{aligned}
\chi_{zz}^{ph+} &= -\mu_B^2 \sum_{\mathbf{k}} \cos^2 \theta_{\mathbf{k}} \left[ \left. \frac{df}{dE} \right|_{E = E_{\mathbf{k}+}} + \left. \frac{df}{dE} \right|_{E = E_{\mathbf{k}-}} \right], \\
\chi_{zz}^{ph-} &= -\mu_B^2 \sum_{\mathbf{k}} \sin^2 \theta_{\mathbf{k}} \left( 1 + \frac{\xi_{\mathbf{k}+} \xi_{\mathbf{k}-}}{E_{\mathbf{k}+} E_{\mathbf{k}-}} + \frac{|\Delta|^2}{E_{\mathbf{k}+} E_{\mathbf{k}-}} \right) \frac{f(E_{\mathbf{k}+}) - f(E_{\mathbf{k}-})}{E_{\mathbf{k}+} - E_{\mathbf{k}-}}, \\
\chi_{zz}^{pp} &= -\mu_B^2 \sum_{\mathbf{k}} \sin^2 \theta_{\mathbf{k}} \left( 1 - \frac{\xi_{\mathbf{k}+} \xi_{\mathbf{k}-}}{E_{\mathbf{k}+} E_{\mathbf{k}-}} - \frac{|\Delta|^2}{E_{\mathbf{k}+} E_{\mathbf{k}-}} \right) \frac{f(E_{\mathbf{k}+}) + f(E_{\mathbf{k}-}) - 1}{E_{\mathbf{k}+} + E_{\mathbf{k}-}}.
\end{aligned}
\end{equation}
\end{widetext}

\section{$p$-wave Pairing States with Only Rashba SOC: Zero-Temperature Spin susceptibility} \label{app:Bog_pwave_soc}

To facilitate the analysis, we introduce local coordinates in $\mathbf{k}$-space, defined by the unit vectors $\hat{n}_{1,2,3}(\mathbf{k})$ as follows:
\begin{equation}\label{eq:n123}
\begin{split}
\hat{n}_3(\mathbf{k}) &= \hat{\mathbf{k}} = (\sin \theta_{\mathbf{k}} \cos \varphi_{\mathbf{k}}, \sin \theta_{\mathbf{k}} \sin \varphi_{\mathbf{k}}, \cos \theta_{\mathbf{k}}), \\
\hat{n}_2(\mathbf{k}) &= \frac{\hat{\mathbf{k}} \times \hat{z}}{|\hat{\mathbf{k}} \times \hat{z}|} = (\sin \varphi_{\mathbf{k}}, -\cos \varphi_{\mathbf{k}}, 0), \\
\hat{n}_1(\mathbf{k}) &= \hat{n}_2(\mathbf{k}) \times \hat{n}_3(\mathbf{k}) = (-\cos \theta_{\mathbf{k}} \cos \varphi_{\mathbf{k}}, -\cos \theta_{\mathbf{k}} \sin \varphi_{\mathbf{k}}, \sin \theta_{\mathbf{k}}).
\end{split}
\end{equation}
With Rashba SOC present, the quasiparticle energy can be expressed as:
\begin{equation}\label{eq:EtripltG2}
E_{\mathbf{k}\pm} = \sqrt{(E_{\mathbf{k}\perp} \pm g k_F)^2 + |d_3(\mathbf{k})|^2},
\end{equation}
where $E_{\mathbf{k}\perp} = \sqrt{\xi_{\mathbf{k}}^2 + |\mathbf{d}_{\mathbf{k}} \times \hat{\mathbf{k}}|^2}$ and $d_3(\mathbf{k}) = \mathbf{d}_{\mathbf{k}} \cdot \hat{n}_3(\mathbf{k})$.

At zero temperature, for $E_{\mathbf{k}\pm} \geq 0$, the spin susceptibility $\chi_{\mu\mu}(T=0)$ arises solely from particle-particle (and hole-hole) processes, given by:
\begin{equation}
\begin{aligned}
&\, \chi_{\mu\mu}(T=0) = \chi_{\mu\mu}^{pp}(T=0) \\
= &\, \mu_B^2 \sum_{\mathbf{k}} \sum_{s_1, s_2} \frac{\left| (u_{\mathbf{k}}^{\dagger} \sigma_{\mu} v_{\mathbf{k}})^{s_1 s_2} - (u_{-\mathbf{k}}^{\dagger} \sigma_{\mu} v_{-\mathbf{k}})^{s_2 s_1} \right|^2}{E_{\mathbf{k} s_1} + E_{-\mathbf{k} s_2}}.   
\end{aligned}
\end{equation}
For the $x$-component $\chi_{xx}(T=0)$, the coefficients are:
\begin{widetext}
\begin{equation}\label{eq:chixxpp+_T0}
\begin{aligned}
\left| (u_{\mathbf{k}}^{\dagger} \sigma_x v_{\mathbf{k}})^{++} - (u_{-\mathbf{k}}^{\dagger} \sigma_x v_{-\mathbf{k}})^{++} \right|^2 &= \left| (u_{\mathbf{k}}^{\dagger} \sigma_x v_{\mathbf{k}})^{--} - (u_{-\mathbf{k}}^{\dagger} \sigma_x v_{-\mathbf{k}})^{--} \right|^2 = \frac{d_{12}^2(\mathbf{k})}{E_{\mathbf{k}\perp}^2} \left( \sin \varphi_{\mathbf{k}} \cos \Omega_{\mathbf{k}} + \cos \theta_{\mathbf{k}} \cos \varphi_{\mathbf{k}} \sin \Omega_{\mathbf{k}} \right)^2,
\end{aligned}
\end{equation}
where $d_{12}(\mathbf{k}) = |\mathbf{d_{\mathbf{k}}}\times \hat{\mathbf{k}}|$, and
\begin{equation}\label{eq:chixxpp-_T0}
\begin{aligned}
& \quad \left| (u_{\mathbf{k}}^{\dagger} \sigma_x v_{\mathbf{k}})^{+-} - (u_{-\mathbf{k}}^{\dagger} \sigma_x v_{-\mathbf{k}})^{-+} \right|^2 = \left| (u_{\mathbf{k}}^{\dagger} \sigma_x v_{\mathbf{k}})^{-+} - (u_{-\mathbf{k}}^{\dagger} \sigma_x v_{-\mathbf{k}})^{+-} \right|^2 \\
&= \frac{1}{4} \left\{ \left[ s_{\mathbf{k}3} (a_{\mathbf{k}1+} a_{\mathbf{k}2-} + a_{\mathbf{k}1-} a_{\mathbf{k}2+}) (\sin \varphi_{\mathbf{k}} \sin \Omega_{\mathbf{k}} - \cos \theta_{\mathbf{k}} \cos \varphi_{\mathbf{k}} \cos \Omega_{\mathbf{k}}) - (a_{\mathbf{k}1+} a_{\mathbf{k}1-} - a_{\mathbf{k}2+} a_{\mathbf{k}2-}) \frac{d_{12}(\mathbf{k})}{E_{\mathbf{k}\perp}} \cos \varphi_{\mathbf{k}} \sin \theta_{\mathbf{k}} \right]^2 \right. \\
&\quad \left. + \frac{\xi_{\mathbf{k}}^2}{E_{\mathbf{k}\perp}^2} (a_{\mathbf{k}1+} a_{\mathbf{k}2-} + a_{\mathbf{k}1-} a_{\mathbf{k}2+})^2 (\sin \varphi_{\mathbf{k}} \cos \Omega_{\mathbf{k}} + \cos \theta_{\mathbf{k}} \cos \varphi_{\mathbf{k}} \sin \Omega_{\mathbf{k}})^2 \right\}.
\end{aligned}
\end{equation}
For the $z$-component $\chi_{zz}(T=0)$, the coefficients are:
\begin{equation}\label{eq:chizzpp+_T0}
\begin{aligned}
\left| (u_{\mathbf{k}}^{\dagger} \sigma_z v_{\mathbf{k}})^{++} - (u_{-\mathbf{k}}^{\dagger} \sigma_z v_{-\mathbf{k}})^{++} \right|^2 &= \left| (u_{\mathbf{k}}^{\dagger} \sigma_z v_{\mathbf{k}})^{--} - (u_{-\mathbf{k}}^{\dagger} \sigma_z v_{-\mathbf{k}})^{--} \right|^2 = \frac{d_{12}^2(\mathbf{k})}{E_{\mathbf{k}\perp}^2} \sin^2 \theta_{\mathbf{k}} \sin^2 \Omega_{\mathbf{k}},
\end{aligned}
\end{equation}
and
\begin{equation}\label{eq:chizzpp-_T0}
\begin{aligned}
&\quad \left| (u_{\mathbf{k}}^{\dagger} \sigma_z v_{\mathbf{k}})^{+-} - (u_{-\mathbf{k}}^{\dagger} \sigma_z v_{-\mathbf{k}})^{-+} \right|^2 = \left| (u_{\mathbf{k}}^{\dagger} \sigma_z v_{\mathbf{k}})^{-+} - (u_{-\mathbf{k}}^{\dagger} \sigma_z v_{-\mathbf{k}})^{+-} \right|^2 \\
&= \frac{1}{4} \left\{ \left[ s_{\mathbf{k}3} (a_{\mathbf{k}1+} a_{\mathbf{k}2-} + a_{\mathbf{k}1-} a_{\mathbf{k}2+}) \sin \theta_{\mathbf{k}} \cos \Omega_{\mathbf{k}} - \frac{d_{12}(\mathbf{k})}{E_{\mathbf{k}\perp}} (a_{\mathbf{k}1+} a_{\mathbf{k}1-} - a_{\mathbf{k}2+} a_{\mathbf{k}2-}) \cos \theta_{\mathbf{k}} \right]^2 + \frac{\xi_{\mathbf{k}}^2}{E_{\mathbf{k}\perp}^2} (a_{\mathbf{k}1+} a_{\mathbf{k}2-} + a_{\mathbf{k}1-} a_{\mathbf{k}2+})^2 \sin^2 \theta_{\mathbf{k}} \sin^2 \Omega_{\mathbf{k}} \right\}.
\end{aligned}
\end{equation}

The coefficients are:
\begin{equation}
\begin{aligned}
a_{\mathbf{k}1+} a_{\mathbf{k}2-} + a_{\mathbf{k}1-} a_{\mathbf{k}2+} &= \frac{1}{\sqrt{E_{\mathbf{k}+} E_{\mathbf{k}-}}} \left[ \sqrt{(E_{\mathbf{k}+} + E_{\mathbf{k}\perp} + g |\mathbf{k}|)(E_{\mathbf{k}-} - E_{\mathbf{k}\perp} + g |\mathbf{k}|)} + \sqrt{(E_{\mathbf{k}-} + E_{\mathbf{k}\perp} - g |\mathbf{k}|)(E_{\mathbf{k}+} - E_{\mathbf{k}\perp} - g |\mathbf{k}|)} \right], \\
a_{\mathbf{k}1+} a_{\mathbf{k}1-} - a_{\mathbf{k}2+} a_{\mathbf{k}2-} &= \frac{1}{\sqrt{E_{\mathbf{k}+} E_{\mathbf{k}-}}} \left[ \sqrt{(E_{\mathbf{k}+} + E_{\mathbf{k}\perp} + g |\mathbf{k}|)(E_{\mathbf{k}-} + E_{\mathbf{k}\perp} - g |\mathbf{k}|)} - \sqrt{(E_{\mathbf{k}+} - E_{\mathbf{k}\perp} - g |\mathbf{k}|)(E_{\mathbf{k}-} - E_{\mathbf{k}\perp} + g |\mathbf{k}|)} \right].
\end{aligned}
\end{equation}
Squaring these gives:
\begin{equation}
\begin{aligned}
(a_{\mathbf{k}1+} a_{\mathbf{k}2-} + a_{\mathbf{k}1-} a_{\mathbf{k}2+})^2 &= 2 \left[ 1 + \frac{d_3^2(\mathbf{k}) - (E_{\mathbf{k}\perp} + g |\mathbf{k}|)(E_{\mathbf{k}\perp} - g |\mathbf{k}|)}{E_{\mathbf{k}+} E_{\mathbf{k}-}} \right] = \frac{(E_{\mathbf{k}+} + E_{\mathbf{k}-})^2 - 4 E_{\mathbf{k}\perp}^2}{E_{\mathbf{k}+} E_{\mathbf{k}-}}, \\
(a_{\mathbf{k}1+} a_{\mathbf{k}1-} - a_{\mathbf{k}2+} a_{\mathbf{k}2-})^2 &= 2 \left[ 1 - \frac{d_3^2(\mathbf{k}) - (E_{\mathbf{k}\perp} + g |\mathbf{k}|)(E_{\mathbf{k}\perp} - g |\mathbf{k}|)}{E_{\mathbf{k}+} E_{\mathbf{k}-}} \right] = \frac{4 E_{\mathbf{k}\perp}^2 - (E_{\mathbf{k}+} - E_{\mathbf{k}-})^2}{E_{\mathbf{k}+} E_{\mathbf{k}-}}, \\
s_{\mathbf{k}3} (a_{\mathbf{k}1+} a_{\mathbf{k}2-} + a_{\mathbf{k}1-} a_{\mathbf{k}2+}) (a_{\mathbf{k}1+} a_{\mathbf{k}1-} - a_{\mathbf{k}2+} a_{\mathbf{k}2-}) &= \frac{4 E_{\mathbf{k}\perp} d_3(\mathbf{k})}{E_{\mathbf{k}+} E_{\mathbf{k}-}}.
\end{aligned}
\end{equation}

The parameters for different unitary $p$-wave pairing states are summarized in Table~\ref{tb:parameters}.
\begin{table}[tb]
\centering
\caption{Parameterization for unitary p-wave pairing states.}
\label{tb:parameters}
\renewcommand\arraystretch{2.0}
\begin{tabular}{c|c|c|c|c|c|c}
\hline
\hline
&\multicolumn{2}{c|}{OSP}&\multicolumn{4}{c}{ESP}\\
\hline
Notation & $(k_x+ik_y)\hat{z}$ & $k_z\hat{z}$ & $k_x\hat{x}+k_y\hat{y}$ & $k_y\hat{x}-k_x\hat{y}$ & $k_x\hat{x}-k_y\hat{y}$ & $k_y\hat{x}+k_x\hat{y}$ \\
\hline
$\mathbf{d}(\mathbf{k})$ & $\Delta\sin\theta_{\mathbf{k}}\mathrm{e}^{i\varphi_{\mathbf{k}}}\hat{z}$& $\Delta\cos\theta_{\mathbf{k}}\hat{z}$ &\tabincell{c}{$\Delta\sin\theta_\mathbf{k}$\\$(\cos\varphi_{\mathbf{k}}\hat{x}+\sin\varphi_{\mathbf{k}}\hat{y})$}  &\tabincell{c}{$\Delta\sin\theta_\mathbf{k}$\\$(\sin\varphi_{\mathbf{k}}\hat{x}-\cos\varphi_{\mathbf{k}}\hat{y})$}  &\tabincell{c}{$\Delta\sin\theta_\mathbf{k}$\\$(\cos\varphi_{\mathbf{k}}\hat{x}-\sin\varphi_{\mathbf{k}}\hat{y})$}  &\tabincell{c}{$\Delta\sin\theta_\mathbf{k}$\\$(\sin\varphi_{\mathbf{k}}\hat{x}+\cos\varphi_{\mathbf{k}}\hat{y})$}  \\
\hline
$\eta_\mathbf{k}$&$\varphi_{\mathbf{k}}$& 0 &0&0&0&0\\
\hline
$\gamma_\mathbf{k}$&$\pi$ & $\pi-\varphi_{\mathbf{k}}$ &$\pi-\varphi_{\mathbf{k}}$&$\pi-\varphi_{\mathbf{k}}$&$\pi-\varphi_{\mathbf{k}}$&$\pi-\varphi_{\mathbf{k}}$\\
\hline
$d_1(\mathbf{k})$&$\Delta\sin^2\theta_{\mathbf{k}}$ & $\Delta\cos\theta_{\mathbf{k}}\sin\theta_{\mathbf{k}}$ &$-\Delta\sin\theta_{\mathbf{k}}\cos\theta_{\mathbf{k}}$&0&$-\Delta\sin\theta_{\mathbf{k}}\cos\theta_{\mathbf{k}}\cos{2\varphi_{\mathbf{k}}}$&$\Delta\sin\theta_{\mathbf{k}}\cos\theta_{\mathbf{k}}\sin{2\varphi_{\mathbf{k}}}$\\
\hline
$d_2(\mathbf{k})$&0&0&0&$\Delta\sin\theta_{\mathbf{k}}$&$\Delta\sin\theta_{\mathbf{k}}\sin{2\varphi_{\mathbf{k}}}$&$-\Delta\sin\theta_{\mathbf{k}}\cos{2\varphi_{\mathbf{k}}}$\\
\hline
$d_3(\mathbf{k})$&$\Delta\sin\theta_{\mathbf{k}}\cos\theta_{\mathbf{k}}$& $\Delta\cos^2\theta_{\mathbf{k}}$&$\Delta\sin^2\theta_{\mathbf{k}}$&0&$\Delta\sin^2\theta_{\mathbf{k}}\cos{2\varphi_{\mathbf{k}}}$&$\Delta\sin^2\theta_{\mathbf{k}}\sin2\varphi_{\mathbf{k}}$\\
\hline
$\cos\Omega_{\mathbf{k}}$ & 1 & $\mbox{Sgn}(\cos\theta_{\mathbf{k}})$ & $\mbox{Sgn}(-\cos\theta_{\mathbf{k}})$&0&$d_1(\mathbf{k})/d_{12}(\mathbf{k})$&$d_1(\mathbf{k})/d_{12}(\mathbf{k})$\\
\hline
$\sin\Omega_{\mathbf{k}}$&0&0&0&1&$d_2(\mathbf{k})/d_{12}(\mathbf{k})$&$d_2(\mathbf{k})/d_{12}(\mathbf{k})$\\
\hline
$d_{12}(\mathbf{k})$&$\Delta\sin^2\theta_{\mathbf{k}}$&$\Delta\sin\theta_{\mathbf{k}}|\cos\theta_{\mathbf{k}}|$&$|\Delta\sin\theta_{\mathbf{k}}\cos\theta_{\mathbf{k}}|$&$\Delta\sin\theta_{\mathbf{k}}$&\tabincell{c}{$\Delta\sin\theta_{\mathbf{k}}[\sin^22\varphi_{\mathbf{k}}$\\$+\cos^2\theta_{\mathbf{k}}\cos^22\varphi_{\mathbf{k}}]^{\frac{1}{2}}$}&\tabincell{c}{$\Delta\sin\theta_{\mathbf{k}}[\cos^22\varphi_{\mathbf{k}}$\\$+\cos^2\theta_{\mathbf{k}}\sin^22\varphi_{\mathbf{k}}]^{\frac{1}{2}}$}\\
\hline
\hline
\end{tabular}
\end{table}
\end{widetext}

\section{$p$-wave Pairing States with Strong Rashba SOC: Zero-Temperature Spin Susceptibility}\label{app:strongSOC_limit}

In this appendix, we demonstrate why the zero-temperature spin susceptibility components $\chi_{xx}(T=0)$ and $\chi_{zz}(T=0)$ approach $\frac{2}{3} \chi_N$ in the strong Rashba SOC limit. When Rashba SOC is sufficiently strong, the superconducting transition temperature $T_c$ and the gap function $\Delta(T=0)$ are significantly suppressed. This allows us to approximate the energy dispersion by neglecting gap-induced corrections, leading to:
\begin{equation}
\begin{aligned}
E_{\mathbf{k}\pm} &= \sqrt{[E_{\mathbf{k}\perp} \pm g k_F]^2 + d_3^2(\mathbf{k})} \approx |\xi_{\mathbf{k}} \pm g k_F|.
\end{aligned}
\end{equation}
Given that contributions to spin susceptibility primarily originate from wave vectors near the Fermi surface, we focus on the regime $-g k_F < \xi_{\mathbf{k}} < g k_F$, which becomes increasingly valid as SOC strength increases. Thus, the energy simplifies to:
\begin{equation}
\begin{aligned}
E_{\mathbf{k}\pm} &= \begin{cases}
g k_F \pm \xi_{\mathbf{k}}, & \xi_{\mathbf{k}} > 0, \\
g k_F \mp \xi_{\mathbf{k}}, & \xi_{\mathbf{k}} < 0.
\end{cases}
\end{aligned}
\end{equation}

For $\xi_{\mathbf{k}} > 0$, the coefficients are approximated as:
\begin{equation}
\begin{aligned}
a_{\mathbf{k}1\pm} &= \sqrt{\frac{E_{\mathbf{k}\pm} + E_{\mathbf{k}\perp} \pm g |\mathbf{k}|}{E_{\mathbf{k}\pm}}} \approx \begin{cases}
\sqrt{2}, & +, \\
0, & -,
\end{cases} \\
a_{\mathbf{k}2\pm} &= \sqrt{\frac{E_{\mathbf{k}\pm} - E_{\mathbf{k}\perp} \mp g |\mathbf{k}|}{E_{\mathbf{k}\pm}}} \approx \begin{cases}
0, & +, \\
\sqrt{2}, & -.
\end{cases}
\end{aligned}
\end{equation}
Substituting these into the expressions for $\chi_{xx}(T=0)$ and $\chi_{zz}(T=0)$, we obtain:
\begin{widetext}
\begin{equation}
\begin{aligned}
\chi_{xx}(T=0) &= \mu_B^2 \sum_{\mathbf{k}} \frac{d_{12}^2(\mathbf{k})}{2 E_{\mathbf{k}\perp}^2} (\sin \varphi_{\mathbf{k}} \cos \Omega + \cos \varphi_{\mathbf{k}} \sin \Omega \cos \theta_{\mathbf{k}})^2 \left[ \frac{1}{E_{\mathbf{k}+}} + \frac{1}{E_{\mathbf{k}-}} \right] \\
&\quad + \frac{1}{2} \left\{ \left[ s_{\mathbf{k}} (a_{\mathbf{k}1+} a_{\mathbf{k}2-} + a_{\mathbf{k}1-} a_{\mathbf{k}2+}) (\sin \varphi_{\mathbf{k}} \sin \Omega_{\mathbf{k}} - \cos \theta_{\mathbf{k}} \cos \varphi_{\mathbf{k}} \cos \Omega_{\mathbf{k}}) - (a_{\mathbf{k}1+} a_{\mathbf{k}1-} - a_{\mathbf{k}2+} a_{\mathbf{k}2-}) \frac{d_{12}(\mathbf{k})}{E_{\mathbf{k}\perp}} \cos \varphi_{\mathbf{k}} \sin \theta_{\mathbf{k}} \right]^2 \right. \\
&\quad \left. + \frac{\xi_{\mathbf{k}}^2}{E_{\mathbf{k}\perp}^2} (a_{\mathbf{k}1+} a_{\mathbf{k}2-} + a_{\mathbf{k}1-} a_{\mathbf{k}2+})^2 (\sin \varphi_{\mathbf{k}} \cos \Omega_{\mathbf{k}} + \cos \theta_{\mathbf{k}} \cos \varphi_{\mathbf{k}} \sin \Omega_{\mathbf{k}})^2 \right\} \frac{1}{E_{\mathbf{k}+} + E_{\mathbf{k}-}} \\
&\approx \mu_B^2 \sum_{\mathbf{k}} \frac{1}{2} \left[ 4 (\sin \varphi_{\mathbf{k}} \sin \Omega_{\mathbf{k}} - \cos \theta_{\mathbf{k}} \cos \varphi_{\mathbf{k}} \cos \Omega_{\mathbf{k}})^2 + 4 (\sin \varphi_{\mathbf{k}} \cos \Omega_{\mathbf{k}} + \cos \theta_{\mathbf{k}} \cos \varphi_{\mathbf{k}} \sin \Omega_{\mathbf{k}})^2 \right] \frac{1}{2 g k_F} \\
&= \mu_B^2 \sum_{\mathbf{k}} (\sin^2 \varphi_{\mathbf{k}} + \cos^2 \theta_{\mathbf{k}} \cos^2 \varphi_{\mathbf{k}}) \frac{1}{g k_F} \\
&\approx \frac{2}{3} \chi_N,
\end{aligned}
\end{equation}
and similarly for $\chi_{zz}(T=0)$:
\begin{equation}
\begin{aligned}
\chi_{zz}(T=0) &= \mu_B^2 \sum_{\mathbf{k}} \frac{d_{12}^2(\mathbf{k})}{E_{\mathbf{k}\perp}^2} \sin^2 \theta_{\mathbf{k}} \sin^2 \Omega_{\mathbf{k}} \left[ \frac{1}{E_{\mathbf{k}+}} + \frac{1}{E_{\mathbf{k}-}} \right] \\
&\quad + \frac{1}{2} \left\{ \left[ (a_{\mathbf{k}1+} a_{\mathbf{k}2-} + a_{\mathbf{k}1-} a_{\mathbf{k}2+}) s_{\mathbf{k}} \cos \Omega_{\mathbf{k}} \sin \theta_{\mathbf{k}} - \frac{d_{12}(\mathbf{k})}{E_{\mathbf{k}\perp}} (a_{\mathbf{k}1+} a_{\mathbf{k}1-} - a_{\mathbf{k}2+} a_{\mathbf{k}2-}) \cos \theta_{\mathbf{k}} \right]^2 \right. \\
&\quad \left. + \frac{\xi_{\mathbf{k}}^2}{E_{\mathbf{k}\perp}^2} (a_{\mathbf{k}1+} a_{\mathbf{k}2-} + a_{\mathbf{k}1-} a_{\mathbf{k}2+})^2 \sin^2 \Omega_{\mathbf{k}} \sin^2 \theta_{\mathbf{k}} \right\} \frac{1}{E_{\mathbf{k}+} + E_{\mathbf{k}-}} \\
&\approx \mu_B^2 \sum_{\mathbf{k}} \frac{1}{2} \left[ 4 \sin^2 \theta_{\mathbf{k}} \cos^2 \Omega_{\mathbf{k}} + 4 \sin^2 \theta_{\mathbf{k}} \sin^2 \Omega_{\mathbf{k}} \right] \frac{1}{2 g k_F} \\
&\approx \frac{2}{3} \chi_N.
\end{aligned}
\end{equation}
\end{widetext}
Therefore, both $\chi_{zz}(T=0)$ and $\chi_{xx}(T=0)$ approach $\frac{2}{3} \chi_N$ in the strong Rashba SOC limit. Given the significant suppression of the gap function ($\Delta(T=0, g) \ll \Delta(T=0, g=0)$), the intra-band term $\chi_{\mu\mu}^{pp+}(T=0)$ can be neglected, as it scales with $\Delta^2(T=0, g)$ according to Eqs.~\eqref{eq:chixxpp+_T0} and \eqref{eq:chizzpp+_T0}. Consequently, $\chi_{\mu\mu}(T=0)$ is dominated by the inter-band term $\chi_{\mu\mu}^{pp-}$. Since this derivation is largely independent of the specific pairing symmetry, the result holds for general unitary $p$-wave pairing states.

\section{OSP States with Only Rashba SOC: $\chi_{xx}(T=0)$}\label{app:osp_chixx_soc}

\begin{figure*}[tb]
\centering
\subfigure[]{
\includegraphics[width=0.48\linewidth]{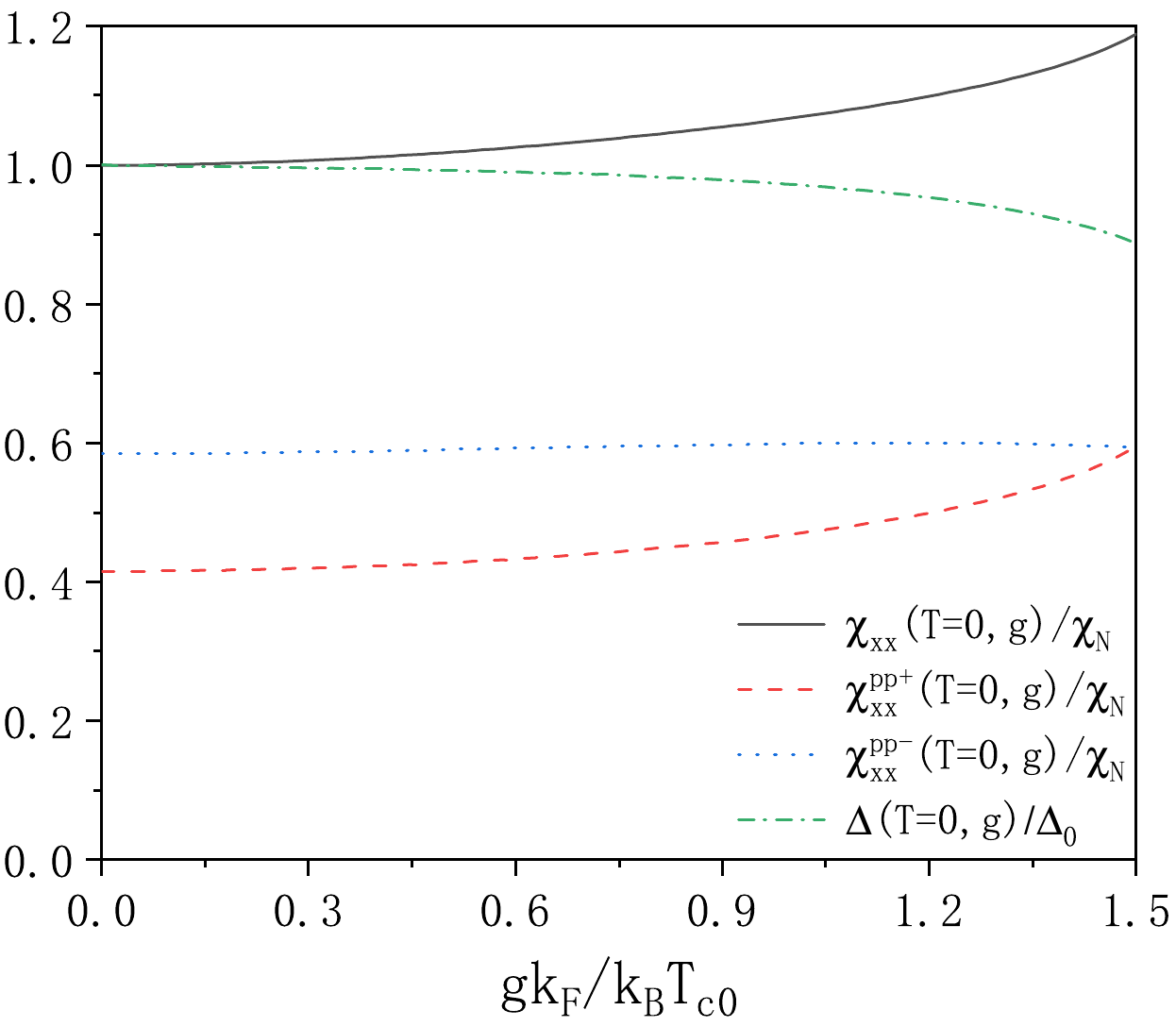}
}
\subfigure[]{
\includegraphics[width=0.48\linewidth]{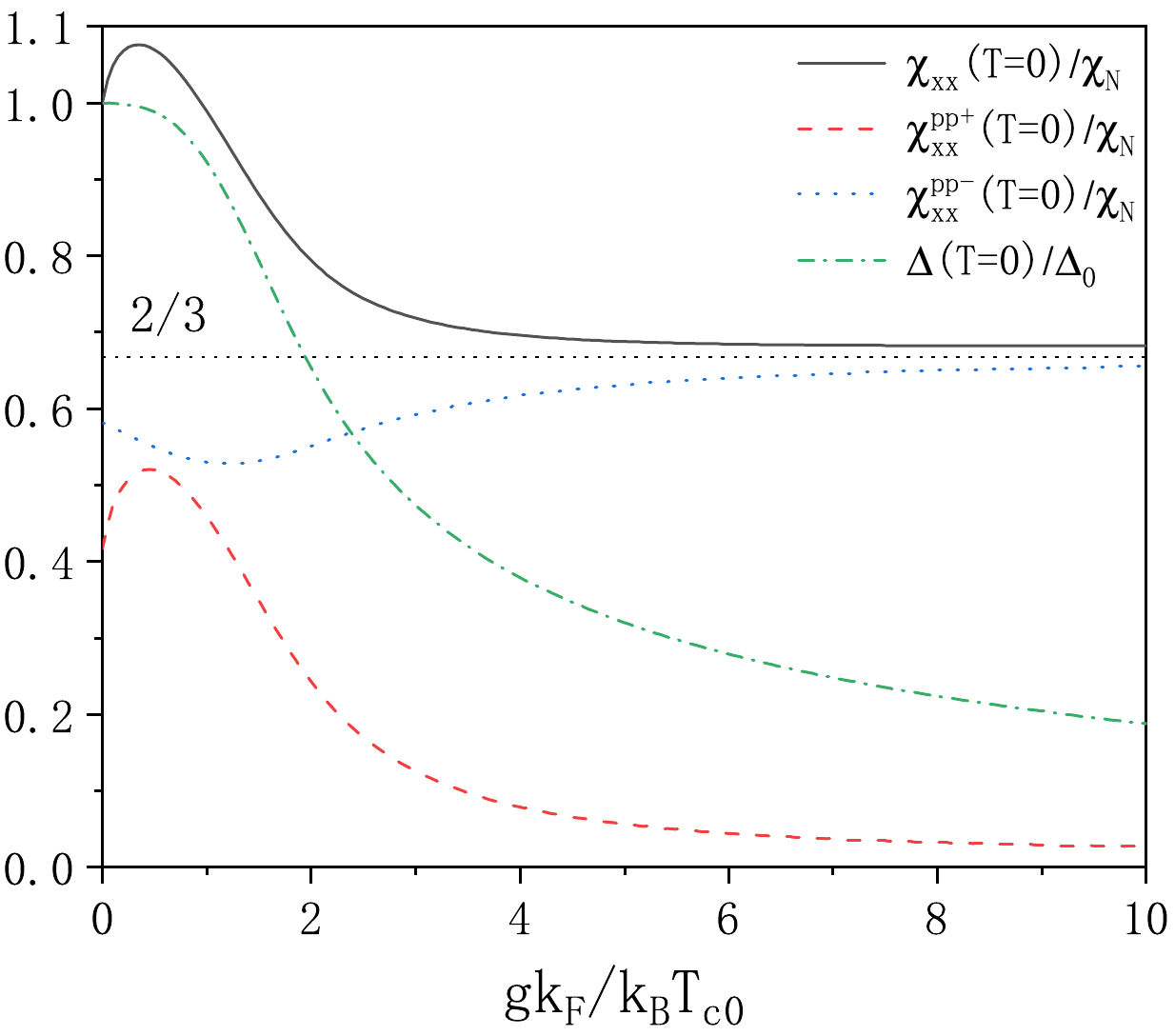}
}
\caption{The effect of Rashba SOC on spin susceptibility $\chi_{xx}(T=0)$ and gap function $\Delta(T=0)$ at zero temperature: (a) $(k_x + i k_y) \hat{z}$ pairing state, (b) $k_z \hat{z}$ pairing state. The solid lines represent zero-temperature spin susceptibility $\chi_{xx}(T=0)$ in units of $\chi_N$, while dashed and dotted lines denote the intra- and inter-band terms of $\chi_{xx}^{pp}(T=0)$, respectively. The dash-dotted lines show the zero-temperature gap function in units of $\Delta_0$.}
\label{fig:susxx_zero_p_kx+iky_soc}
\end{figure*}

For the $(k_x + i k_y) \hat{z}$ and $k_z \hat{z}$ opposite-spin pairing (OSP) states, Fig.~\ref{fig:susxx_zero_p_kx+iky_soc} illustrates the effects of Rashba SOC on the zero-temperature gap function $\Delta(T=0)$ and spin susceptibility along the $x$-axis, $\chi_{xx}(T=0)$. The numerical results indicate that for small Rashba SOC strengths, the self-consistent gap function $\Delta(T=0, g)$ remains largely unchanged, and the behavior of $\chi_{xx}(T=0)$ is predominantly governed by the intra-band term $\chi_{xx}^{pp+}$. Building on the framework from Appendix~\ref{app:Bog_pwave_soc}, the expression for $\chi_{xx}^{pp+}(T=0)$ is given by:
\begin{widetext}
\begin{equation}
\chi_{xx}^{pp+}(T=0) = \mu_B^2 \sum_{\mathbf{k}} \frac{d_{12}^2(\mathbf{k})}{2 E_{\mathbf{k}\perp}^2} \left( \sin \varphi_{\mathbf{k}} \cos \Omega_{\mathbf{k}} + \cos \theta_{\mathbf{k}} \cos \varphi_{\mathbf{k}} \sin \Omega_{\mathbf{k}} \right)^2 \cdot \left( \frac{1}{E_{\mathbf{k}+}} + \frac{1}{E_{\mathbf{k}-}} \right).
\end{equation}
\end{widetext}
Under the assumption that $\Delta(T=0)$ is approximately constant for weak Rashba SOC, the dependence of $\chi_{xx}^{pp+}(T=0)$ on $g$ stems solely from the factor $\frac{1}{E_{\mathbf{k}+}} + \frac{1}{E_{\mathbf{k}-}}$. To analyze this dependence, we compute the derivative with respect to $g |\mathbf{k}|$:
\begin{equation}\label{eq:ekpekm}
\frac{\partial}{\partial g |\mathbf{k}|} \left( \frac{1}{E_{\mathbf{k}+}} + \frac{1}{E_{\mathbf{k}-}} \right) = \frac{E_{\mathbf{k}\perp} - g |\mathbf{k}|}{E_{\mathbf{k}-}^3} - \frac{E_{\mathbf{k}\perp} + g |\mathbf{k}|}{E_{\mathbf{k}+}^3}.
\end{equation}
Furthermore, the derivative of the related term is:
\begin{equation}
\frac{\partial}{\partial g |\mathbf{k}|} \frac{E_{\mathbf{k}\perp} + g |\mathbf{k}|}{E_{\mathbf{k}+}^3} = \frac{-3 (E_{\mathbf{k}\perp} + g |\mathbf{k}|)^2 + E_{\mathbf{k}+}^2}{E_{\mathbf{k}+}^5}.
\end{equation}
Since this derivative assumes negative values for most wave vectors when $g |\mathbf{k}|$ is small, the factor $\frac{1}{E_{\mathbf{k}+}} + \frac{1}{E_{\mathbf{k}-}}$ increases as $g$ grows from zero, resulting in an overall increase in spin susceptibility for weak Rashba SOC.

\section{OSP States in a Perpendicular Zeeman Field: Temperature-Independent Spin Susceptibility Component}\label{app:chixx_osp}

In this section, we analyze the temperature-independent behavior of the spin susceptibility component $\chi_{xx}(T)$ for opposite-spin pairing (OSP) states under a perpendicular Zeeman field. Specifically, we focus on the $(k_x + i k_y) \hat{z}$ and $k_z \hat{z}$ pairing states, demonstrating that $\chi_{xx}(T) = \chi_N$ for all temperatures, as derived from the decomposition in Eq.~\eqref{eq:chikd}, which yields $\chi_{xx}^{ph-} = 0$ and $\chi_{xx}^{ph+} + \chi_{xx}^{pp} = \chi_N$.

The numerical and theoretical effects are illustrated in Fig.~\ref{fig:susxx_zero_p_kx+iky_soc}, which shows the influence of Rashba SOC on $\chi_{xx}(T=0)$ and the gap function $\Delta(T=0)$. For a Zeeman field along the $x$-axis, the Bogoliubov transformation for a general OSP state is given by:
\begin{subequations}
\begin{equation}
u_{\mathbf{k}} = \begin{pmatrix}
a_{\mathbf{k}+} & a_{\mathbf{k}-} \\
a_{\mathbf{k}+} & -a_{\mathbf{k}-}
\end{pmatrix}, \quad v_{\mathbf{k}} = \begin{pmatrix}
-b_{\mathbf{k}+} & b_{\mathbf{k}-} \\
-b_{\mathbf{k}+} & -b_{\mathbf{k}-}
\end{pmatrix},
\end{equation}
where the coefficients are:
\begin{equation}
\begin{aligned}
a_{\mathbf{k}\pm} &= \frac{E_{\mathbf{k}\pm} + \xi_{\mathbf{k}\pm}}{2 [E_{\mathbf{k}\pm} (E_{\mathbf{k}\pm} + \xi_{\mathbf{k}\pm})]^{1/2}}, \\
b_{\mathbf{k}\pm} &= \frac{d_z(\mathbf{k})}{2 [E_{\mathbf{k}\pm} (E_{\mathbf{k}\pm} + \xi_{\mathbf{k}\pm})]^{1/2}},
\end{aligned}
\end{equation}
and the energy terms are defined as:
\begin{equation}
\xi_{\mathbf{k}\pm} = \xi_{\mathbf{k}} \pm \mu_B H_z, \quad E_{\mathbf{k}\pm} = \sqrt{\xi_{\mathbf{k}\pm}^2 + |d_z(\mathbf{k})|^2}.
\end{equation}
\end{subequations}

From these, the relevant matrix elements are:
\begin{equation}
\begin{split}
[(u_{\mathbf{k}}^{\dagger} \sigma_x u_{\mathbf{k}})^{ss} - (v_{-\mathbf{k}}^{\dagger} \sigma_x v_{-\mathbf{k}})^{ss}]^2 &= \frac{\xi_{\mathbf{k}s}^2}{E_{\mathbf{k}s}^2}, \\
|(u_{\mathbf{k}}^{\dagger} \sigma_x v_{\mathbf{k}})^{ss} - (u_{-\mathbf{k}}^{\dagger} \sigma_x v_{-\mathbf{k}})^{ss}|^2 &= 1 - \frac{\xi_{\mathbf{k}s}^2}{E_{\mathbf{k}s}^2} = \frac{|d_z(\mathbf{k})|^2}{E_{\mathbf{k}s}^2},
\end{split}
\end{equation}
and the off-diagonal elements vanish:
\begin{equation}
(u_{\mathbf{k}}^{\dagger} \sigma_x u_{\mathbf{k}})^{s\bar{s}} = (v_{\mathbf{k}}^{\dagger} \sigma_x v_{\mathbf{k}})^{s\bar{s}} = (u_{\mathbf{k}}^{\dagger} \sigma_x v_{\mathbf{k}})^{s\bar{s}} = 0,
\end{equation}
where $s = \pm$ and $\bar{s}$ is the opposite of $s$. Consequently, the spin susceptibility components are:
\begin{equation}
\begin{split}
\chi_{xx}^{ph+} &= -\mu_B^2 \sum_{\mathbf{k}} \sum_{s = \pm} \frac{\xi_{\mathbf{k}s}^2}{E_{\mathbf{k}s}^2} \frac{df(E_{\mathbf{k}s})}{dE_{\mathbf{k}s}}, \\
\chi_{xx}^{ph-} &= 0, \\
\chi_{xx}^{pp} &= -\mu_B^2 \sum_{\mathbf{k}} \sum_{s} \frac{|d_z(\mathbf{k})|^2}{E_{\mathbf{k}s}^2} \frac{f(E_{\mathbf{k}s}) + f(E_{-\mathbf{k}s}) - 1}{E_{\mathbf{k}s} + E_{-\mathbf{k}s}}.
\end{split}
\end{equation}
Given that $E_{\mathbf{k}s} = E_{-\mathbf{k}s}$, we can rewrite the components as:
\begin{subequations}
\begin{equation}
\begin{aligned}
\chi_{xx}^{ph+}(T) &= 4 N(0) \mu_B^2 \int \frac{d\Omega}{4\pi} \int_{|d_z(\mathbf{k})|}^{\infty} dE \,  \frac{1}{\sqrt{E^2 - |d_z(\mathbf{k})|^2}}\\
&\quad\times\frac{|d_z(\mathbf{k})|^2}{E^2} f(E),
\end{aligned}
\end{equation}
and
\begin{equation}
\begin{aligned}
\chi_{xx}^{pp}(T) &= 4 N(0) \mu_B^2 \int \frac{d\Omega}{4\pi} \int_{|d_z(\mathbf{k})|}^{\infty} dE \frac{E}{\sqrt{E^2 - |d_z(\mathbf{k})|^2}} \\
&\quad\times\frac{|d_z(\mathbf{k})|^2}{E^2} \frac{1 - 2f(E)}{2E},
\end{aligned}
\end{equation}
\end{subequations}
where $\Omega$ is the solid angle in $\mathbf{k}$-space. Summing these contributions, we obtain:
\begin{equation}
\begin{split}
\chi_{xx}(T) &= 2 N(0) \mu_B^2 \int \frac{d\Omega}{4\pi} \int_{|d_z(\mathbf{k})|}^{\infty} dE \frac{1}{\sqrt{E^2 - |d_z(\mathbf{k})|^2}} \frac{|d_z(\mathbf{k})|^2}{E^2} \\
&= 2 N(0) \mu_B^2 = \chi_N.
\end{split}
\end{equation}
This result confirms that $\chi_{xx}(T)$ is temperature-independent and equal to $\chi_N$ for the OSP states under a perpendicular Zeeman field.

\section{Pairing States $k_x \hat{x} - k_y \hat{y}$ and $k_y \hat{x} + k_x \hat{y}$: Divergence in $\chi_{zz}(T=0)$}\label{app_diverge}

In this section, we analyze the divergence of the zero-temperature spin susceptibility $\chi_{zz}(T=0)$ for the $k_x \hat{x} - k_y \hat{y}$ and $k_y \hat{x} + k_x \hat{y}$ pairing states when $g k_F = \Delta(T=0, g)$. This divergence arises near nodal lines, as derived from Eq.~\eqref{eq:xT0} and Appendix~\ref{app:Bog_pwave_soc}. The expression for $\chi_{zz}(T=0)$ is:
\begin{equation}
\begin{aligned}
\chi_{zz}(T=0) &= \chi_{zz}^{pp}(T=0) = \chi_{zz}^{pp+}(T=0) + \chi_{zz}^{pp-}(T=0),
\end{aligned}
\end{equation}
where the intra-band term is:
\begin{equation}
\chi_{zz}^{pp+}(T=0) = \mu_B^2 \sum_{\mathbf{k}} \frac{d_{12}^2}{E_{\mathbf{k}\perp}^2} \sin^2 \theta_{\mathbf{k}} \sin^2 \Omega_{\mathbf{k}} \cdot \frac{1}{2} \left( \frac{1}{E_{\mathbf{k}+}} + \frac{1}{E_{\mathbf{k}-}} \right).
\end{equation}
Numerical results indicate that the divergence is dominated by the intra-band term rather than the inter-band term, as $E_{\mathbf{k}+} > 0$. Thus, we focus on the integral near nodal lines:
\begin{equation}
\begin{aligned}
\mu_B^2 \sum_{\mathbf{k} \text{ near } \mathbf{k}_0} \frac{d_{12}^2}{E_{\mathbf{k}\perp}^2} \sin^2 \theta_{\mathbf{k}} \sin^2 \Omega_{\mathbf{k}} \cdot \frac{1}{2 E_{\mathbf{k}-}},
\end{aligned}
\end{equation}
where $\mathbf{k}_0$ lies on a nodal line defined by:
\begin{equation}
\begin{aligned}
\cos 2 \varphi_0 &= 0, \\
\xi_0^2 + \Delta^2 \sin^2 \theta_0 &= g^2 k_F^2.
\end{aligned}
\end{equation}

For wave vectors near $\mathbf{k}_0$, we apply the following approximations:
\begin{equation}
\begin{aligned}
\sin(\theta_0 + \delta \theta) &\approx \sin \theta_0 + \delta \theta \cdot \cos \theta_0, \\
\cos(\theta_0 + \delta \theta) &\approx \cos \theta_0 - \delta \theta \cdot \sin \theta_0, \\
\sin(2 \varphi_0 + 2 \delta \varphi) &\approx \sin 2 \varphi_0 \approx \pm 1, \\
\cos(2 \varphi_0 + 2 \delta \varphi) &\approx \cos 2 \varphi_0 - 2 \delta \varphi \approx -2 \delta \varphi,
\end{aligned}
\end{equation}
and for the energy terms:
\begin{equation}
\begin{aligned}
E_{\mathbf{k}\perp} &\approx \sqrt{g^2 k_F^2 + 2 \xi_0 \delta \xi + 2 \Delta^2 \sin \theta_{\mathbf{k}} \cos \theta_{\mathbf{k}} \delta \theta} \\
&\approx g k_F + \frac{\xi_0}{g k_F} \delta \xi + \frac{\Delta^2 \sin \theta_0 \cos \theta_0}{g k_F} \delta \theta, \\
E_{\mathbf{k}-} &\approx \sqrt{\left( \frac{\xi_0}{g k_F} \delta \xi + \frac{\Delta^2 \sin \theta_0 \cos \theta_0}{g k_F} \delta \theta \right)^2 + 4 \Delta^2 \sin^4 \theta_0 \delta \varphi^2}.
\end{aligned}
\end{equation}
Substituting these approximations, the integral simplifies to:
\begin{widetext}
\begin{equation}
\begin{aligned}
&\mu_B^2 \sum_{\mathbf{k} \text{ near } \mathbf{k}_0} \frac{d_{12}^2}{E_{\mathbf{k}\perp}^2} \sin^2 \theta_{\mathbf{k}} \sin^2 \Omega_{\mathbf{k}} \cdot \frac{1}{2 E_{\mathbf{k}-}} \\
&= \frac{N(0) \mu_B^2}{8 \pi} \cdot \frac{\Delta^2 \sin^4 \theta_0}{g^2 k_F^2} \int_0^{\delta_1} d(\delta \xi) \int_0^{\delta_2} \sin \theta_0 \, d(\delta \theta) \int_0^{\delta_3} d(\delta \varphi) \frac{1}{\sqrt{\left( \frac{\xi_0}{g k_F} \delta \xi  +  \frac{\Delta^2 \sin \theta_0 \cos \theta_0}{g k_F} \delta \theta \right)^2 + 4 \Delta^2 \sin^4 \theta_0 \delta \varphi^2}}.
\end{aligned}
\end{equation}
\end{widetext}
This integral can be recognized as:
\begin{equation*}
\int_0^{\delta_1} dx \int_0^{\delta_2} dy \int_0^{\delta_3} dz \frac{1}{\sqrt{(x + y)^2 + z^2}},
\end{equation*}
which converges for all points on the nodal lines. However, setting $\xi_0 = 0$ yields:
\begin{equation*}
\int_0^{\delta_1} dx \int_0^{\delta_2} dy \int_0^{\delta_3} dz \frac{1}{\sqrt{y^2 + z^2}},
\end{equation*}
which remains convergent. At $g k_F = \Delta$, setting $\xi_0 = 0$ and $\theta_0 = \pi/2$ results in:
\begin{equation*}
\int_0^{\delta_1} dx \int_0^{\delta_2} dy \int_0^{\delta_3} dz \frac{1}{z},
\end{equation*}
indicating divergence.

To identify the specific form of divergence near $g k_F = \Delta$, we set $\xi_0 = 0$ and:
\begin{equation}
\begin{aligned}
\sin^2 \theta_0 &= \frac{g^2 k_F^2}{\Delta^2}, \quad \cos^2 \theta_0 = 1 - \frac{g^2 k_F^2}{\Delta^2}.
\end{aligned}
\end{equation}
The integral near this point becomes:
\begin{widetext}
\begin{equation}
\begin{aligned}
&\int_0^{\delta_2} d(\delta \theta) \int_0^{\delta_3} d(\delta \varphi) \frac{1}{\sqrt{(\Delta^2 - g^2 k_F^2) \delta \theta^2 + 4 \Delta^2 \delta \varphi^2}} \\
&= \frac{1}{2 \Delta} \int_0^{\delta_2} d(\delta \theta) \ln \left( \frac{2 \Delta \delta_3 + \sqrt{4 \Delta^2 \delta_3^2 + (\Delta^2 - g^2 k_F^2) \delta \theta^2}}{\sqrt{(\Delta^2 - g^2 k_F^2) \delta \theta^2}} \right) \\
&= \frac{1}{2 \Delta \sqrt{\Delta^2 - g^2 k_F^2}} \left[ \sqrt{\Delta^2 - g^2 k_F^2} \delta_2 \ln \left( \frac{2 \Delta \delta_3 + \sqrt{4 \Delta^2 \delta_3^2 + (\Delta^2 - g^2 k_F^2) \delta_2^2}}{\sqrt{\Delta^2 - g^2 k_F^2} \delta_2} \right) + 2 \Delta \delta_3 \ln \left( \frac{\sqrt{\Delta^2 - g^2 k_F^2} \delta_2 + \sqrt{4 \Delta^2 \delta_3^2 + (\Delta^2 - g^2 k_F^2) \delta_2^2}}{2 \Delta \delta_3} \right) \right].
\end{aligned}
\end{equation}
\end{widetext}
As $g k_F \to \Delta$, this diverges as:
\begin{equation}
- \frac{\delta_2}{4 \Delta} \ln (\Delta^2 - g^2 k_F^2).
\end{equation}
A similar divergence occurs for $g k_F = \Delta + 0^+$, proportional to $-\ln (g^2 k_F^2 - \Delta^2)$. Thus, $\chi_{zz}(T=0)$ diverges as:
\begin{equation}
\chi_{zz}(T=0) \propto - \ln | \Delta^2 - g^2 k_F^2 |.
\end{equation}

For finite temperatures ($T > 0$), the factor $\frac{1}{E_{\mathbf{k}-}}$ is replaced by $\frac{1 - 2f(E_{\mathbf{k}-})}{E_{\mathbf{k}-}}$. In the limit $E_{\mathbf{k}-} \to 0$:
\begin{equation}
\lim_{E_{\mathbf{k}-} \to 0} \frac{1 - 2f(E_{\mathbf{k}-})}{E_{\mathbf{k}-}} = \frac{1}{2 k_B T},
\end{equation}
implying that when $g k_F = \Delta$, $\chi_{zz}(T) \propto \frac{1}{T}$ as $T \to 0$.

\bibliography{mag-soc}
	
\end{document}